\documentclass{elsart}

\usepackage{graphicx}

\usepackage{amssymb}
\usepackage{amsmath}
\usepackage{booktabs}

\newcommand{\fm}{\mathrm{fm}}
\newcommand{\MeV}{\mathrm{MeV}}
\newcommand{\GeV}{\mathrm{GeV}}

\renewcommand{\vec}[1]{\boldsymbol{{#1}}}
\newcommand{\unitvec}[1]{\hat{\vec{#1}}}

\newcommand{\I}{\mathrm{i}}

\newcommand{\eqdot}{\; .}
\newcommand{\eqcomma}{\; ,}

\newcommand{\bra}[1]{\big< \,{#1}\, \big|}

\newcommand{\ket}[1]{\big| \,{#1}\, \big> }

\newcommand{\braket}[2]{\big< \,{#1}\, \big| \,{#2}\, \big> }

\newcommand{\expect}[1]{\big< \, {#1} \, \big>}

\newcommand{\matrixe}[3]{\big< \,{#1}\, \big| \,{#2}\, 
\big| \,{#3}\, \big> }

\newcommand{\matrixeaa}[3]{{}_a\matrixe{#1}{#2}{#3}{}_a}
\newcommand{\matrixered}[3]{\big<\,{#1}
  \,\big|\big|\,{#2}\,\big|\big|\,{#3}\,\big>}

\newcommand{\comm}[2]{\bigl[ {#1}, {#2} \bigr]_{-} }

\newcommand{\partd}[2]{\ensuremath{ \frac{\partial #1}
{\partial #2} }}

\newcommand{\partdd}[2]{\ensuremath{ \frac{\partial^2 #1}
{\partial #2^2} }}

\newcommand{\intd}[1]{\int\!\mathrm{d}{#1}\;}
\newcommand{\intdt}[1]{\int\!\mathrm{d}^3{#1}\;}

\newcommand{\hermit}[1]{{{#1}}^{\dag}}

\newcommand{\op}[1]{{#1}}
\newcommand{\hop}[1]{\hermit{\op{#1}}}

\newcommand{\conop}[1]{\op{#1}^{\dagger}}
\newcommand{\desop}[1]{\op{#1}^{\phantom{\dagger}}}

\newcommand{\rhoone}{\rho^{(1)}}
\newcommand{\rhotwo}{\rho^{(2)}}

\newcommand{\eqrrep}{\overset{\scriptstyle{\vec{r}}}{\Rightarrow}}

\newcommand{\couple}[3]{\left\{ {#1} \: {#2} \right\}^{(#3)}}
\newcommand{\coupletensor}[3]{\left\{ {#1} \otimes {#2} \right\}^{(#3)}}

\newcommand{\couplev}[3]{({#1}\,{#2})^{(#3)}}

\newcommand{\sixj}[6]{\left\{ \begin{matrix} {#1} & {#3} & {#5}\\ {#2}
& {#4} & {#6} \end{matrix} \right\}}
\newcommand{\ninej}[9]{\left\{ \begin{matrix} {#1} & {#4} & {#7}\\ {#2}
& {#5} & {#8} \\ {#3} & {#6} & {#9} \end{matrix} \right\}}
\newcommand{\cg}[6]{\mathrm{C}\Biggl(\,\begin{matrix} {#1} & {#3} \\ {#2}
& {#4} \end{matrix}\, \Biggr|\Biggl.\, \begin{matrix} {#5} \\
{#6} \end{matrix} \,\Biggr)}

\newcommand{\talmi}[4]{\left< \begin{matrix} {#1} \\ {#2} \end{matrix}
    \right|\left. \begin{matrix} {#3} \\ {#4} \end{matrix} \right>}

\newcommand{\corr}[1]{\hat{#1}}

\newcommand{\cop}[1]{\op{\corr{#1}}}

\newcommand{\opC}{\op{C}}
\newcommand{\hopC}{\hop{C}}
\newcommand{\opCr}{\op{C}_{r}^{\phantom{\dagger}}}
\newcommand{\hopCr}{\op{C}_{r}^{\dagger}}
\newcommand{\opCom}{\op{C}_{\Omega}^{\phantom{\dagger}}}
\newcommand{\hopCom}{\op{C}_{\Omega}^{\dagger}}
\newcommand{\opG}{\op{G}}
\newcommand{\opGr}{\op{G}_{r}}
\newcommand{\opGom}{\op{G}_{\Omega}}

\newcommand{\sopLom}{\mathsf{L}_{\Omega}}

\newcommand{\opc}{\op{c}}
\newcommand{\hopc}{\hop{c}}
\newcommand{\opcr}{\op{c}_{r}^{\phantom{\dagger}}}
\newcommand{\hopcr}{\op{c}_{r}^{\dagger}}
\newcommand{\opcom}{\op{c}_{\Omega}^{\phantom{\dagger}}}
\newcommand{\hopcom}{\op{c}_{\Omega}^{\dagger}}
\newcommand{\opg}{\op{g}}
\newcommand{\opgr}{\op{g}_{r}}
\newcommand{\opgom}{\op{g}_{\Omega}}

\newcommand{\opone}[1]{\op{#1}^{[1]}}
\newcommand{\copone}[1]{\cop{#1}^{[1]}}

\newcommand{\coptwo}[1]{\cop{#1}^{[2]}}

\newcommand{\icm}{\mathrm{cm}}
\newcommand{\irel}{\mathrm{rel}}
\newcommand{\intr}{\mathrm{intr}}

\newcommand{\vecr}{\vec{r}}
\newcommand{\opvecr}{\op{\vec{r}}}
\newcommand{\vecx}{\vec{x}}
\newcommand{\opvecx}{\op{\vec{x}}}
\newcommand{\vecX}{\vec{X}}

\newcommand{\vecp}{\vec{p}}
\newcommand{\opvecp}{\op{\vec{p}}}

\newcommand{\vecl}{\vec{l}}
\newcommand{\opvecl}{\op{\vec{l}}}

\newcommand{\vecpr}{\vec{p}_{r}}
\newcommand{\opvecpr}{\op{\vec{p}}_{r}}
\newcommand{\pr}{p_{r}}
\newcommand{\oppr}{\op{p}_{r}}

\newcommand{\pom}{p_{\Omega}}
\newcommand{\vecpom}{\vec{p}_{\Omega}}
\newcommand{\opvecpom}{\op{\vec{p}}_{\Omega}}

\newcommand{\lsq}{\vecl^2}
\newcommand{\oplsq}{\op{\vecl}^2}

\newcommand{\srpom}{s_{\!12}(\vecr,\vecpom)}
\newcommand{\opsrpom}{\op{s}_{\!12}(\vecr,\vecpom)}

\newcommand{\Rp}{R_{+}}
\newcommand{\Rm}{R_{-}}
\newcommand{\Rpm}{R_{\pm}}
\newcommand{\Rmp}{R_{\mp}}

\newcommand{\cmur}{\hat{\mu}_{r}}
\newcommand{\cmuom}{\hat{\mu}_{\Omega}}

\newcommand{\opSone}{\op{S}^{(1)}}
\newcommand{\opStwo}{\op{S}^{(2)}}

\newcommand{\ls}{\vecl\!\cdot\!\vec{s}}
\newcommand{\lssq}{(\ls)^2}
\newcommand{\srr}{s_{\!12}(\unitvec{r},\unitvec{r})}
\newcommand{\sll}{s_{\!12}(\vecl,\vecl)}

\newcommand{\opPinot}{\op{\Pi}_{0}}
\newcommand{\opPione}{\op{\Pi}_{1}}
\newcommand{\opls}{\op{\vecl}\!\cdot\!\op{\vec{s}}}
\newcommand{\oplssq}{(\opls)^2}
\newcommand{\opsrr}{\op{s}_{\!12}(\unitvec{r},\unitvec{r})}
\newcommand{\opsll}{\op{s}_{\!12}(\vecl,\vecl)}
\newcommand{\opspompom}{\op{s}_{\!12}(\vecpom,\vecpom)}
\newcommand{\opsbarpompom}{\op{\bar{s}}_{\!12}(\vecpom,\vecpom)}

\newcommand{\opLsq}{\op{\vec{L}}^2}

\newcommand{\lone}{\op{l}^{(1)}}

\newcommand{\rpomtwo}{\couplev{\op{r}}{\op{\pom}}{2}}

\newcommand{\crhoone}{\corr{\rho}^{(1)}}
\newcommand{\crhotwo}{\corr{\rho}^{(2)}}

\newcommand{\Vlowk}{\ensuremath{V_{\mathit{low-k}}}}

\newcommand{\Hefour}{\nuc{4}{He}}
\newcommand{\Osixteen}{\nuc{16}{O}}
\newcommand{\Cafourty}{\nuc{40}{Ca}}

\newcommand{\tautau}{\vec{\tau}\! \cdot\! \vec{\tau}}
\newcommand{\sigsig}{\vec{\sigma}\! \cdot \! \vec{\sigma}}

\begin{document}

\begin{frontmatter}

\title{Tensor correlations\\ in the Unitary Correlation Operator Method}

\author{T. Neff\thanksref{tn} and H. Feldmeier\thanksref{hf}}

\thanks[tn]{email: t.neff@gsi.de, http://theory.gsi.de/\~{}tneff/}
\thanks[hf]{email: h.feldmeier@gsi.de, http://www.gsi.de/\~{}feldm/}

\address{Gesellschaft f\"ur Schwerionenforschung Darmstadt mbH,
Planckstra\ss e 1,\\ D-64291 Darmstadt, Germany} %

\begin{abstract}
  
  We present a unitary correlation operator that explicitly induces
  into shell model type many-body states short ranged two-body
  correlations caused by the strong repulsive core and the pronounced
  tensor part of the nucleon-nucleon interaction.  Alternatively an
  effective Hamiltonian can be defined by applying this unitary
  correlator to the realistic nucleon-nucleon interaction.  The
  momentum space representation shows that realistic interactions
  which differ in their short range behaviour are mapped on the same
  correlated Hamiltonian, indicating a successful provision for the
  correlations at high momenta.  Calculations for {\Hefour} using the
  one- and two-body part of the correlated Hamiltonian compare
  favorably with exact many-body methods.  For heavier nuclei like
  {\Osixteen} and {\Cafourty} where exact many-body calculations are
  not possible we compare our results with other approximations.  The
  correlated single-particle momentum distributions describe the
  occupation of states above the Fermi momentum. The Unitary
  Correlation Operator Method (UCOM) can be used in mean-field and
  shell model configuration spaces that are not able to describe these
  repulsive and tensor correlations explicitly.

\end{abstract}


\end{frontmatter}

\section{Introduction and summary}

Quantum Chromo Dynamics (QCD) is the fundamental theory of the strong
interaction and the nucleons represent bound systems of quark and
gluon degrees of freedom. Nuclear physics in the low energy regime is
considered as an effective theory where the center of mass positions,
the spins and the isospins of the nucleons are the essential degrees
of freedom, whose interaction can be described by a nucleon-nucleon
force. In the QCD picture the force between the color neutral nucleons
is a residual interaction like the van-der-Waals force between
electrically neutral atoms. Therefore it is expected that the nuclear
interaction is not a simple local potential but has a rich operator
structure in spin and isospin and in many-body systems may also
include genuine three- and higher-body forces.

There are attempts to derive the nucleon-nucleon force using Chiral
Perturbation Theory \cite{entem01}. However this approach cannot
compete yet with the so-called realistic interactions that reproduce
the nucleon-nucleon scattering data and the deuteron properties. The
realistic nucleon-nucleon forces are essentially phenomenological. The
Bonn interactions \cite{machleidt89,machleidt01} are based on
meson-exchange that is treated in a relativistic nonlocal fashion.
The Argonne interactions \cite{wiringa84,wiringa95} on the other hand
describe the pion exchange in a local approximation and use a purely
phenomenological parameterization of the nuclear interaction at short
and medium range.

It is a central challenge of nuclear physics to describe the
properties of nuclear many-body systems in terms of such realistic
nuclear interactions. However in mean-field and shell-model
approaches, typically employed to describe the properties of finite
nuclei, realistic interactions cannot be used directly.

Both, relativistic and nonrelativistic mean-field calculations are
successful in describing the ground state energies and mass and charge
distributions for all nuclei in the nuclear chart but the lightest
ones.  As the short-range radial and tensor correlations induced by
realistic forces cannot be represented by the Slater determinants of
the Hartree-Fock method, direct parameterizations of the
energy-density or an effective finite range force are used instead.

In shell-model calculations with configuration mixing a two-body
Hamiltonian is used in the vector space spanned by the many-body
states that represent particle and/or hole configurations in the
selected shells. The solution of the energy eigenvalue problem in the
high-dimensional many-body space yields detailed information on the
spectra of nuclei, the transitions between the states, electromagnetic
moments, charge and mass distributions, $\beta$-decay etc. However one
has to use an effective interaction in the two-body Hamiltonian.
Although one often starts with a G-matrix derived from realistic
interactions the two-body matrix elements of the Hamiltonian have to
be modified and adjusted to a large set of ground state and excitation
energies for an accurate description of the data. This has to be done
separately for each region of the nuclear chart.

Only recently it became possible to perform \emph{ab initio}
calculations of the nuclear many-body problem with realistic
interactions.  In Green's Function Monte Carlo (GFMC) calculations
\cite{pieper01mp} the exact ground-state wave function is calculated
by approximating the many-body Green's functions in a Monte Carlo
approach.  The GFMC calculations of light nuclei up to $A=8-10$ with
the Argonne interaction demonstrate the necessity of additional
three-body forces in order to reproduce the experimental nuclear
binding energies and radii as well as the spectra.

Another \emph{ab initio} approach for nuclei up to $A=12$ is the
large-basis no core shell model \cite{barrett00}. All nucleons are
treated as active in a large shell-model basis. Despite the large
basis it is necessary to treat the short range correlations
separately. An effective interaction for the model space is derived in
a G-Matrix procedure in two-body approximation.

\emph{Ab inito} calculations for the doubly magic nuclei $\Osixteen$
and $\Cafourty$ are performed with the Correlated Basis Function (CBF)
method \cite{fabrocini98,fabrocini00}. Here a perturbation expansion
on a complete set of correlated basis functions is performed. The
evaluation of expectation values is done in the Fermi Hypernetted
Chain (FHNC) method where the Single Operator Chain (SOC)
approximation is used.

\subsection{Aim}
  
Our aim is to perform \emph{ab initio} calculations of larger nuclei
with realistic interactions in a mean-field or a shell-model many-body
approach. To make this possible we introduce a unitary correlation
operator $\opC$ that takes care of the short-range radial and tensor
correlations. The correlations are not expressed in a certain basis of
a model space but are given analytically in terms of operators of
relative distance, relative momentum and the spins and isospins of the
nucleons.  The correlated interaction
\begin{equation}
  \cop{H} = \hopC \op{H} \opC
\end{equation}
we obtain by applying the correlation operator to the realistic
interaction is therefore not restricted to the model space of a
certain many-body theory but can be used for example in a Hartree-Fock
calculation as well as in shell-model calculations with configuration
mixing. The fact, that the correlations are expressed in a basis-free
manner in coordinate space, makes it also easier to understand the
physics of the radial and tensor correlations.

Furthermore the unitary correlation operator provides not only a
correlated interaction, that can be considered as an effective
interaction, but other operators can be correlated as well and the
physical implications of the short-range correlations on other
observables, for example on the nucleon momentum distributions or the
spectroscopic factors, can be evaluated.

The correlated interaction is used successfully with simple shell
model and Fermionic Molecular Dynamics \cite{fmd00,neff:diss} Slater
determinants. This allows us to perform calculations for all nuclei up
to about $A=50$. Although a single Slater determinant as many-body
trial state is the most simple ansatz we obtain results very close to
those of the quasi-exact methods.

\subsection{Procedure}

The repulsive core and the strong tensor force of the nuclear
interaction induce strong short-range radial and tensor correlations
in the nuclear many-body system. These correlations cannot be
represented with Slater determinants of single-particle states
\begin{equation}
  \ket{\Psi} = \mathcal{A} \left\{ \ket{\psi_1} \otimes \ldots \otimes
    \ket{\psi_A} \right\} \eqcomma
\end{equation}
which are used as many-body states in Hartree-Fock or a shell-model
calculations. $\mathcal{A}$ denotes the antisymmetrization operator
and $\ket{\psi_i}$ the single-particle states.

We describe the radial and tensor correlations by a unitary
correlation operator that is the product of a radial correlator
$\opCr$ and a tensor correlator $\opCom$.
\begin{equation} \label{eq:ComCr}
  \opC = \opCom \opCr
\end{equation}

The radial correlator $\opCr$ (described in detail in \cite{ucom98})
shifts a pair of particles in the radial direction away from each
other so that they get out of the range of the repulsive core.  To
perform the radial shifts the generator of the radial correlator uses
the radial momentum operator $\oppr$ together with a shift function
$s(r)$ that depends on the distance of the two nucleons.
\begin{equation}
  \opCr = \exp \Bigl\{ -\I \sum_{i<j} \frac{1}{2}
  \bigl( \oppr{}_{ij} s(r_{ij}) + s(r_{ij}) \oppr{}_{ij} \bigr) \Bigr\} 
  \label{eq:Cr}
\end{equation}
The shift will be strong for short distances and will vanish at large
distances.

\begin{figure}[htb]
  \includegraphics[width=0.48\textwidth]{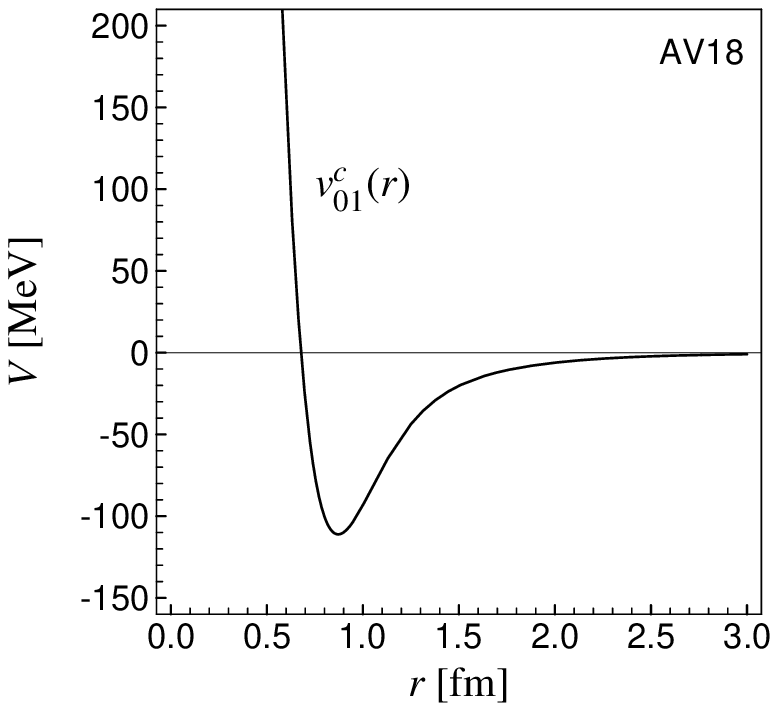}\hfil
  \includegraphics[width=0.48\textwidth]{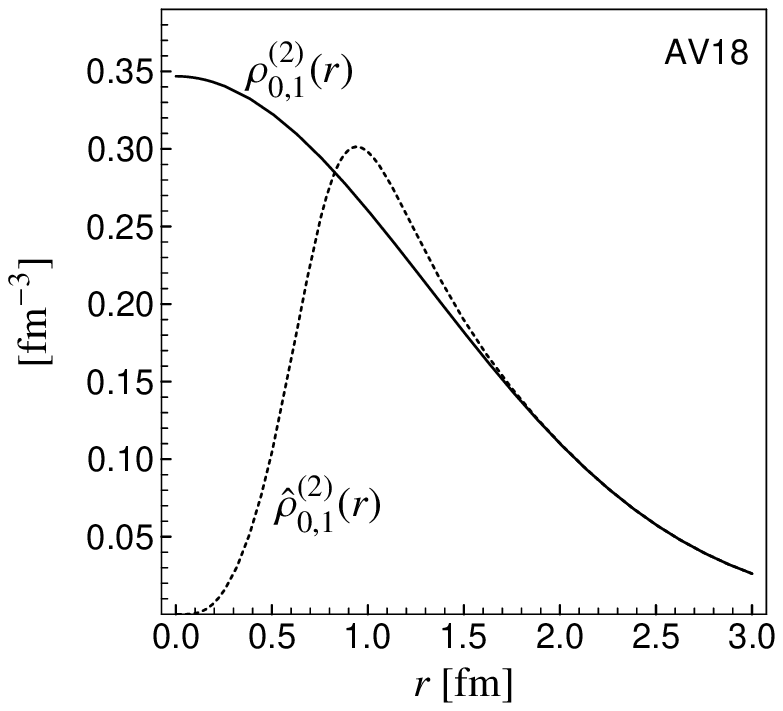}
  \caption{The Argonne~V18 potential in the $S,T=0,1$ channel
    is plotted on the left hand side. Its strong repulsion at short
    distances causes a pronounced depletion in the correlated two-body
    density $\crhotwo_{0,1}$ when compared with the uncorrelated
    two-body density $\rhotwo_{0,1}$ of the $\Hefour$ shell-model
    state.} 
  \label{fig:correlationhole}
\end{figure}

To illustrate the short-range radial correlations in the nucleus the
two-body density of the $\Hefour$ nucleus and the corresponding
potential in the $S,T=0,1$ channel are shown in
Fig.~\ref{fig:correlationhole}. (In the $S=0$ channels $\opCom =
\op{1}$.)  The two-body density $\crhotwo_{0,1}$ of the correlated
trial state, plotted as a function of the distance $r$ between two
nucleons, is suppressed in the region of the repulsive core of the
interaction and shifted outwards. This correlation hole is completely
absent in the two-body density $\rhotwo_{0,1}$ of the uncorrelated
shell-model trial state. The radial correlations are explained in
Sec.~\ref{sec:centralcorrelations}.

The tensor force in the $S=1$ channels of the nuclear interaction
depends on the spins and the spatial orientation
$\unitvec{r}=(\vec{r}_1-\vec{r}_2)/(|\vec{r}_1-\vec{r}_2|)$ of the
nucleons via the tensor operator
\begin{equation}
  S_{\!12}(\unitvec{r},\unitvec{r}) = 
  3 (\vec{\sigma}_1\cdot\unitvec{r})(\vec{\sigma}_2\cdot\unitvec{r}) -
  (\vec{\sigma}_1\cdot\vec{\sigma}_2) = 
  2\ \bigl( 3 (\vec{S}\cdot\unitvec{r})^2 - \vec{S}^2 \bigr) \eqdot
\end{equation}
An alignment of $\unitvec{r}$ with the direction of total spin
$\vec{S}= \frac{1}{2}(\vec{\sigma}_1+\vec{\sigma}_2)$ is favored
energetically.
The tensor correlator $\opCom$, defined as
\begin{equation} 
  \opCom = \exp \Bigl\{ -\I \sum_{i<j} \vartheta(r_{ij})
  \frac{3}{2}\bigl(
  (\vec{\sigma}_i \vec{p}_{\Omega ij})(\vec{\sigma}_j \vecr_{ij}) +
  (\vec{\sigma}_i \vecr_{ij})(\vec{\sigma}_j \vec{p}_{\Omega ij})
  \bigr) \Bigr\} \eqcomma
  \label{eq:Com}
\end{equation}
achieves this alignment by shifts perpendicular to the relative
orientation $\unitvec{r}_{ij}$.  For that the generator of the tensor
correlator uses a tensor operator constructed with the orbital part of
the momentum operator $\vecpom = \vecp - \vecpr$. The $r$-dependent
strength of the tensor correlations is controlled by $\vartheta(r)$.
The correlator $\opCom$ acts on the $S=1$ part of a pair in such a way
that probability is shifted towards regions where $\vec{r}_{ij}$ and
$\vec{S}$ are aligned which implies more binding from the tensor
interaction.

\begin{figure}[tb]
 \includegraphics[width=\textwidth]{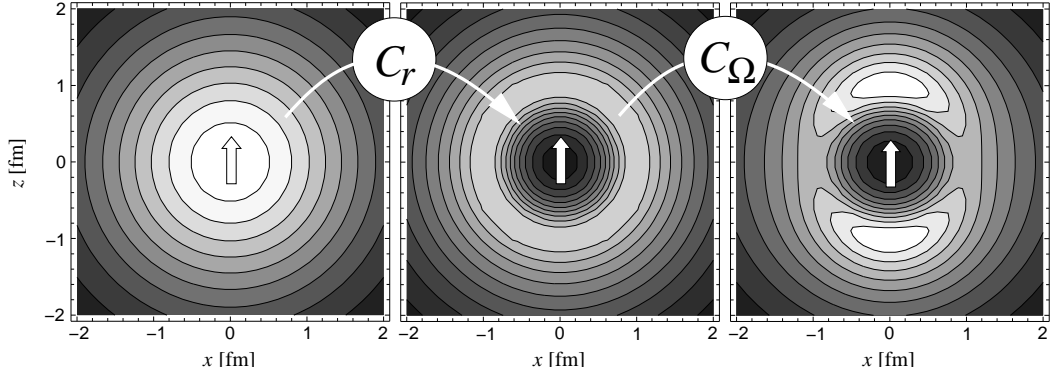}
  \caption{Two-body density $\rhotwo_{S,M_S,T,M_T}(\vec{r})$ in the
    $S,M_S=1,1$; $T,M_T=0,0$ channel of $\Hefour$.  Left: uncorrelated
    trial state $\ket{\Psi}$, middle: after radial correlation
    $C_r \ket{\Psi}$, and right: with additional tensor correlations
    $\ket{\hat{\Psi}}=C_{\Omega} C_r \ket{\Psi}$.  The arrow indicates
    the direction of the spin $\vec{S}$.  The correlation operator is
    determined for the Argonne~V18 interaction.}
  \label{fig:CrCom}
\end{figure}

In Fig.~\ref{fig:CrCom} the actions of $\opCr$ and $\opCom$ are
illustrated in the $S=1, M_S=1; T=0$ channel of the $\Hefour$ two-body
density. The arrow indicates the direction of the total spin
$\vec{S}$. The contour plots show on the left the density calculated
with a shell-model state with four nucleons in the $s$-shell. It
exhibits a maximum at zero distance where the interaction is most
repulsive. The action of the radial correlator $\opCr$ (middle frame)
corrects this unphysical property by shifting probability radially
outwards in order to accommodate the repulsive core of the two-body
potential. The subsequent application of $\opCom$ (see
Eq.~(\ref{eq:ComCr})) results in the tensor correlations as shown in
the third frame. 

The graph also demonstrates that $\opCom$ as defined in
Eq.~(\ref{eq:Com}) moves probability perpendicular to $\vec{r}$ from
the ``equator'' to the ``poles''. The spherical distribution
transforms into an axially symmetric one with enhanced probability in
regions where $\vec{r}$ and $\vec{S}$ are parallel. An even stronger
alignment would bring more binding from the tensor force but costs at
the same time kinetic energy because the nucleons are more localized.

The correlated many-body trial state
\begin{equation}
  \ket{\corr{\Psi}} = \opC \ket{\Psi}
\end{equation}
consists of two parts, the correlator $\opC$ and the uncorrelated
trial state $\ket{\Psi}$. In the sense of the Ritz variational
principle both can be varied. The optimal correlator will however
depend on the restrictions imposed on $\ket{\Psi}$.  Or in other
words: The more variational freedom is in $\ket{\Psi}$ the less
remains for $\opC$.  It is important to note that for trial states
$\ket{\Psi}$ consisting of a superposition of few or many or even very
many Slater determinants the corresponding correlators $\opC$ differ
only in their long range behavior and are very similar at short
distances.

The correlator is the exponential of a two-body operator and therefore
the correlated Hamilton operator contains not only one- and two-body
but in principle also higher-order contributions.  If the correlators
are of short-range and the densities are not too high, i.e. the mean
distance of the nucleons is larger than the correlation length, the
two-body approximation, where only the one- and two-body contributions
are taken into account, is well justified.  In this two-body
approximation the radial and the tensor correlations are evaluated
analytically in the angular momentum representation without further
approximations in Sec.~\ref{sec:tensorinangmomeigenstates}.

In Sec.~\ref{sec:bonnargonnecorrelators} we apply the unitary
correlation operator method (UCOM) to the Bonn-A and Argonne~V18
interaction which agree in their phase shifts but differ with respect
to short-range repulsion and the strength of the tensor interaction.
Based on this we perform in Sec.~\ref{sec:manybody} \emph{ab initio}
calculations for the doubly magic nuclei $\Hefour$, $\Osixteen$ and
$\Cafourty$ where we use a single Slater determinant of harmonic
oscillator shell-model states.
\begin{figure}[tb]
  \begin{center}
    \includegraphics[width=0.9\textwidth]{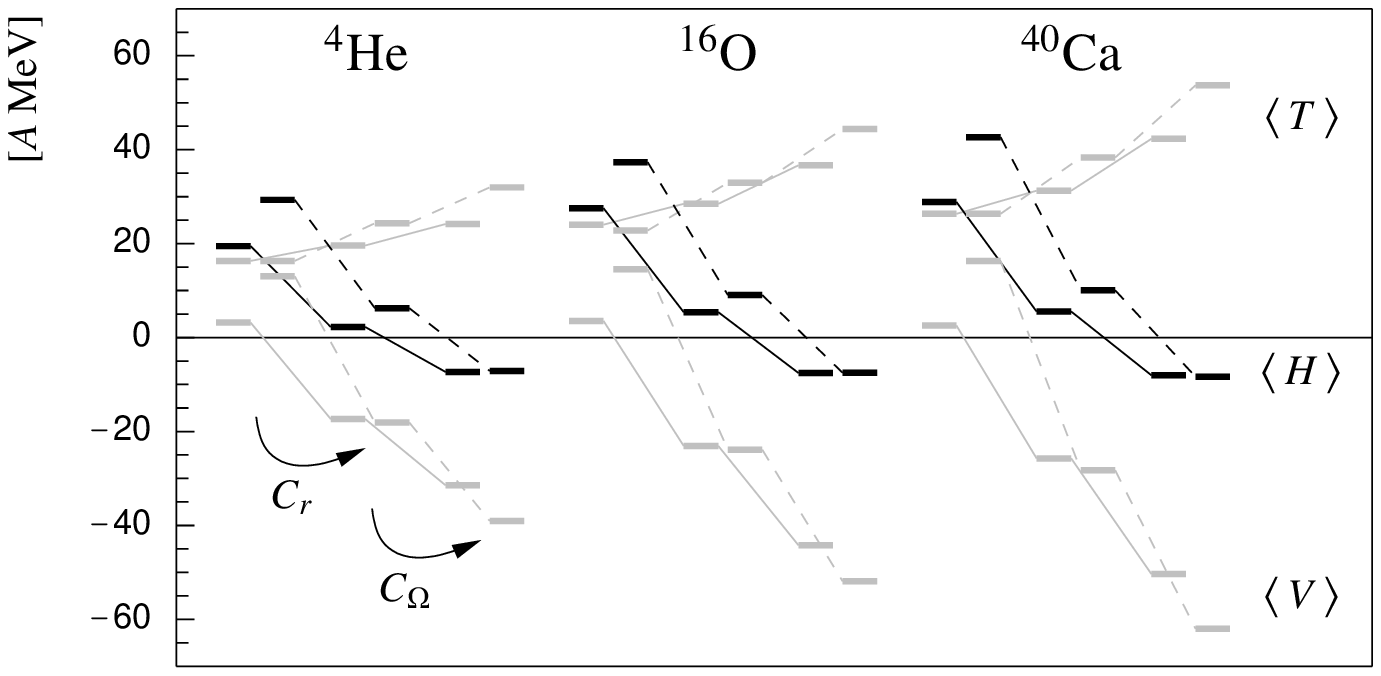}
    \caption{Kinetic $\expect{\op{T}}$, potential $\expect{\op{V}}$
      and total energies $\expect{\op{H}}$ obtained with the Bonn-A
      (left bars in each column connected by full lines) and the
      Argonne~V18 (right bars connected by dashed lines) interaction
      for the doubly-magic nuclei $\Hefour$, $\Osixteen$ and
      $\Cafourty$. For each nucleus the energies obtained with
      uncorrelated, radially correlated and finally with radially and
      tensorially correlated trial states are displayed.}
    \label{fig:energies}
  \end{center}
\end{figure}
The effect of the unitary correlator on the kinetic and potential
energy is summarized in Fig.~\ref{fig:energies}. In case of the Bonn-A
interaction the radial correlator $\opCr$ that tames the repulsive
core reduces the potential energy by about $20-30\;\MeV$ per nucleon
in all three nuclei compared to the expectation value of the bare
interaction. At the same time the kinetic energy rises only by about
$5\;\MeV$ per nucleon. The nuclei are however still unbound after
inclusion of the radial correlations (depicted in
Fig.~\ref{fig:CrCom}). The introduction of the tensor correlations,
i.e. the alignment of spins along the distance vector between particle
pairs, leads to an increase in binding by about $15-25\;\MeV$ per
nucleon while the kinetic energy goes up by $5-10\;\MeV$. Now the
nuclei are bound at about $-8\;\MeV$ per nucleon. The Argonne~V18
induces stronger correlations than the Bonn-A interaction and the
radial correlations lower the potential energy by about $30-40\;\MeV$
per nucleon, the tensor correlations give additional $20-30\;\MeV$ per
nucleon. At the same time the kinetic energy is increased by
$6-12\:\MeV$ (radial correlations) and $8-15\:\MeV$ (tensor
correlations). Despite the large differences in potential and kinetic
energies the net binding energies obtained by the Bonn-A and the
Argonne~V18 interaction are almost identical.

We demonstrate in Sec.~\ref{sec:momentumspace} the effects of the
correlator in coordinate and momentum space representation and find
that the correlated Bonn-A and Argonne~V18 interactions agree very
well in glaring contrast to the bare interactions.  The correlated
interactions in momentum space are also compared with the $\Vlowk$
potential \cite{kuo01} that is obtained by integrating out the high
momentum modes. Although derived in a quite different scheme the
$\Vlowk$ potential is very similar to our correlated potential.

The case of $\Hefour$, where exact results are available, is discussed
in detail in Sec.~\ref{sec:manybody}, $\Osixteen$ and $\Cafourty$ are also
calculated.  Despite the oversimplified trial state the energy and
radii compare very favorably with other much more expensive methods.
We also observe that the correlated Argonne~V18 and Bonn-A interaction
interactions give almost identical results for the nuclei although the
corresponding uncorrelated potentials (see Sec.~\ref{sec:potentials})
and their expectation values differ greatly.  As in the GFMC
calculations for light nuclei and the CBF calculations, that are
numerically feasible only for $\Osixteen$ and $\Cafourty$, when
compared to experimental data the nuclei are not bound enough with
realistic interactions.

In Sec.~\ref{sec:momentumdistribution} we calculate the
single-particle momentum distribution of the correlated trial states
for $\Hefour$ and $\Osixteen$. We find good agreement with variational
Monte-Carlo \cite{pieper92} and spectral function analysis results
\cite{benhar94}.

We conclude from these observations that the unitary correlator
extracts the common low-momentum behaviour of the Bonn-A and the
Argonne~V18 interactions. With the unitary correlator we therefore
have successfully performed a separation of scales. The high-momentum
scale of the short-range correlations is covered by the unitary
correlator, the low-momentum or long-range behavior is described by
the uncorrelated many-body trial state

However the tensor correlations are longer in range than the radial
correlations and the separation in short-range or high-momentum
components and long-range or low-momentum components as it is possible
for the radial correlations is not as clear cut for the tensor
correlations.  We can ensure the validity of the two-body
approximation by using the unitary correlator only for the shorter
part of the tensor correlations and use an improved many-body
description for the long ranged part of the tensor correlations.
Nevertheless a decent description is also possible with very simple
many-body trial states and a long ranged tensor correlator when we
tolerate larger uncertainties due to three-body contributions to the
correlated Hamiltonian.

The fact that realistic two-body forces alone give not enough binding
in the many-body system is unfortunate because it implies that
additional genuine three-body forces have to be added. In principle
Chiral Perturbation Theory should be able to provide a derivation of
three-body forces that are consistent with the two-body forces.
Because of the complexity of this approach a more phenomenological
ansatz for three-body forces is used \cite{pieper01} whose parameters
are adjusted to reproduce the many-body properties. The three-body
contributions are small compared to the two-body ones. Because of the
large cancellations between kinetic and potential energy the
three-body forces are nevertheless important for an accurate
description of nuclei.

\subsection{Summary}

The unitary correlator provides a transparent and powerful method to
use realistic nucleon-nucleon interactions in \emph{ab initio}
calculations for larger nuclei, but with comparatively little
numerical effort.  The big advantage of realistic interactions for
nuclear structure calculations is that the spin and isospin dependence
of the correlated two-body force is fixed.  Different from effective
interactions or mean-field parameterizations of the energy-density
there are no free parameters in the two-body force.  This is
especially important for the predictions concerning exotic nuclei with
large isospins and low density tails.

Our results show in agreement with GFMC calculations of light nuclei
that the realistic two-body forces alone cannot successfully reproduce
the experimental binding energies of the nuclei.

The \emph{ab initio} calculations for $\Hefour$, $\Osixteen$ and
$\Cafourty$ demonstrate nevertheless that a correlated realistic
interaction is a very good starting point. The main problem, the short
range repulsion and the short range part of the tensor correlations is
successfully tackled by the unitary correlator. What remains to do is
the inclusion of three-body forces and improvements in the
uncorrelated trial state that take better account of the long range
correlations.

\section{Nucleon-nucleon interaction}
\label{sec:potentials}

It has been proposed by Weinberg \cite{weinberg90,weinberg91} that in
the low energy regime of nuclear physics ($\leq 1\;\GeV$) where the
appropriate degrees of freedom are the nucleons and the pions one can
describe those by an effective field theory based on broken chiral
symmetry.  In such a scheme the general structure of the
nucleon-nucleon interaction is obtained by including all terms up to a
certain order that are compatible with the symmetries of the
$\pi$-$N$-Lagrangian.  The parameters in this effective field theory
are determined from the low-energy observables like scattering data.
Work is under way to derive the short range part of the
nucleon-nucleon potential following these ideas \cite{entem01}.
Another encouraging aspect of this approach is that it allows to
determine the structure of the three-body forces on the same footing
-- at least in principle. Up today this promising method cannot yet
compete with the ``established'' nucleon-nucleon interactions in
reproducing the well known scattering data.

Two prominent interactions of the 80's, that fit the nucleon-nucleon
scattering data up to $300\;\MeV$ and the deuteron properties, are the
Bonn-A \cite{machleidt89} and the Argonne~V14 \cite{wiringa84}
interactions. With more precise and better analyzed scattering data
improved versions of these interactions which include additional
charge independence breaking and charge symmetry breaking components
have been presented in the 90's, namely the Bonn-CD \cite{machleidt01}
and the Argonne~V18 \cite{wiringa95} potential.

In the Bonn approach the nuclear interaction is described in the
meson-exchange picture. This includes also higher orders like the
correlated two pion-exchange. In the end the interaction is
parameterized by one-boson exchange in a relativistic treatment. The
nonrelativistic variant of the Bonn interaction has nonlocal momentum
dependent terms which appear as relativistic corrections. The
nonrelativistic Argonne interaction on the other hand is rather
phenomenological. The long range part of the interaction is given as
in the Bonn potential by pion exchange, the short-range part is
modeled differently, where the Bonn interaction generates repulsion by
momentum-dependent interaction terms the Argonne interaction has
repulsive local contributions. Nevertheless both describe the measured
phase shifts equally well.  The potentials are therefore identical
on-shell but they differ in their off-shell behavior which matters for
nucleons interacting inside bound nuclei. For example, three-body
forces that are added to reproduce the experimental binding energies
and radii of light nuclei \cite{pudliner97,wiringa00,pieper01} depend
on the choice of the two-body force.

Thus the nuclear force is not uniquely determined by the scattering
data.  We want to remark that the Unitary Correlation Operator Method
presented in this work allows the construction of a manifold of
two-body interactions $\cop{H} = \hopC \op{H} \opC$ which are all
phase shift equivalent because our correlator is of finite range, i.e.
$\opC (r\rightarrow \infty) = \op{1}$.

\subsection{Bonn potentials}

\begin{figure}[b]
  \includegraphics[width=0.48\textwidth]{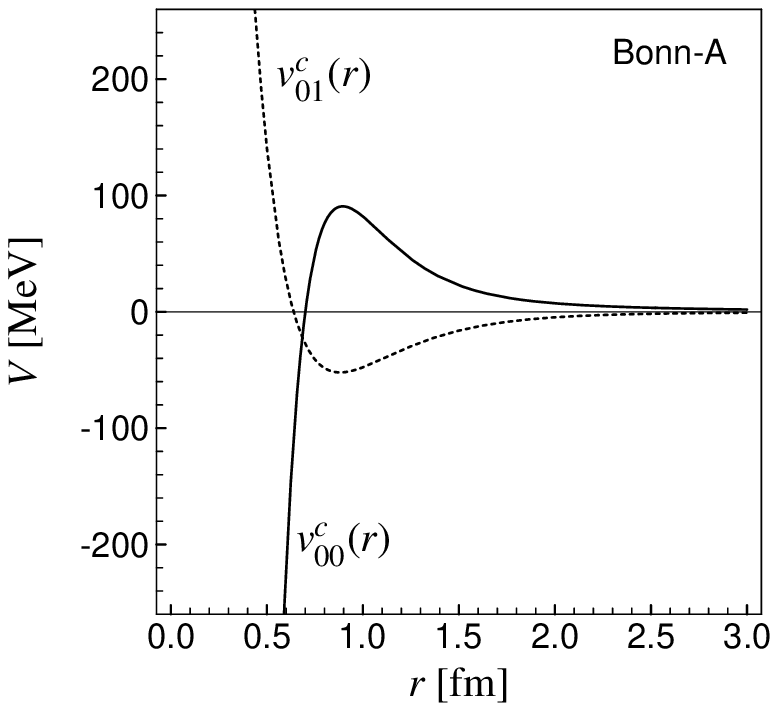}\hfil
  \includegraphics[width=0.48\textwidth]{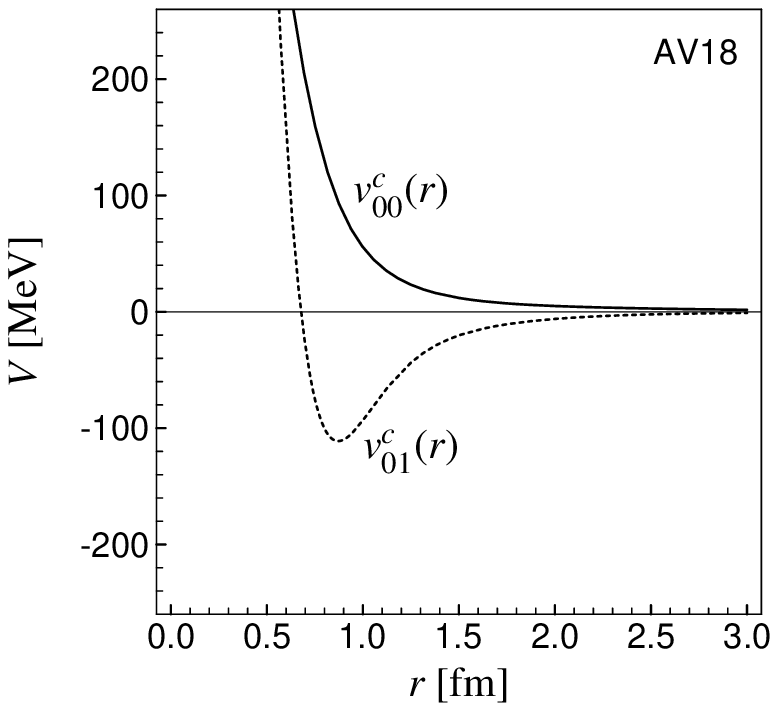}\vspace{1ex}
  \includegraphics[width=0.48\textwidth]{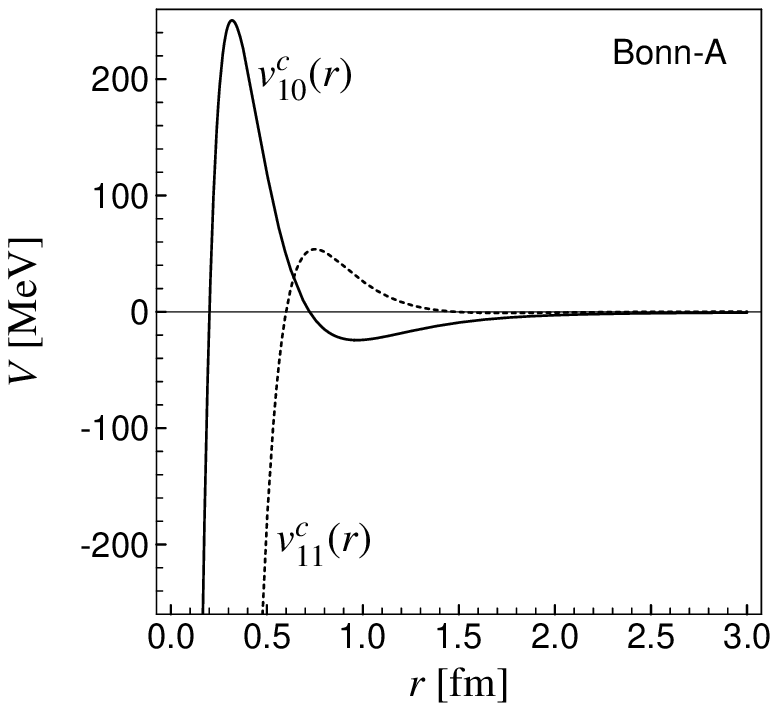}\hfil
  \includegraphics[width=0.48\textwidth]{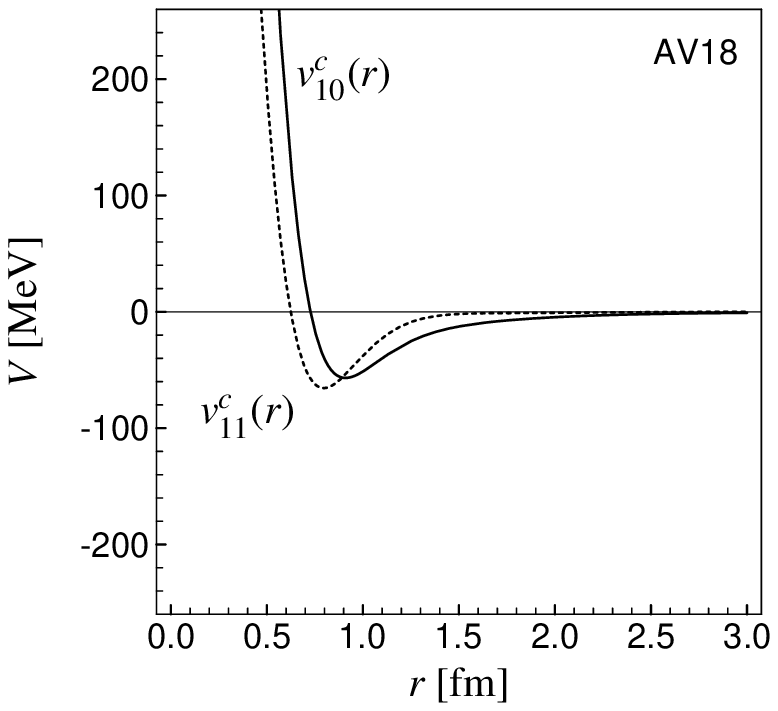}
  \caption{Central potential in the $S=0$ and the $S=1$
    channels for the Bonn-A interaction (\ref{eq:bonnarepresentation})
    (left hand side) and the Argonne~V18 interaction
    (\ref{eq:av18representation}) (right hand side).}
  \label{fig:centralpot}
\end{figure}

\begin{figure}[b]
  \includegraphics[width=0.48\textwidth]{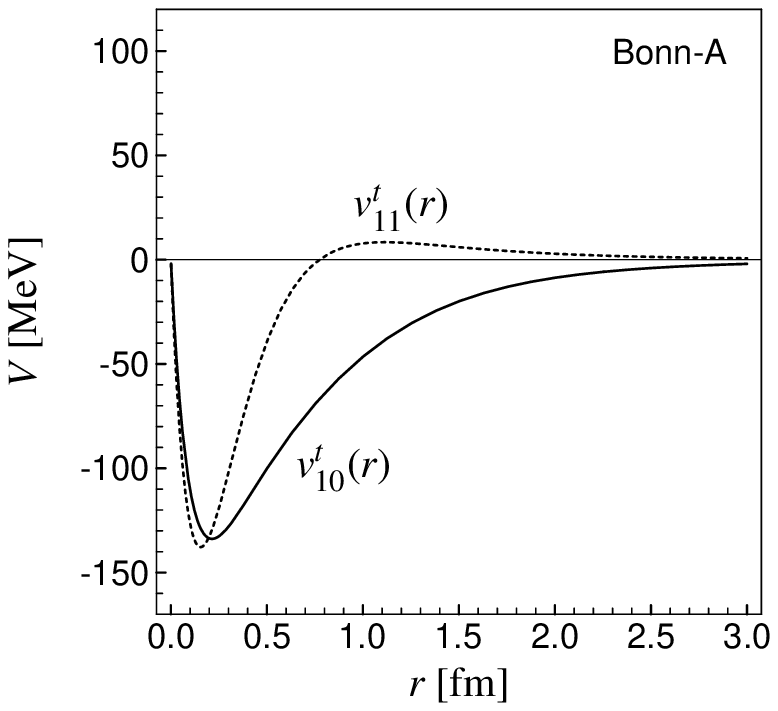}\hfil
  \includegraphics[width=0.48\textwidth]{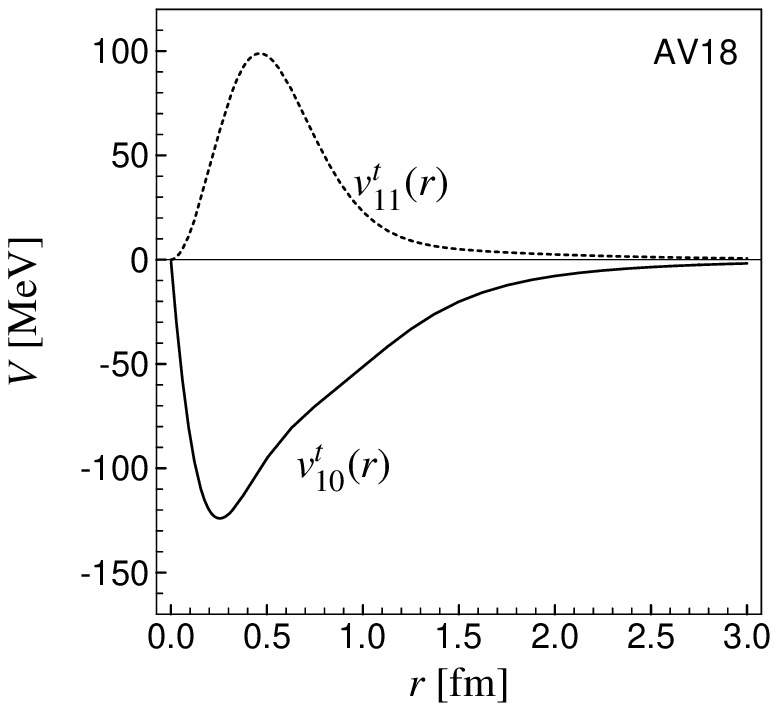}
  \caption{Tensor potential for the Bonn-A interaction
    (\ref{eq:bonnarepresentation}) (left hand side) and the
    Argonne~V18 interaction (\ref{eq:av18representation}) (right hand
    side).}
  \label{fig:tensorpot}
\end{figure}

\begin{figure}[b]
  \includegraphics[width=0.48\textwidth]{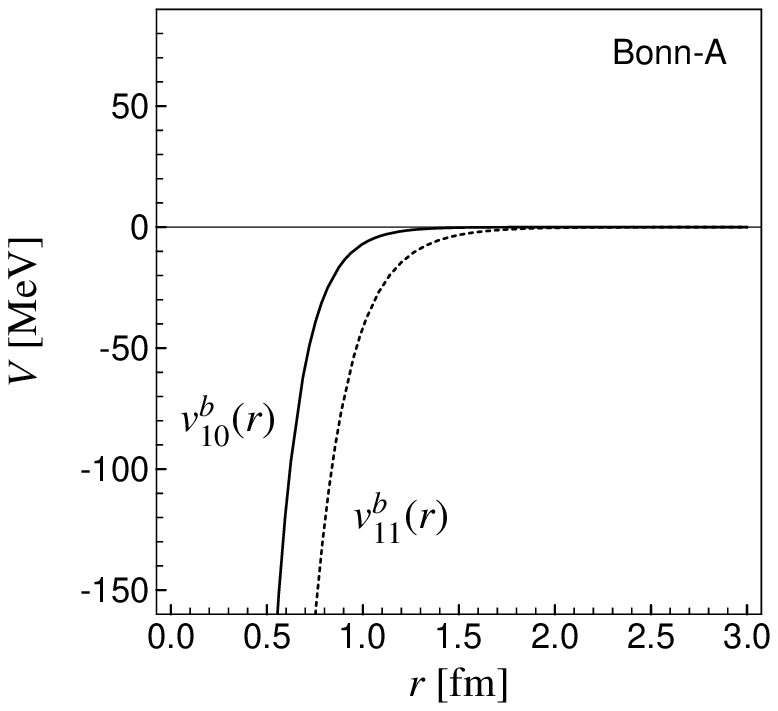}\hfil
  \includegraphics[width=0.48\textwidth]{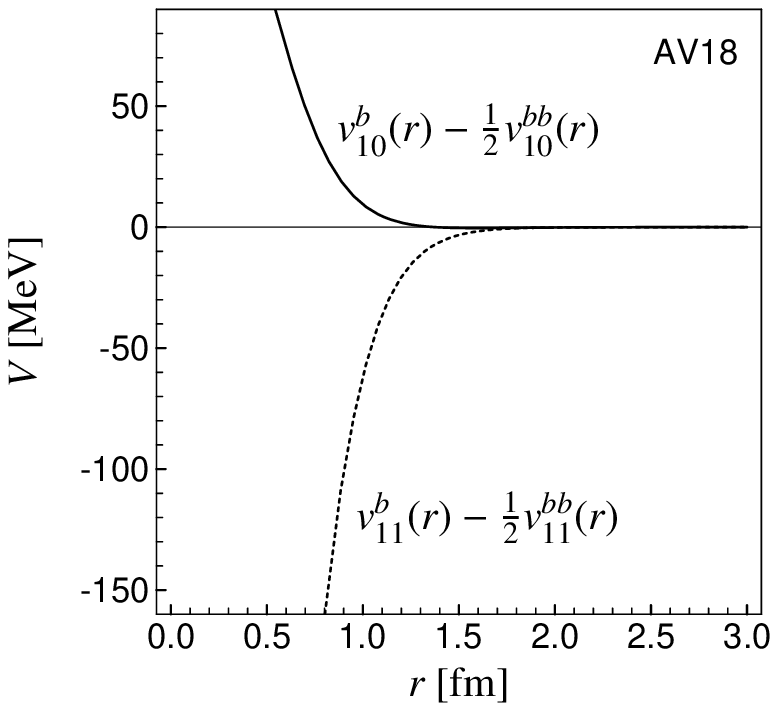}
  \caption{Spin-orbit potential for the Bonn-A interaction
    (\ref{eq:bonnarepresentation}) (left hand side) and the
    Argonne~V18 interaction (\ref{eq:av18representation}) (right hand
    side). The spin-orbit part contained in the $\lssq$ potential has
    been added for the Argonne~V18 potential.}
  \label{fig:spinorbitpot}
\end{figure}

\begin{figure}[b]
    \includegraphics[width=0.48\textwidth]{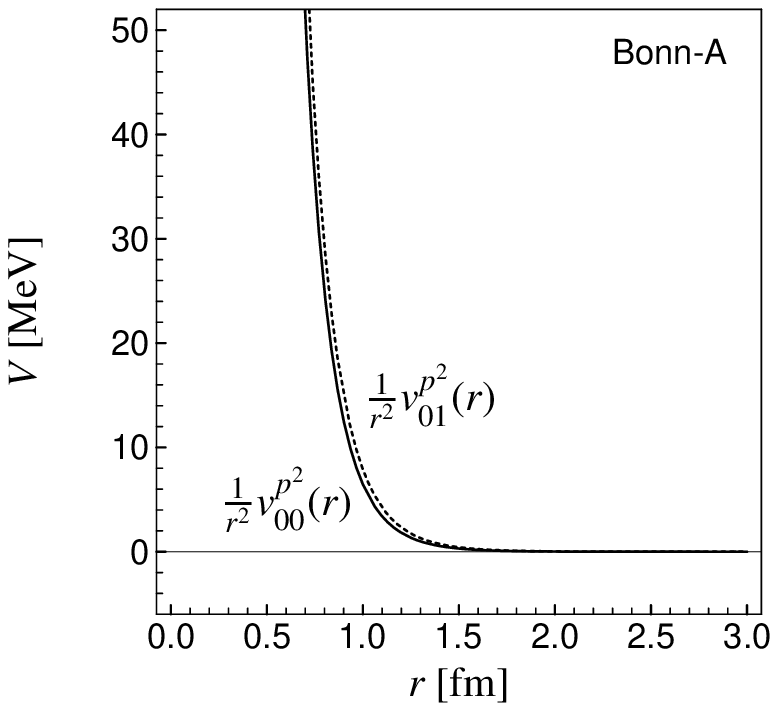}\hfil
    \includegraphics[width=0.48\textwidth]{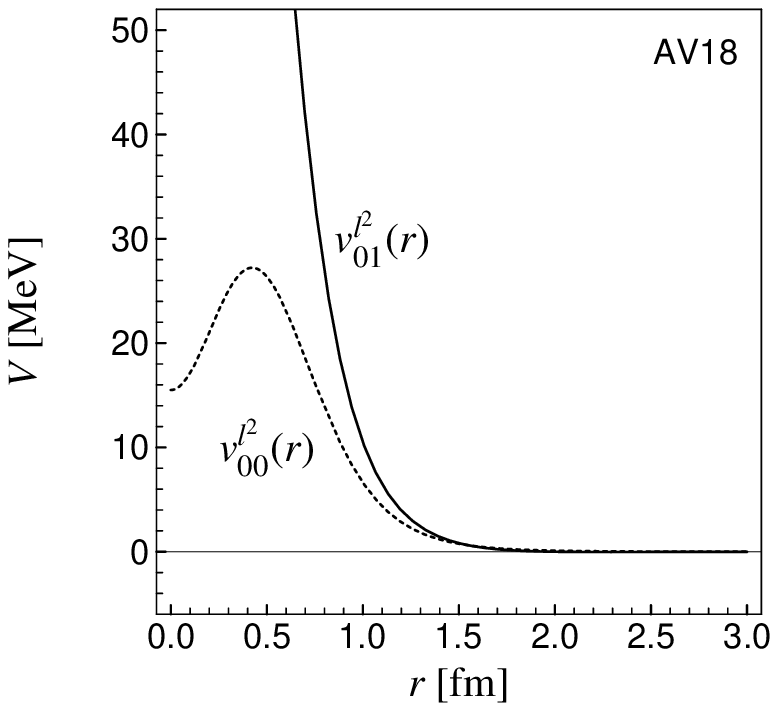}\vspace{1ex}
    \includegraphics[width=0.48\textwidth]{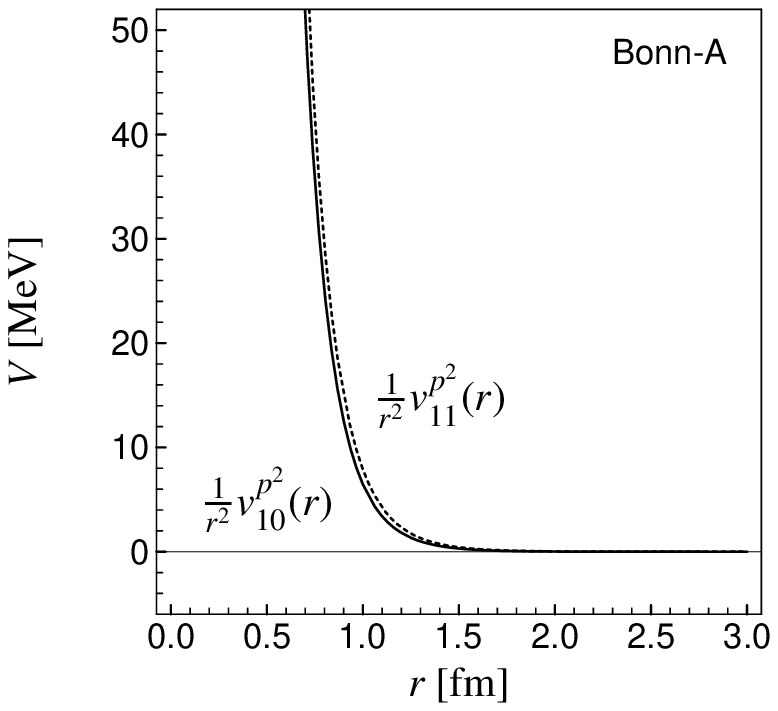}\hfil
    \includegraphics[width=0.48\textwidth]{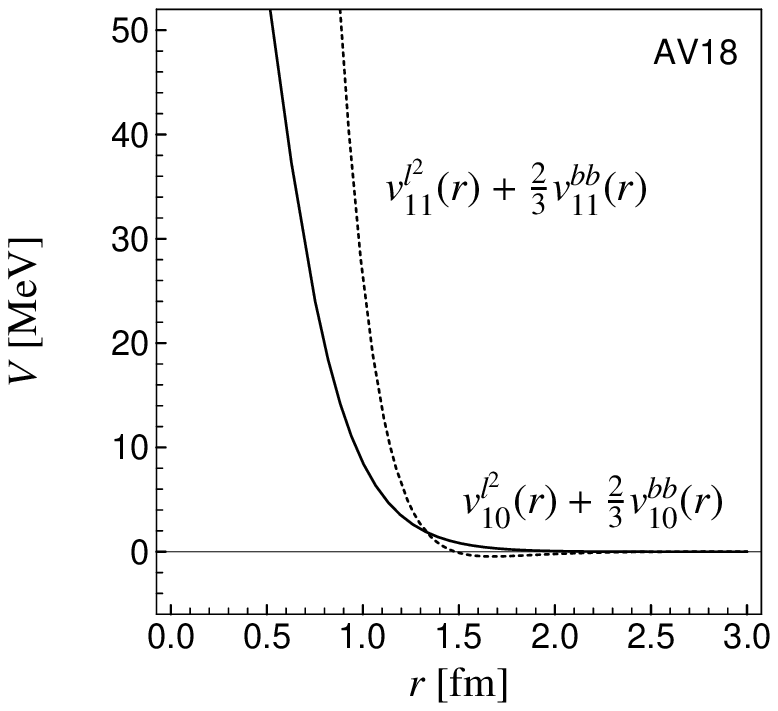}\vspace{1ex}
    \begin{minipage}[b]{0.48\textwidth}
      \caption{Momentum-dependent potential ($p^2$) of the Bonn-A
        interaction (\ref{eq:bonnarepresentation}) (left hand side).
        For better comparison with the $l^2$ potential of the
        Argonne~V18 interaction (\ref{eq:av18representation}) the
        radial dependence of the potential is divided by $r^2$.  In
        the $S=1$ channels the $\lsq$ part of the $\lssq$
        potential has been added for the Argonne~V18. Only the
        Argonne~V18 potential has an $\sll$ potential (right) from the
        decomposition of the $\lssq$ operator.}
      \label{fig:momentumpot}
      \vspace{3ex}
    \end{minipage}\hfil
    \includegraphics[width=0.48\textwidth]{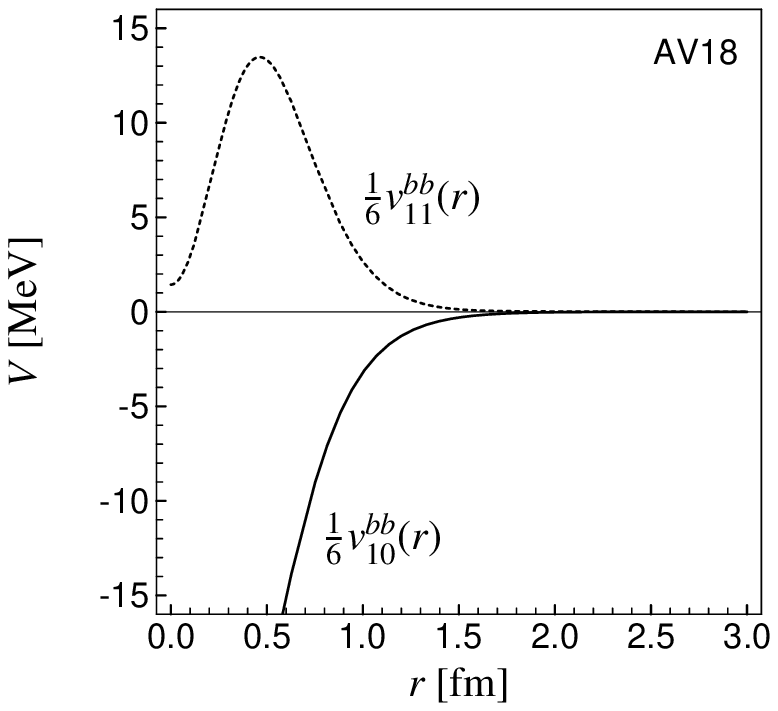}
\end{figure}

Out of the family of the Bonn potentials we will use the
(nonrelativistic) Bonn-A interaction \cite{machleidt89} which has
strong nonlocal contributions and a relatively weak tensor force.
Because of the technical problems arising from the nonlocal terms
there are only a few many-body calculations using the Bonn-A
potential in the literature.

We use the more convenient representation of the interaction that projects
the potential on spin- and isospin-channels instead of the $\sigsig$
and $\tautau$ or the exchange operators
\begin{eqnarray} \label{eq:bonnarepresentation}
  \op{v} =& \sum_{S,T} v^{c}_{ST}(r)\, \Pi_{ST} +
  \sum_{T} v^{t}_{1T}(r)\, \srr\, \Pi_{1T} +
  \sum_{T} v^{b}_{1T}(r)\, \ls\, \Pi_{1T} \nonumber \\
   &+ \sum_{S,T} \frac{1}{2} \left(\vec{p}^2 v^{p^2}_{ST}(r) +
    v^{p^2}_{ST}(r) \vec{p}^2 \right) \Pi_{ST} \eqdot
\end{eqnarray}
The operators $\Pi_{ST}$ project on the respective $S,T$ channels.
The nonlocal $\vec{p}^2$ dependent terms of the Bonn potential are
actually independent of the spin. 

The radial dependences in this parameterization are shown in
Figs.~\ref{fig:centralpot}--\ref{fig:momentumpot} together with the
plots for the Argonne~V18 interaction.

The Bonn-CD interaction \cite{machleidt01}, that has been refitted to
the new set of scattering data and includes additional charge
independence breaking and charge symmetry breaking parts, is
unfortunately not given in a nonrelativistic parameterization.
Therefore we shall use only the Bonn-A potential in our
nonrelativistic approach.

\subsection{Argonne potentials}

Unlike the Bonn interaction the Argonne~V14 interaction
\cite{wiringa84} is designed to be as local as possible which has
technical advantages in GFMC calculations. However momentum dependence
is needed to reproduce the phase-shifts. In the Argonne potentials
this is included via angular momentum dependence in form of $\lsq$ and
$\lssq$ terms in contrast to the $\vecp^2$ terms of the Bonn
interaction.

The Argonne~V14 potential can be written in the spin- and isospin
channel representation as
\begin{eqnarray}
  \op{v} = & \sum_{S,T} v^{c}_{ST}(r) \, \Pi_{ST} +
  \sum_{T} v^{t}_{1T}(r) \, \srr\, \Pi_{1T} +
  \sum_{T} v^{b}_{1T}(r) \,\ls\, \Pi_{1T} \nonumber \\
  & + \sum_{S,T} v^{l^2}_{ST}(r)\, \lsq\, \Pi_{ST} +
  \sum_{T} v^{bb}_{1T}(r)\, \lssq\, \Pi_{1T} \eqdot
  \label{eq:av18representation}
\end{eqnarray}

We decompose the $\lssq$ operator into irreducible tensor operators
\begin{equation}
  \lssq = \frac{2}{3} \lsq \, \Pi_{S=1} - \frac{1}{2}\ls + \frac{1}{6}
  \sll
\end{equation}
and for a proper comparison add the $\lsq$ and $\ls$ terms to the
respective potential terms in Fig.~\ref{fig:spinorbitpot} and
\ref{fig:momentumpot}.

The improved Argonne~V18 interaction \cite{wiringa95} fitted to the
improved scattering data includes the electro-magnetic interaction
beyond the static approximation and contains terms that break charge
independence and charge symmetry. These additional terms give only
minor corrections and we will use the Argonne~V18 interactions without
those.


A simplified version of the Argonne~V18 is the Argonne~V8' interaction
\cite{pudliner97}. In the Argonne~V8' the $\lsq$ and $\lssq$ parts of
the full Argonne~V18 interaction are projected onto the central,
tensor and spin-orbit parts in such a way that the interaction is
unchanged in the $s$- and $p$- waves and in the deuteron channel.

\begin{alignat}{2}
  v^{c}_{00}(r) & \longleftarrow v^{c}_{00}(r) + 2 v^{l^2}_{00}(r) \qquad
  & v^{c}_{01}(r) & \longleftarrow v^{c}_{01}(r) \notag \\
  v^{c}_{10}(r) & \longleftarrow v^{c}_{10}(r) \qquad
  & v^{c}_{11}(r) & \longleftarrow v^{c}_{11}(r) + 2 v^{l^2}_{11}(r)+
  \frac{4}{3} v^{bb}_{11}(r) \notag \\
  v^{t}_{10}(r) & \longleftarrow v^{t}_{10}(r) \qquad
  & v^{t}_{11}(r) & \longleftarrow v^{t}_{11}(r) - \frac{5}{12}
  v^{bb}_{11}(r) \notag \\
  v^{b}_{10}(r) & \longleftarrow v^{b}_{10}(r) -2 v^{l^2}_{10}(r) - 
  3 v^{bb}_{10}(r) \qquad
  & v^{b}_{11}(r) & \longleftarrow v^{b}_{11}(r) - \frac{1}{2}
  v^{bb}_{11}(r)
  \label{eq:argonneprojection}
\end{alignat}


\section{Unitary correlation operator method}
\label{sec:UCOMmethod}

With the Unitary Correlation Operator Method (UCOM) we want to bring
together realistic nuclear interactions and the simple many-body
states of a mean-field or shell-model calculation. The short-range or
high-momentum properties of the many-body state are treated by the
unitary correlator which is almost independent on the low energy scale
of the long-range correlations that can be successfully described in a
mean-field approach.

The correlated many-body states are constructed by applying the
unitary correlation operator $\opC$ to the uncorrelated many-body
state $\ket{\Psi}$ that may for example be a Slater determinant of
harmonic oscillator states as used in shell model calculations or a
Slater determinant of Gaussian wave packets as used in the Fermionic
Molecular Dynamics (FMD) model
\begin{equation}
  \ket{\corr{\Psi}} = \opC \ket{\Psi} \eqdot
\end{equation}
Alternatively we can apply the correlations to the operators and
define correlated operators
\begin{equation}
  \cop{A} = \hopC \op{A} \opC \eqdot
\end{equation}
Due to the unitarity of the correlation operator we can evaluate
matrix elements either using uncorrelated operators and correlated
states or using correlated operators and uncorrelated states:
\begin{equation}
  \matrixe{\Phi}{\hopC \op{A} \opC}{\Psi} =
  \matrixe{\corr{\Phi}}{\op{A}}{\corr{\Psi}} =
  \matrixe{\Phi}{\cop{A}}{\Psi} \eqdot
\end{equation}

As the correlation operator $\opC$ should be unitary and describe
two-body correlations we will use hermitian two-body generators
$\opGom$ and $\opGr$
\begin{equation}
  \opC = \opCom \opCr \qquad \textrm{with} \quad
  \opCom =  \exp \bigl\{ -\I \opGom \bigr\} \qquad 
  \textrm{and} \quad
  \opCr = \exp \bigl\{ -\I \opGr \bigr\} \eqdot
\end{equation}
The correlation operator $\opC$ itself is not a two-body operator
because the repetitive application of the generator produces operators
of increasing order in particle number.  If we want to describe
genuine three-body correlations we would have to use three-body
operators in the generators $\opG$.

The correlated Hamilton operator should posses the same symmetries
with respect to global transformations like translation, rotation, or
boost as the uncorrelated one. Therefore the generator can only depend
on the relative coordinates and momenta of the two particles and it
has to be a scalar operator with respect to rotations. Furthermore the
correlations should fulfill the cluster decomposition property, which
implies that observables in subsystems, that are mutually outside the
range of the interaction, are not affected by the other subsystems.
The correlations therefore have to be of finite range.

Before we look at the explicit form of the generators for radial and
tensor correlators we discuss the application of correlated operators
in many-body systems, which is performed in the sense of the cluster
expansion.

\subsection{Cluster expansion}
\label{sec:clusterexpansion}

As the correlation operator $\opC$ is the exponential of a two-body
operator, the correlation operator itself and correlated operators
have irreducible contributions of higher particle orders. The Fock
space representation of the correlated operator $\cop{A}$ is given by
the cluster expansion
\begin{equation}
  \cop{A} = \hopC \op{A} \opC = \sum_{i=1} \cop{A}^{[i]} \eqcomma
\end{equation}
where $i$ denotes the irreducible particle number.

Using an orthonormal one-body basis
$\bigl\{ \ket{k} = \conop{a}_k \ket{0} \bigr\}$ we get
\begin{align}
  \copone{A} & = \sum_{k, k'} \matrixe{k}{\hopC \op{A} \opC}{k'} \;
  \conop{a}_{k} \desop{a}_{k'} = \sum_{k, k'} \matrixe{k}{\op{A}}{k'} \;
  \conop{a}_{k} \desop{a}_{k'} \label{eq:onebody}\\
  \coptwo{A} & = \frac{1}{4} \sum_{\substack{k_1,k_2 \\ k_1',k_2'}}
  \matrixeaa{k_1,k_2}{\hopC \op{A}
    \opC - \copone{A}}{k_1',k_2'} \; \conop{a}_{k_1} \conop{a}_{k_2}
  \desop{a}_{k_2'} \desop{a}_{k_1'} \label{eq:twobody}\\
  & \vdots \notag \\
  \cop{A}^{[n]} & = \frac{1}{(n!)^2} \sum_{\substack{k_1, \ldots,
    k_n\\k_1',\ldots,k_n'}}
  \matrixeaa{k_1, \ldots, k_n}{\hopC \op{A}
    \opC - \sum_{i=1}^{n-1} \cop{A}^{[i]}}{k_1', \ldots, k_n'} \;
  \conop{a}_{k_1} \cdots \conop{a}_{k_n} \,
  \desop{a}_{k_n'} \cdots \desop{a}_{k_1'} \eqcomma
\end{align}
where at each order of the cluster expansion the contributions of the
lower particle orders have to be subtracted.

In practice we would like to restrict the calculations to the two-body
level as the three-body contributions are already very involved. We
should therefore use generators $\opG$ which cause only small
three-body contributions. The importance of three-body contributions
will increase with the range of the correlator and the density of the
system.

We will use the notation
\begin{equation}
  \left[ \hopC \op{A} \opC \right]^{C2} = \copone{A} + \coptwo{A}
\end{equation}
to indicate the two-body approximation.

\subsection{Two-body system}
\label{sec:twobody}

For calculating many-body matrix elements in two-body approximation
one has to evaluate matrix elements of correlated operators in one-
and two-body space only.  As defined in Eqs.~(\ref{eq:onebody}) and
(\ref{eq:twobody}) those matrix elements then define the one- and
two-body operators in Fock space, respectively.

In contrast to Fock space operators which are denoted by uppercase
letters (e.g. $\opC$) we will use lowercase letters for operators in
one- or two-body space (e.g. $\opc$ for the correlation operator in
two-body space).

The correlator affects only the relative and not the center of mass
motion of two nucleons. Therefore relative and center of mass
variables of two nucleons are introduced:
\begin{align}
  \vecr & = \vecx_1 - \vecx_2 \eqcomma & \qquad
  \vecx_{\icm} & = \frac{1}{2} ( \vecx_1 + \vecx_2 ) \eqcomma \\
  \opvecp & = \frac{1}{2} ( \opvecp_1 - \opvecp_2 ) \eqcomma & \qquad
  \opvecp_{\icm} & = \opvecp_1 + \opvecp_2
  \eqdot
\end{align}
For example in the kinetic energy
\begin{equation} 
  \op{t} = \op{t}_{\irel} + \op{t}_{\icm}  \qquad \mathrm{with}\qquad
  \op{t}_{\irel} = \frac{1}{m} \opvecp^2  \eqcomma \qquad
  \op{t}_{\icm} = \frac{1}{4m} \opvecp_{\icm}^2
  \label{eq:tcmtrel}
\end{equation}
only the kinetic energy $\op{t}_{\irel}$ of the relative motion has to
be correlated.  The orbital angular momentum of the two-body system
can also be decomposed into the orbital angular momentum of the
relative motion and the center of mass one
\begin{equation}
  \opvecl_1 + \opvecl_2 =
  \opvecl_{\icm} + \opvecl \qquad \mathrm{with}\qquad
  \opvecl  = \opvecr \times \opvecp \eqcomma \qquad
  \opvecl_{\icm} = \opvecx_{\icm} \times \opvecp_{\icm}\eqdot
\end{equation}

As we want to study correlations induced by the nuclear force, which
in two-body space does not connect states of different total spin and
isospin, we use in the following the basis states
$\ket{n;(LS)JM;TM_T}$ with quantum numbers $L,M_L$ for relative
angular momentum, $S,M_S$ for total spin, $J,M$ for total angular
momentum, and $T,M_T$ for total isospin. With the help of
Clebsch-Gordan coefficients their coordinate representation is given
by
\begin{multline}
  \braket{\vec{x}_1,\vec{x}_2}{n;(LS)JM;TM_T} = \\
  \varphi^n_{(LS)J,T}(r)\!\sum_{M_L,M_S} \cg{L}{M_L}{S}{M_S}{J}{M}
  Y_{LM_L}(\hat{\vecr})\ket{SM_S}\ket{TM_T}\,\phi_{\icm}(\vecx_{\icm})
  \eqcomma
  \label{eq:twobodybasis}
\end{multline}
where $Y_{LM_L}(\hat{\vecr})$ denotes spherical harmonics that depend
on the unit vector of the relative coordinate. The quantum numbers of
the center of mass motion are indicated only when needed.

The transformation back to one-body variables with spin components
$\chi_i$ and isospin components $\xi_i$ is obtained from
\begin{multline}
  \braket{\vecx_1 \chi_1 \xi_1; \vecx_2 \chi_2 \xi_2}{\phi} = \\
  \sum_{S,M_S} \cg{\half}{\chi_1}{\half}{\chi_2}{S}{M_S}
  \sum_{T,M_T} \cg{\half}{\xi_1}{\half}{\xi_2}{T}{M_T}
  \braket{\vecr\,\vecx_{\icm}; S M_S; T M_T}{\phi} \eqdot
\end{multline}

\subsection{Spin-isospin dependent correlators}

As the nuclear force depends strongly on spin and isospin the
correlations will be distinct in the different spin-isospin
channels.  In two-body space we use the ansatz
\begin{equation}
  \opg = \sum_{ST} \opg_{ST} \, \op{\Pi}_{ST}
\end{equation}
for a spin-isospin dependent generator with the
projectors $\op{\Pi}_{ST}$ on the spin-isospin channels. Exploiting
the projector properties
we can rewrite the correlation operator
\begin{equation}
  \opc = \exp \big\{ -\I \opg \bigr\} =
  \sum_{ST} \exp \bigl\{ -\I \op{g}_{ST} \bigr\} \, \op{\Pi}_{ST} \eqcomma
\end{equation}
and as the nuclear interaction does not connect the different spin and
isospin channels\footnote{This is not the case for terms that break
  charge symmetry and charge independence in the Argonne~V18 and
  Bonn~CD interactions. However, in this work we do not consider these
  terms, which give only minor corrections.}  we obtain the correlated
potential in two-body space as
\begin{equation}
  \cop{v} = \hopc \op{v} \opc = 
  \sum_{ST} \exp \bigl\{ \I \op{g}_{ST} \bigr\} \: \op{v}_{ST} \:
  \exp \bigl\{ - \I \op{g}_{ST} \bigr\} \, \op{\Pi}_{ST} \eqdot
\end{equation}
We have thus the important result that the correlations in the
different spin-isospin channels decouple and the correlation operators
can be determined for each spin-isospin channel independently.

\subsection{Correlated densities}

The short-range radial and tensor correlations in the nucleus can be
studied best by inspecting the one- and two-body density matrices. The
density matrices of the many-body state $\ket{\Phi}$ in coordinate
representation are defined as
\begin{equation}
  \rhoone(\vecx_1\chi_1\xi_1;\vecx_1'\chi_1'\xi_1') =
  \matrixe{\Phi}{ \conop{\Psi}_{\chi_1'\xi_1'}(\vecx_1')
    \desop{\Psi}_{\chi_1\xi_1}(\vecx_1) }{\Phi}
  \label{eq:rhoone}
\end{equation}
and
\begin{multline}
  \rhotwo(\vecx_1\chi_1\xi_1,\vecx_2\chi_2\xi_2;
  \vecx_1'\chi_1'\xi_1',\vecx_2'\chi_2'\xi_2') = \\
  \matrixe{\Phi}{ \conop{\Psi}_{\chi_1'\xi_1'}(\vecx_1')
    \conop{\Psi}_{\chi_2'\xi_2'}(\vecx_2')
    \desop{\Psi}_{\chi_2\xi_2}(\vecx_2)
    \desop{\Psi}_{\chi_1\xi_1}(\vecx_1) }{\Phi} \eqcomma
  \label{eq:rhotwo}
\end{multline}
where $\desop{\Psi}_{\chi\xi}(\vecx)$ is the field operator for
a nucleon with spin component $\chi$ and isospin component $\xi$ at
position $\vecx$.

The two-body correlations can be visualized best with the help of the
two-body density matrix $\rhotwo_{SM_S,TM_T}(\vecr)$ which describes
the probability density to find two nucleons at a distance $\vecr$ in
the $S,T$ channel with spin and isospin orientations $M_S,M_T$. The
center of mass coordinate is integrated out.
\begin{multline}
  \rhotwo_{SM_S,TM_T}(\vecr) =
  \sum_{\chi_1,\chi_2}
  \cg{\half}{\chi_1}{\half}{\chi_2}{S}{M_S}
  \sum_{\xi_1,\xi_2} \cg{\half}{\xi_1}{\half}{\xi_2}{T}{M_T} \\
  \times \intdt{X} \rhotwo
  (\vecX\!+\!\half\vecr\chi_1\xi_1,\vecX\!-\!\half\vecr\chi_2\xi_2;
  \vecX\!+\!\half\vecr\chi_1\xi_1,\vecX\!-\!\half\vecr\chi_2\xi_2)
  \label{eq:rhotwodiag}
\end{multline}
The information about the short-range radial correlations is contained
in the radial dependence of this two-nucleon correlation function. The
tensor correlations manifest themselves by the angular and the spin
dependence.

As with other operators the density matrices of correlated many-body
states $\ket{\corr{\Phi}}=\opC \ket{\Phi}$ are calculated in the
cluster expansion
\begin{equation}
  \begin{split}
  \crhoone(\vecx_1\chi_1\xi_1;\vecx_1'\chi_1'\xi_1')
  & = \matrixe{\Phi}{ \hopC \conop{\Psi}_{\chi_1'\xi_1'}(\vecx_1')
    \desop{\Psi}_{\chi_1\xi_1}(\vecx_1) \opC }{\Phi} \\
  & = \sum_{n=1}^{A} \matrixe{\Phi}{ \bigl[ \hopC \conop{\Psi}_{\chi_1'\xi_1'}(\vecx_1')
    \desop{\Psi}_{\chi_1\xi_1}(\vecx_1) \opC \bigr]^{[n]} }{\Phi} \eqcomma
  \end{split}
  \label{eq:crhoone}
\end{equation}
\begin{multline}
  \crhotwo(\vecx_1\chi_1\xi_1, \vecx_2\chi_2\xi_2;
  \vecx_1'\chi_1'\xi_1', \vecx_2'\chi_2'\xi_2') = \\
  \begin{split}
  & = \matrixe{\Phi}{ \hopC \conop{\Psi}_{\chi_1'\xi_1'}(\vecx_1')
    \conop{\Psi}_{\chi_2'\xi_2'}(\vecx_2')
    \desop{\Psi}_{\chi_2\xi_2}(\vecx_2)
    \desop{\Psi}_{\chi_1\xi_1}(\vecx_1)
    \opC }{\Phi} \\
  & = \sum_{n=2}^{A}
  \matrixe{\Phi}{ \bigl[ \hopC \conop{\Psi}_{\chi_1'\xi_1'}(\vecx_1')
    \conop{\Psi}_{\chi_2'\xi_2'}(\vecx_2')
    \desop{\Psi}_{\chi_2\xi_2}(\vecx_2)
    \desop{\Psi}_{\chi_1\xi_1}(\vecx_1)
    \opC \bigr]^{[n]} }{\Phi} \eqdot
  \end{split}
  \label{eq:crhotwo}
\end{multline}

In the two-body approximation the expansions will be truncated  after
the second order. One should notice that density matrices calculated
in a truncated cluster expansion fulfill the reduction property of the
exact density matrices
\begin{equation}
  \sum_{\chi_2,\xi_2} \intdt{x_2}
  \crhotwo(\vecx_1\chi_1\xi_1, \vecx_2\chi_2\xi_2;
  \vecx_1'\chi_1'\xi_1', \vecx_2\chi_2\xi_2) =
  (A-1) \, \crhoone(\vecx_1\chi_1\xi_1;\vecx_1'\chi_1'\xi_1')
  \label{eq:rhoreduction}
\end{equation}
only approximately. If the truncation at second order is justified,
Eq.~(\ref{eq:rhoreduction}) is well approximated.

The nucleon density distribution $\hat{\rho}(\vecx)$ is given by the
diagonal part of the one-body density matrix which has been summed
over spin and isospin indices
\begin{equation}
  \hat{\rho}(\vecx)=
  \sum_{\chi,\xi} \crhoone(\vecx\chi\xi;\vecx\chi\xi) \ .
\end{equation}

The nucleon momentum distribution $\hat{n}(\vec{k})$, displayed for
$\Hefour$ and $\Osixteen$ in Sec.~\ref{sec:momentumdistribution}
is evaluated in momentum-space in two-body approximation
\begin{multline}
  \hat{n}(\vec{k}) = \sum_{\chi,\xi} \Bigl(
  \matrixe{\Phi}{\conop{a}_{\chi\xi}(\vec{k})
    \desop{a}_{\chi\xi}(\vec{k})}{\Pi} + \\
  \intdt{k'} \sum{\chi',\xi'} \matrixe{\Phi}{\bigl[ \hopC
    \conop{a}_{\chi\xi}(\vec{k}) \conop{a}_{\chi'\xi'}(\vec{k}')
    \desop{a}_{\chi'\xi'}(\vec{k}') \desop{a}_{\chi\xi}(\vec{k}) \opC
    \bigr]^{[2]}}{\Phi} \Bigr)
\end{multline}
and can
be related to the Fourier transform of the off-diagonal matrix
elements of the one-body density matrix
\begin{equation}
  \hat{n}(\vec{k}) = \intdt{x_1} \intdt{x_2} \sum_{\chi\xi}
  \crhoone(\vecx_1\chi\xi;\vecx_2\chi\xi) \: e^{i \vec{k} \cdot
    (\vecx_1-\vecx_2)} \eqdot
  \label{eq:Fouriermomentum}
\end{equation}

\section{Correlations}
\label{sec:correlations}

All realistic nucleon-nucleon interactions, like the Bonn and the
Argonne potentials discussed in Sec.~\ref{sec:potentials}, possess a
short ranged repulsive core and a tensor force.  The repulsion causes
an attenuation of the two-body density at short distances while the
tensor interaction induces a strong correlation between the spatial
orientation of the nucleons and the orientation of their spins. It
operates only in the $S=1$ channels but is crucial for a successful
description of the nucleus, without the tensor part of the potential
nuclei are not bound.

We introduce unitary transformations to create the short range
correlations induced by the interaction that are missing in the
uncorrelated many-body trial state. This is done by moving the
particles subject to their relative distance and their spin
directions.  It turns out that a separation into shifts parallel and
perpendicular to the relative distance leads to a concise operator
structure.

\begin{figure}[htb]
  \begin{minipage}[b]{0.48\textwidth}
    \caption{Decomposition of relative momentum $\op{\vec{p}}_{ij}$ parallel
    and perpendicular to relative distance $\op{\vec{r}}_{ij}$.}
    \label{fig:rprpom}
    \vspace{3ex}
  \end{minipage}\hfil
  \includegraphics[width=0.48\textwidth]{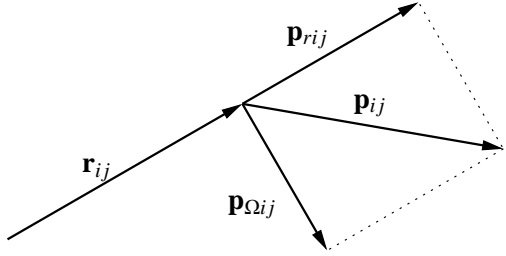}
\end{figure}

In order to achieve shifts in the direction of
$\opvecr_{ij}=\opvecr_i-\opvecr_j$ and perpendicular to it we
decompose the relative momentum operator
$\opvecp_{ij}=\frac{1}{2}(\opvecp_i-\opvecp_j)$ into a component
$\opvecp_{rij}$ that is parallel and a component $\opvecp_{\Omega ij}$
that is orthogonal
\begin{equation}
  \opvecp_{ij} = \opvecp_{rij} + \opvecp_{\Omega ij}
  \label{eq:opp}
\end{equation}
as indicated in Fig.~\ref{fig:rprpom}.  The relative momentum of the
pair $(ij)$ in the radial direction is defined by the projection
\begin{equation}
  \opvecp_{rij}= \frac{\vecr_{ij}}{r_{\!ij}}\ \opvecp_{rij} \eqcomma
\end{equation}
where the hermitean operator $\op{p}_{ij}$ for the radial relative
momentum is given by \footnote{The symbol $\eqrrep$ is used to denote
  the coordinate space representation of an operator.}

\begin{equation}
  \op{p}_{rij}  = \frac{1}{2}\biggl(
  \frac{\vecr_{ij}}{r_{ij}}\opvecp_{ij}+
  \opvecp_{ij}\frac{\vecr_{ij}}{r_{\!ij}}\biggr) \eqrrep
  \biggl( \frac{1}{r_{\!ij}}\frac{\partial}{\I\partial r_{\!ij}} r_{\!ij}
  \biggr) \eqdot   
  \label{eq:oppr}
\end{equation}
As $\opvecp_{ij}$ and $\opvecr_{ij}$ are operators and do not commute
we take the hermitean combination of the classical expressions.
The remaining part
$\opvecp_{\Omega ij}$ that is perpendicular to $\vecr_{ij}$ is
\begin{equation}
  \opvecp_{\Omega ij}  =  \frac{1}{2 r_{\!ij}} \biggl(
  \op{\vecl_{ij}}\times\frac{\op{\vecr}_{ij}}{\op{r}_{\!ij}} -
  \frac{\vecr_{ij}}{r_{\!ij}} \times \vecl_{ij} \biggr) \eqdot
  \label{eq:opvecpom}
\end{equation}
We call $\opvecp_{\Omega ij}$ relative \emph{orbital momentum}. It
should not be confused with the relative orbital angular momentum
$\opvecl_{ij}=\opvecr_{ij}\times\opvecp_{ij}$. Both, $\opvecp_{rij}$
and $\opvecp_{\Omega ij}$ are hermitian. We have summarized some
properties of these operators in Appendix \ref{app:radialmomentum}.

The radial relative momentum $\op{p}_{rij}$ will be used in $\opCr$ as
the generator of radial shifts that move the particle pair $(ij)$ out
of their mutual repulsive interaction area, see Eq.~(\ref{eq:Cr}). The
orbital relative momentum $\opvecp_{\Omega ij}$ in combination with
the spin operators $\op{\vec{\sigma}}_i$ and $\op{\vec{\sigma}}_j$
will be used in the unitary transformation $\opCom$
(Eq.~(\ref{eq:Com})) to relocate the angular position of the pair to
regions where the tensor interaction is attractive.

The correlator $\opC=\opCom\opCr$ has to be of finite range for the
application in the many-body system in order not to destroy the
cluster decomposition property. The task of the unitary correlation
operators is to introduce the short-range repulsive and tensor
correlations into the many-body state. Possible long range
correlations should be described by the many-body trial state and not
by the correlation operator. If analyzed in momentum space the
correlator describes the high momentum components of the state while
the low momentum part is taken care of by the model space.

In the following we restrict the investigations to the two-body space
as this is sufficient for a cluster expansion up to two-body
contributions.  We use lower case letters for the operators to
indicate that and omit the indices $(ij)=(12)$ for the particles.

\subsection{Radial correlations}
\label{sec:centralcorrelations}

Radial correlations within the UCOM framework have already been
studied in detail in Refs.~\cite{ucom98,roth:doktor}. In this section
we shall only provide a short summary of the ideas and techniques. For
the illustrations we use the Argonne~V18 potential in the $S,T=0,1$
channel (for $L=0$ identical to the Argonne~V8' potential) where we
do not have to deal with additional correlations from tensor
interactions.

The strong repulsion at short distances is suppressing the two-body
density in the range of the core as already shown in Fig.
\ref{fig:correlationhole}. To describe these short-range correlations
the radial correlator
\begin{equation}
  \opcr =   \exp\bigl\{-\I \, \op{g}_r \bigr\}=
  \exp\Bigl\{-\I \frac{1}{2}\bigl( \oppr s(\op{r}) +
    s(\op{r})\oppr \bigr) \Bigr\}
\end{equation}
(c.f. Eq.~(\ref{eq:Cr})) shifts nuleons that are in the range of the
repulsive core radially outwards while those being already further out
are not affected. To achieve this the generator $\opgr$ contains the
radial momentum operator $\oppr$ and the shift function $s(r)$ which
tends to zero for large $r$.  (see also Appendix
\ref{app:radialmomentum} for further properties)

It is advantageous to introduce the correlation
functions $\Rp(r)$ and $\Rm(r)$ which are related to the shift
function $s(r)$ by
\begin{equation}
  \pm 1 = \int_r^{\Rpm(r)} \frac{d\xi}{s(\xi)} \eqdot
\end{equation}
The correlation functions are mutually inverse to each other
\begin{equation}
  \Rpm(\Rmp(r)) = r \eqcomma
\end{equation}
which reflects the unitarity of the correlation operator.
For small shifts one has approximately
\begin{equation}
  \Rpm(r) \approx r \pm s(r) \eqdot
\end{equation}

\begin{figure}[tb]
  \includegraphics[width=0.48\textwidth]{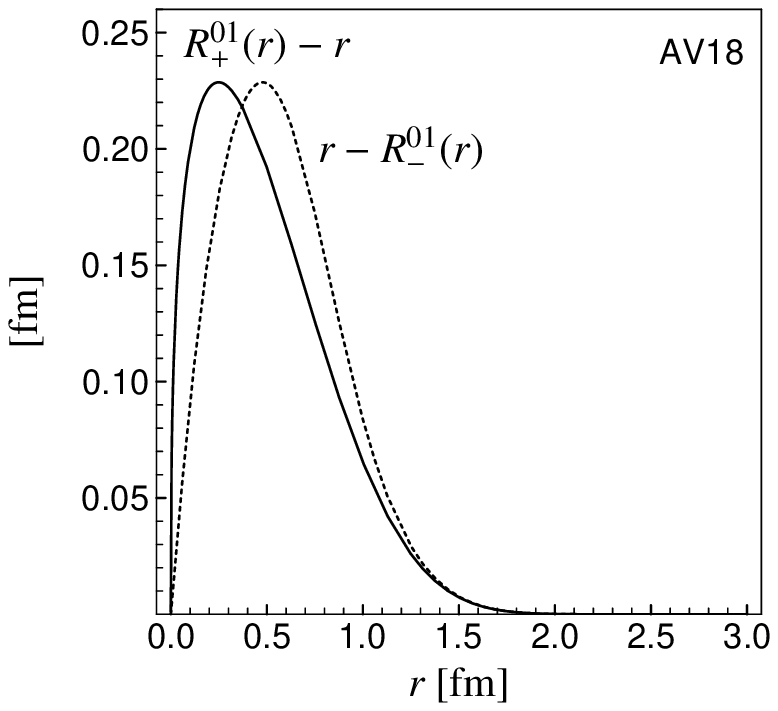}
  \includegraphics[width=0.48\textwidth]{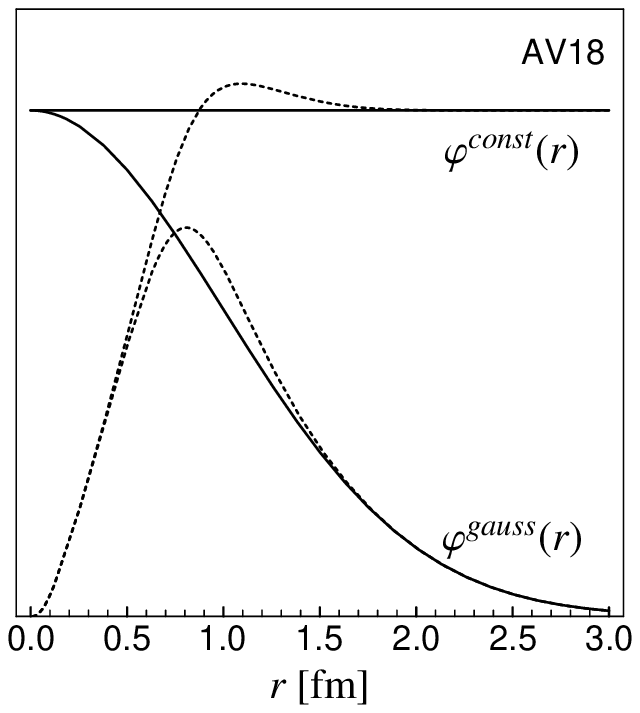}
  \caption{Left: correlation function $\Rp^{01}(r)$ and its inverse
    $\Rm^{01}(r)$ for the Argonne~V18 potential in the $S,T=0,1$
    channel.  Right: Correlated (dotted) and uncorrelated (solid)
    Gaussian and constant wave functions.}
  \label{fig:argonnerprm_corrwave} 
\end{figure}

The correlation functions $\Rp(r)$ and $\Rm(r)$ for the $S,T=0,1$
channel of the Argonne~V18 potential are displayed in
Fig.~\ref{fig:argonnerprm_corrwave}.  The radial shift is strongest in
the range up to about $0.5\;\fm$ and extends to about $1.5\;\fm$.  On
the right hand side of Fig.~\ref{fig:argonnerprm_corrwave} a
correlated Gaussian wave function and a correlated constant wave
function are displayed. The correlated wave functions are almost
identical in the range of the repulsive core of the interaction. As
the correlator conserves the norm of the wave function the hole
created at short distances has to be compensated by an enhancement of
the wave function further out.

In the case of radial correlations it is possible to give closed
expressions for all operators in coordinate representation
\cite{ucom98}, like \footnote {In Ref.~\cite{ucom98} Eq.~(59) is
  misprinted. It should read $\hat{q}_r=\frac{1}{\sqrt{\Rp'(r)}}
  \frac{\Rp(r)}{r} q_r\frac{r}{\Rp(r)} \frac{1}{\sqrt{\Rp'(r)}}.$ }
\begin{alignat}{2}
  \hopcr \, r \, \opcr &= \Rp(r) \eqcomma \qquad\qquad & 
  \hopcr \, \frac{\vec{r}}{r} \, \opcr &= \frac{\vec{r}}{r} \eqcomma \notag \\
  \hopcr \, \oppr \, \opcr &= \frac{1}{\sqrt{\Rp'(\op{r})}} 
  \, \oppr \, \frac{1}{\sqrt{\Rp'(\op{r})}} \eqcomma \qquad\qquad &
  \hopcr \, \opvecpom \, \opcr &= \opvecpom \eqcomma \notag \\
  \hopcr \, \op{\vec{l}} \, \opcr &= \op{\vec{l}} \eqcomma \qquad\qquad &
  \hopcr \, \op{\vec{\sigma}}_{1,2} \, \opcr &= \op{\vec{\sigma}}_{1,2}
  \eqdot \label{eq:crops}
\end{alignat}
Due to the unitary of $\opcr$ any combination of operators can be
transformed by individual transformations.

The calculation of the correlated kinetic energy in two-body
approximation results in a one- and a two-body contribution to the
correlated Hamilton operator. The one-body contribution is again the
kinetic energy because the generator $\opgr$ is a two-body operator
and thus the correlator $\opcr = \exp \{ -\I \opgr \}$ contains
besides the unit operator only two-body or higher terms.
\begin{equation}
  \copone{t} = \hopc \opone{t} \opc = \opone{t}
\end{equation}

For the calculation of the two-body contribution of the correlated kinetic
energy we use the relative and center of mass variables introduced in
Sec.~\ref{sec:twobody}.  The center of mass kinetic energy $t_{\icm}$
is not influenced by the correlator and the kinetic energy of the
relative motion can be decomposed further in a radial and an angular
part %
\begin{equation}
  \op{t}_{\irel} = \op{t}_{r} + \op{t}_{\Omega}
  = \frac{1}{m} \pr^2 + \frac{1}{m}\frac{\lsq}{r^2} \eqcomma
\end{equation}
where the angular part is correlated simply by replacing $\op{r}$ with
$\Rp(\op{r})$ because the relative angular momentum $\op{\vec{l}}$
commutes with $\opcr$ (see Eqs.~(\ref{eq:crops})). Thus one obtains the
two-body contribution to the radially correlated operator as %
\begin{equation}
  \hopcr \op{t}_{\Omega} \opcr - \op{t}_{\Omega}=
  \frac{1}{m} \left(
  \frac{\lsq}{\Rp(r)^2} -  \frac{\lsq}{r^2} \right)=
  \frac{1}{2\cmuom(r)} \frac{\lsq}{r^2}
  \label{eq:ctom}
\end{equation}
with a correlated ``angular mass''
\begin{equation}
  \frac{1}{2\cmuom(r)} = \frac{1}{m} \biggl( \frac{r^2}{\Rp(r)^2} -
  1 \biggr) \eqdot
  \label{eq:ctommass}
\end{equation}

The radially correlated radial part of the kinetic energy leads with
(\ref{eq:crops}) to a momentum dependent potential
\begin{equation}
  \hopcr \op{t}_{r} \opcr - \op{t}_{r}
  = \frac{1}{2} \biggl[ \pr^2 \frac{1}{2\cmur(r)} +
  \frac{1}{2\cmur(r)} \pr^2 \biggr] + \hat{w}(r)
  \label{eq:ctrad}
\end{equation}
similar to the kinetic energy but with a correlated ``radial mass''
\begin{equation}
  \frac{1}{2\cmur(r)} = \frac{1}{m} \biggl( \frac{1}{\Rp'(r)^2} - 1
  \biggr)
  \label{eq:ctradmass}
\end{equation}
and an additional local potential \footnote{We use the representation
  of the momentum dependent part as in the Bonn-A potential
  Eq.~(\ref{eq:bonnarepresentation}), different from Ref.
  \cite{ucom98}.  The transformation rules between different
  parameterizations are given in Appendix \ref{app:radialmomentum}.}
\begin{equation}
  \hat{w}(r) = \frac{1}{m} \biggl(\frac{7 \Rp''(r)^2}{4
    \Rp'(r)^4} - \frac{\Rp'''(r)}{2 \Rp'(r)^3} \biggr) \eqdot
  \label{eq:ctradu}
\end{equation}
The radially correlated kinetic energy has interaction components that
we also find in the uncorrelated Bonn-A interaction
(\ref{eq:bonnarepresentation}), but there ``radial mass'' and
``angular mass'' are the same.  The radial dependence of the $\lsq$
term that is expressed in terms of the ``angular mass''
$\hat{\mu}_{\Omega\,01}(r)$ corresponds to the $v^{l^2}_{01}(r)$
dependence of the Argonne~V18 interaction
(\ref{eq:av18representation}). One recognizes that the radial
correlation of the kinetic energy introduces momentum dependent
interactions of the same type as are already present in the original
interaction.

\begin{figure}[tb]
  \includegraphics[width=0.48\textwidth]{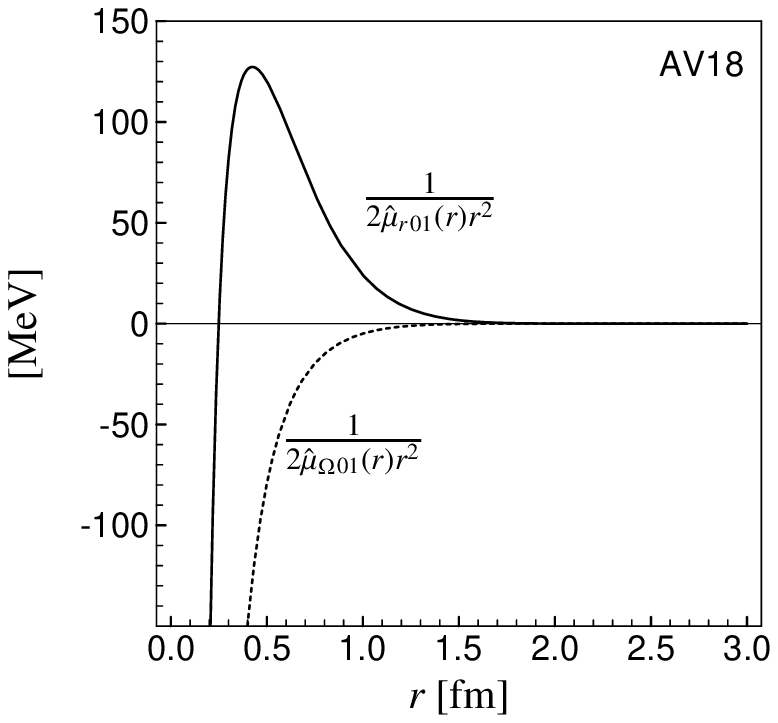}\hfil
  \includegraphics[width=0.48\textwidth]{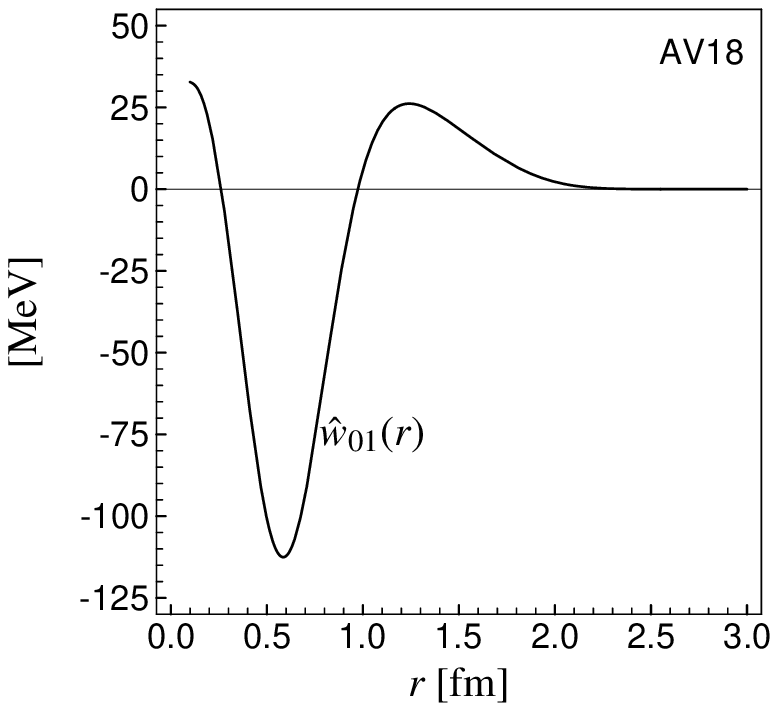}
  \caption{Correlated kinetic energy: inverse radial and angular
    mass divided by $r^2$ (left) and potential part (right) for the
    Argonne~V18 potential in the $S,T=0,1$ channel.}
  \label{fig:ckineticenergy}
\end{figure}

In Fig.~\ref{fig:ckineticenergy} the contributions to the correlated
kinetic energy of the Argonne~V18 interaction are shown. Comparison
with Fig.~\ref{fig:momentumpot} shows that $v^{l^2}_{01}(r)$ is of the
same order as $1/(2\hat{\mu}_{\Omega\,01}(r)r^2)$ but of opposite sign.
Since the correlated and uncorrelated interactions have the same phase
shifts this points already at the ambiguities in trading a local
repulsion for a momentum dependence.

\begin{figure}[tb]
  \begin{minipage}[b]{0.48\textwidth}
    \caption{Correlated central potential of the Argonne~V18 in
      the $S,T=0,1$ channel. The central correlator weakens the core
      of the interaction.}
    \label{fig:ccentralpotential}
    \vspace{3ex}
  \end{minipage}\hfil
  \includegraphics[width=0.48\textwidth]{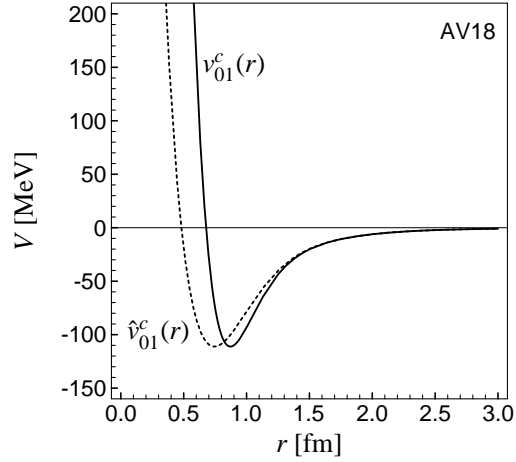}
\end{figure}

The correlated central potential plotted in
Fig.~\ref{fig:ccentralpotential} is expressed easily with help of
relation (\ref{eq:crops}) as the uncorrelated potential with a
transformed radial dependence
\begin{equation}
  \cop{v}^c = \hopcr v^{c}(\op{r}) \opcr = v^{c}(\Rp(r)) \eqdot
\end{equation}

Like for the central interaction, the radially correlated spin-orbit
and tensor potentials only have a transformed radial dependence
($\opcr$ commutes with the operators $\ls$ and $\srr$, see Eqs.
(\ref{eq:crops}))
\begin{align}
 \hopcr \op{v}^{b} \opcr &= v^{b}(\Rp(r)) \, \ls \eqcomma\\
 \hopcr \op{v}^{t} \opcr &= v^{t}(\Rp(r)) \,\srr \eqdot
\end{align}
This holds also in case of the spin-orbit squared and orbital angular
momentum squared potentials
\begin{align}
 \hopcr\op{v}^{l^2}\opcr &= v^{l^2}\!(\Rp(r))\,\lsq \eqcomma\\
 \hopcr\op{v}^{bb} \opcr &= v^{bb}\!(\Rp(r)) \,\lssq \eqdot
\end{align}

The momentum-dependent potential $\op{v}^{p^2}$ occuring in the Bonn
potential can be decomposed into a radial and an angular part
\begin{equation}
  \op{v}^{p^2} =
  \frac{1}{2} \biggl[ \vecp^2 \, v^{p^2}\!(r) + 
  v^{p^2}\!(r) \, \vecp^2 \biggr] =
  \frac{1}{2} \biggl[ \pr^2 \, v^{p^2}\!(r) + v^{p^2}\!(r) \, \pr^2 \biggr]
  + \frac{v^{p^2}\!(r)}{r^2} \, \lsq \eqdot
\end{equation}
The correlated momentum-dependent potential is similar in structure to the
correlated kinetic energy but has an additional term because of the
radial dependence of $v^{p^2}(r)$
\begin{equation}
  \begin{split}
    \hopcr \op{v}^{p^2} \opcr =
    & \frac{1}{2} \biggl[ \pr^2 \, \frac{v^{p^2}\!(\Rp(r))}{\Rp'(r)^2} +
    \frac{v^{p^2}\!(\Rp(r))}{\Rp'(r)^2} \, \pr^2 \biggr] \\
    & + v^{p^2}\!(\Rp(r)) \, \biggl(\frac{7 \Rp''(r)^2}{4
      \Rp'(r)^4} - \frac{\Rp'''(r)}{2 \Rp'(r)^3} \biggr) -
    v^{p^{2}}{}'(\Rp(r)) \frac{\Rp''(r)}{\Rp'(r)^2} \\
    & + \frac{v^{p^2}\!(\Rp(r))}{\Rp(r)^2} \, \lsq \eqdot
  \end{split}
\end{equation}


\subsection{Tensor correlations in the deuteron}
\label{sec:tensorcorrelations}

\begin{figure}[b]
  \includegraphics[width=0.48\textwidth]{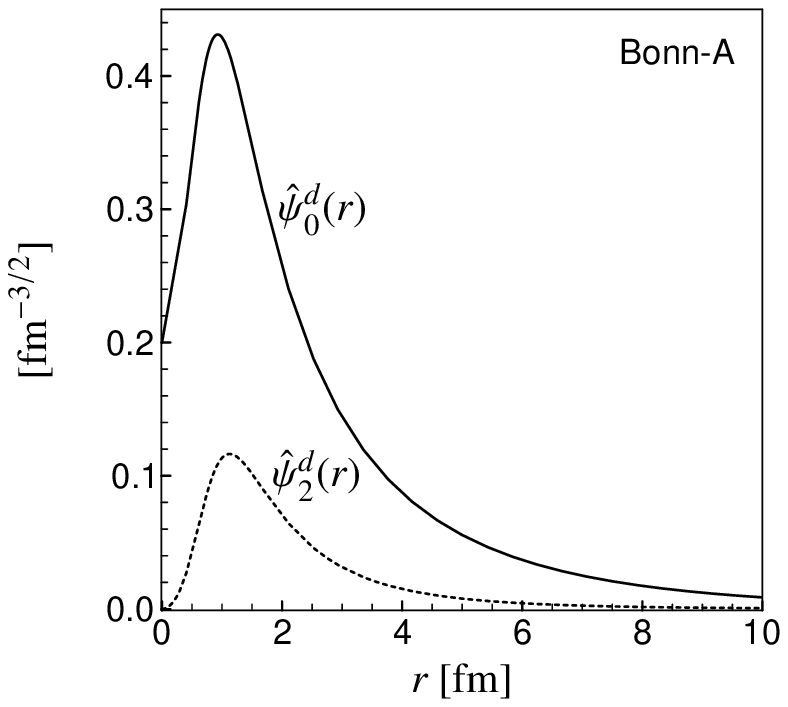}\hfil
  \includegraphics[width=0.48\textwidth]{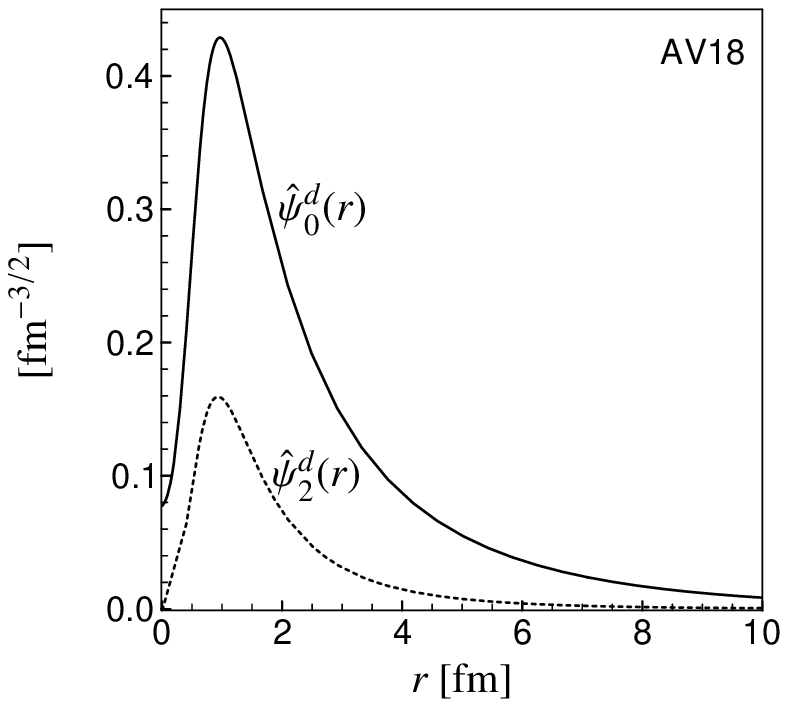}
  \caption{Deuteron wave function: $\hat{\psi}^d_0(r)$
    and $\hat{\psi}^d_2(r)$ for the Bonn-A (left) and the Argonne~V18
    (right) interaction. There is a noticeable difference in the
    $d$-wave admixture, the Argonne~V18 interaction has stronger
    tensor correlations.  The behavior of the wave function at the
    origin shows a larger correlation hole caused by the stronger
    repulsive core of the Argonne~V18 as compared to that of the
    Bonn-A interaction.}
  \label{fig:deuteronwavefunction}
\end{figure}

To get a basic understanding of the tensor correlations we display in
Fig.~\ref{fig:deuteronwavefunction} for two potentials the radial
dependences of the deuteron state $\ket{\corr{d};1M}$
\begin{equation}
  \braket{r}{\corr{d};1M} = \hat{\psi}^d_0(r) \ket{(01)1M}
     +\hat{\psi}^d_2(r) \ket{(21)1M} \eqdot
  \label{eq:deuteronansatz}
\end{equation}

$\ket{(LS)JM}$ is a short hand notation for the angular-spin part of
the basis states defined in Eq.~ (\ref{eq:twobodybasis}) and
$\hat{\psi}^d_L(r)$ stands for the radial dependence
$\hat{\psi}^d_{(LS)J,T}(r)$. The omitted quantum numbers for the
deuteron are $S=1$, $J=1$ and $T=0$.

The correlations due to the repulsive core express themselves as a
depletion of the relative wave functions at distances below $1\;\fm$.
This correlation hole will be created by the unitary radial shift
operator $\opcr$ as explained in section \ref{sec:centralcorrelations}.

\begin{figure}[b]
  \includegraphics[trim=30 50 30 50,width=0.48\textwidth]{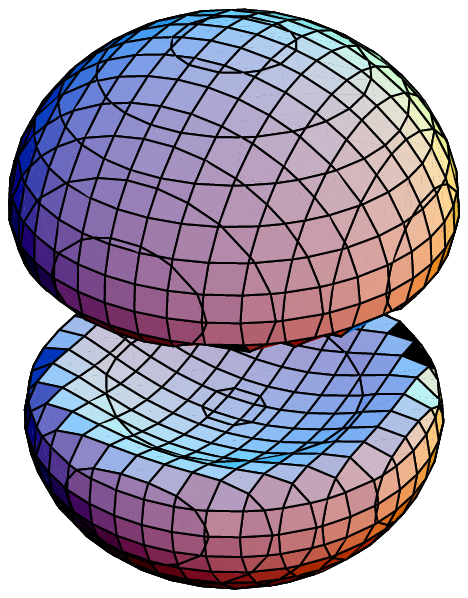}\hfil
  \includegraphics[trim=30 50 30 50,width=0.48\textwidth]{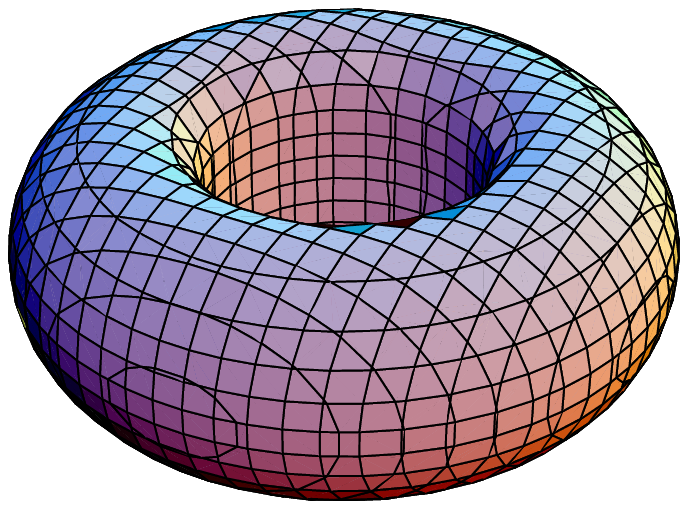}
  \caption{Surfaces of constant density in the deuteron ($\rhotwo_{1M_S} =
    0.005 \fm^{-3}$) for $M_S=\pm 1$ on the left and $M_S=0$ on
    the right. The plots are done for the Argonne~V18 interaction.  }
  \label{fig:deuterondensitiestd}
\end{figure}

The correlation between spins and relative orientation of the nucleons
due to the tensor force is contained in the admixture of a $d$-wave
component $\corr{\psi}^d_2(r) \ket{(21)1M}$, where the orbital angular
momentum $L=2$ is coupled with the spin $S=1$ to the total angular
momentum $J=1$.  In Fig.~\ref{fig:deuterondensitiestd} the two-body
density
\begin{equation}
  \hat{\rho}^{(2)}_{1M_S}(\vec{r})= \frac{1}{3} \sum^1_{M=-1}
  \left|\braket{\vec{r};S=1,M_S}{\corr{d};1M}\right|^2
\end{equation}
is displayed. It is the projection of the deuteron on the $S=1,M_S$
component, averaged over all directions.  Both, the correlation hole
in the center and the alignment of the spatial density with the spin
direction are clearly visible.  Those proton-neutron pairs which have
parallel spins, i.e. $M_S=1$, are located at the ``north'' and
``south'' poles so that their relative distance vector $\vec{r}$ is
aligned with their spins. Pairs with opposite spin, i.e. $M_S=0$, are
found around the ``equator'' where their spins are perpendicular to
$\vec{r}$. This is exactly the pattern one expects for the interaction
of two magnetic dipoles.

A single Slater determinant, where the spin direction of a single
particle state depends at most on the position $\vec{r}_i$ and not on
the relative coordinate $\vec{r}_{ij} = \vec{r}_i - \vec{r}_j$ of two
nucleons, cannot reflect the correlations between spins and relative
orientation of the nucleons.  We face a similar problem as with the
short-range correlations. A single Slater determinant does not have
the proper degrees of freedom needed for the description of the
physical system. Since Slater determinants form a complete basis the
appropriate state can always be written as a superposition of
determinants. One needs however a huge number of those for a
successful description. For example in second order perturbation
theory the tensor interaction scatters to intermediate states with
energies of about $300\;\MeV$ which would mean that about $30 \hbar
\omega$ excitations have to be included in shell-model calculations
\cite{brown71}.

Our aim is, as with the short-range correlations, to include these
tensor correlations into the many-body state by means of a unitary
correlation operator $\opCom$.  The form of the generator is, however,
not self-evident because one needs to correlate the spins of the
nucleons with the relative orientation of the nucleons.

To define our unitary tensor correlator we start with an ansatz
$\ket{d;1M}$ for the uncorrelated deuteron state that has only an
$L=0$ component.
\begin{equation}
  \braket{r}{d;1M} = \psi^d_0(r) \ket{(01)1M}
  \label{eq:deuteronuncorrelatedansatz}
\end{equation}
Here we imply that $\psi^d_0(r)$ already includes the radial
correlations.

We require that the tensor correlator $\opcom$ generates the $L=2$
component of the correlated deuteron state by mapping the uncorrelated
$\ket{d;1M}$ onto the exact solution $\ket{\corr{d};1M}$
\begin{equation}
 \ket{\corr{d};1M}=\opcom\ket{d;1M}=\exp\{-\I \opgom \}\ket{d;1M} \eqdot
\end{equation}

The generator $\opgom$ has to be a scalar operator with respect to
rotations as the quantum numbers $J,M$ of the total angular momentum
are not changed. In coordinate space it has to be a tensor operator of
rank two, there is no other possibility to connect $L=0$ with $L=2$
states. If it is of rank two in coordinate space it also has to be of
rank two in spin space. In the two-body spin space there is only one
tensor of rank two so that only the coordinate space part needs to be
specified.  We restrict the choice by demanding that the correlator
should make shifts only perpendicular to the relative orientation of
the nucleons because radial shifts are already treated by the radial
correlator.  In order to achieve these shifts we use the orbital
momentum operator $\vecpom$, defined in Eq.~(\ref{eq:opvecpom}), to
construct the following tensor
\begin{equation}
 \srpom = \frac{3}{2}\Bigl(
  (\vec{\sigma}_1\!\cdot\!\vecpom)(\vec{\sigma}_2\!\cdot\!\vecr) +
  (\vec{\sigma}_1\!\cdot\!\vecr)(\vec{\sigma}_2\!\cdot\!\vecpom)\Bigr)
\end{equation}
and the generator $\opgom$ for spin-dependent tangential shifts as
\begin{equation}
  \opgom = \vartheta(r) \srpom \eqdot
\end{equation}

This operator has indeed all the required properties. It is a scalar
operator of rank two both in coordinate and spin-space. It does not
shift the relative wave function radially because
$\srpom$ commutes with $\op{r}=|\vecr|$. This can be
shown by using the commutator relation
\begin{equation} 
  \bigl[\,\op{p}_{\Omega k}\, ,\op{r}_l \, \bigr] =
  \I \left(\frac{\op{r}_k \op{r}_l}{r^2}
    -\delta_{kl} \right) \eqcomma
  \label{eq:commrpom}
\end{equation}
where $k,l$ denote the three spatial directions.
With Eq.~(\ref{eq:commrpom}) it is also easy to
proof that $\srpom$ is hermitean and symmetric under particle exchange.

The strength of the tensor correlation can be adjusted with the tensor
correlation function $\vartheta(r)$ for each distance $r$
independently. Like the shift function $s(r)$ in the case of radial
correlations, $\vartheta(r)$ will depend on the potential and on the
system under investigation.

It is very illustrative to look at the generator $\opgom$ in angular
momentum representation. The angular and spin dependence of $\opgom$
is contained in the $\srpom$ operator.  The calculation of its matrix
elements is outlined in appendix~\ref{app:metensorops}. We notice
that all diagonal matrix elements are zero. The operator $\srpom$ and
therefore also the generator $\opgom$ only connects states with
$L-L'=\pm 2$ and the same total angular momentum $J,M$. For example
in the $J=1$ subspace the matrix elements of $\srpom$ are
\begin{center}
\begin{tabular}{@{}l|ccc@{}}
  $\srpom$ & $\ket{(01)1}$ & $\ket{(11)1}$ & $\ket{(21)1}$ \\ \hline
  $\bra{(01)1}$ & $0$ & $0$ & $-3 \I \sqrt{2}$ \\
  $\bra{(11)1}$ & $0$ & $0$ & $0$  \\
  $\bra{(21)1}$ & $3\I\sqrt{2}$ & $0$ & $0$ \\ \hline
\end{tabular}
\end{center}

Using these matrix elements we can immediately write the correlated
deuteron wave function as
\begin{multline}
  \matrixe{r}{\opcom}{d;1M} =
  \matrixe{r}{\exp\bigl\{-\I\,\vartheta^d(r)\srpom\bigr\}}{d;1M} =
  \\
  \psi^d_0(r)\Bigl[ \,
  \cos\!\left(3\sqrt{2}\,\vartheta^d(r)\right) \ket{(01)1M} +
  \sin\!\left(3\sqrt{2}\,\vartheta^d(r)\right) \ket{(21)1M}
  \, \Bigr] \eqdot
\end{multline}

When we compare this to the exact deuteron solution
Eq.~(\ref{eq:deuteronansatz}) we find the deuteron correlation
function
\begin{equation}
  \label{eq:thetadeuteron}
  \vartheta^d(r) = \frac{1}{3 \sqrt{2}} \arctan
  \frac{\corr{\psi}^d_2(r)}{\corr{\psi}^d_0(r)} \eqdot
\end{equation}

\begin{figure}[tb]
  \begin{minipage}[b]{0.48\textwidth}
    \caption{Deuteron correlation functions $\vartheta^d(r)$ for
      the Bonn-A and Argonne~V18 interactions. }
    \label{fig:deuteroncorrelator}
    \vspace{3ex}
  \end{minipage}\hfil
  \includegraphics[width=0.48\textwidth]{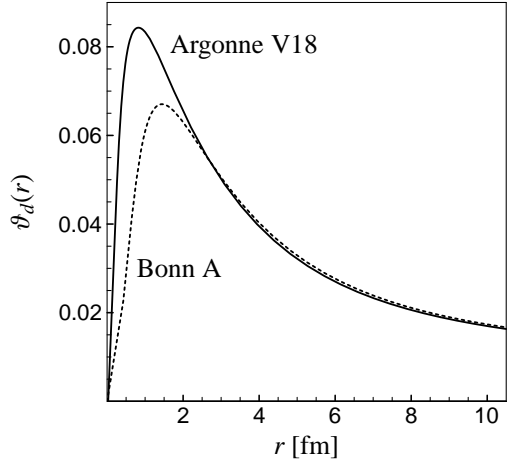}
\end{figure}

The deuteron correlation functions $\vartheta^d(r)$ for the Bonn-A and
the Argonne V18 potential are shown in
Fig.~\ref{fig:deuteroncorrelator}.  The correlations are stronger in
case of the Argonne~V18 interaction at short distances and show a
different behavior for $r \rightarrow 0$ indicating the stronger
tensor force in the AV18 potential.

The uncorrelated deuteron trial state
Eq.~(\ref{eq:deuteronuncorrelatedansatz}) has no $L=2$ admixture at
all. Thus the deuteron correlator which maps the trial state onto the
exact deuteron solution has to generate the entire radial wave
function $\corr{\psi}^d_2(r)$. The correlator is therefore very long
ranged and will induce three- and higher-body correlations in
many-body states.  In order to avoid the higher-body terms one should
not put all responsibility for creating long-ranged or low-momentum
$L=2$ admixtures on the correlator, but should include those into the
degrees of freedom of the uncorrelated many-body trial state. Only the
short-range or high-momentum part of the tensor correlations should be
taken care of by the correlator $\opCom$.  Such a correlator can then
be used efficiently in many-body calculations because the two-body
approximation is valid when the correlation range is short enough.

In the discussion of tensor correlations in the deuteron the radial
correlations have not been considered explicitly but were already
included in the ``tensor-uncorrelated'' state $\ket{d;1M}$. In
the following sections we combine both correlations.

\subsection{Correlated operators in coordinate space representation}

Using the unitarity of $\opcr$ and the transformation properties
(\ref{eq:crops}) it is easy to see that
\begin{equation}
  \hopcr \, \opvecpom \, \opcr =
  \frac{r}{\Rp(r)}\  \opvecpom \qquad \mathrm{and} \qquad
  \hopcr \, \srpom \, \opcr =  \srpom \eqdot
\end{equation}
From that follows immediately that the radially correlated
tensor correlator
\begin{equation}
  \hopcr \opcom \opcr
  = \hopcr \exp \bigl\{ -\I\,\vartheta(r) \srpom \bigr\} \opcr
  = \exp \bigl\{ -\I\,\vartheta(\Rp(r)) \srpom \bigr\}
\end{equation}
differs only by a transformed radial dependence of the tensor
correlation function $\vartheta(r)$. The technical advantage of
unitary correlations is quite obvious in this example.

For tensor correlated operators
\begin{equation}
  \hopcom \, \op{a} \, \opcom  = e^{\I \opgom}\, \op{a}\, e^{-\I \opgom}
\end{equation}
it is no longer possible to give a closed form for the most general
operator $\op{a}$ in coordinate representation.  We can use the
Baker-Campbell-Hausdorff formula
\begin{equation}
  \begin{split}
  \hopcom \, \op{a} \, \opcom
    & = \op{a} + \I \comm{\opgom}{\op{a}} + \frac{i^2}{2}
  \comm{\opgom}{\comm{\opgom}{\op{a}}} + \ldots \\
  & = e^{\sopLom} \op{a} \eqcomma
  \end{split}
  \label{eq:tensorcorrop}
\end{equation}
and the superoperator $\sopLom = \I\,\comm{\opgom}{\circ}$ to evaluate
the expression in two-body space.  In the case of the Hamilton
operator for example we have to to evaluate the repeated application
of $\sopLom$ on the kinetic energy and all components of the
potential.

The tensor correlation of the radial momentum $\oppr$ and distance
$\op{r}$ are special cases where the power series
(\ref{eq:tensorcorrop}) terminates
\begin{equation}
  \begin{split}
    \hopcom \, \op{r}  \opcom & =  \op{r} \\
    \hopcom \, \oppr  \opcom & =  \oppr - \vartheta'(r) \srpom \eqdot
  \end{split}
  \label{eq:tensorcorrpr}
\end{equation}
$\vartheta'(r)$ denotes the first derivative of $\vartheta(r)$. With
Eq.~(\ref{eq:tensorcorrpr}) the tensor correlated radial kinetic energy 
is readily calculated as
\begin{equation} 
  \begin{split}
    \hopcom \, \frac{\oppr^2}{m}  \opcom  &=  \frac{1}{m}
    \bigl(\oppr - \vartheta'(r) \srpom \bigr)^2 \\
    &= \frac{\oppr^2}{m}-
    \bigl(\oppr \vartheta'(r)+ \vartheta'(r)\oppr\bigr)\srpom
    +\bigl(\vartheta'(r)\bigr)^2 \srpom^2 \eqcomma
  \end{split} 
  \label{eq:tensorcorrtr}
\end{equation}
where $\srpom^2$ can also be written as (see Eqs.~(\ref{eq:lssq})
and (\ref{eq:srpomsq}))
\begin{equation}
\srpom^2=9\ \bigl(\op{\vec{s}}^2 + 3\ \ls + (\ls)^2\bigr) \eqdot
\end{equation}
One sees that the tensor correlation creates again structures that
exist already in the potential, see Eq.~(\ref{eq:av18representation}).

For other parts of the interaction, like the spin-orbit $\ls$ or the
tensor one $\srr$, the power series (\ref{eq:tensorcorrop}) does not
terminate but results in ever higher powers of angular momentum and
tensor operators.  The coordinate space representation is discussed in
\cite{neff:diss} and will be pursued further in a forthcoming paper.
As it is no longer possible to give closed expressions for the most
general tensor correlated operator $\hopcom \op{a} \opcom $ in
coordinate representation, in the following sections we shall use the
angular momentum representation where it is easy to state explicitly
the correlated states and correlated interaction.

\subsection{Tensor correlated angular momentum eigenstates}
\label{sec:tensorinangmomeigenstates}

The action of the tensor correlator can be evaluated explicitly in
two-body angular momentum eigenstates $\left|(LS)JM\right>$ that are
used for example in shell-model calculations.  The operator $\srpom$
is a scalar two-body operator that is a tensor of rank two in
coordinate and spin space. Therefore it connects only states with the
same total angular momentum $J,M$ and spin $S=1$. In addition it is
constructed to have only off-diagonal matrix elements for $|L'-L|=2$
so that for each $J$-subspace it is only a $2\times2$ matrix and thus
easy to exponentiate.

The matrix elements of $\srpom$ are (see App.~\ref{app:mesrpomangmom})
\begin{multline} 
  \matrixe{(L'S')J'M'}{\srpom}{(LS)JM} =\\ 
  \I\,3\,\sqrt{J(J+1)} \bigl(\delta_{L',J+1}\delta_{L,J-1} - 
  \delta_{L'\!,J-1}\delta_{L,J+1}\bigr)
  \delta_{S'\!,1}\delta_{S,1} \delta_{J'\!,J}\delta_{M'\!,M} \eqdot
  \label{eq:srpommatr}
\end{multline}
For the sake of convenience we introduce the angle
\begin{equation}
 \theta^{(J)}(r)= 3\, \sqrt{J(J+1)} \, \vartheta(r) \eqdot
 \label{eq:thetaJ}
\end{equation}
With (\ref{eq:srpommatr}) and (\ref{eq:thetaJ})
states with $L=J\mp 1$ are tensor correlated as
\begin{multline}
  \opcom \ket{(J\!\mp\!1,1)JM} =  \\
  \cos \bigl( \theta^{(J)}(r) \bigr) \ket{(J\!\mp\!1,1)JM} \pm
  \sin \bigl( \theta^{(J)}(r) \bigr) \ket{(J\!\pm\!1,1)JM} \eqcomma
  \label{eq:corrangmomstate}
\end{multline}
while states with $L=J$  remain unchanged for both,
$S=0$ and $S=1$,
\begin{equation}
  \opcom \ket{(JS)JM} = \ket{(JS)JM} \eqdot
  \label{eq:corrangmomstatenull}
\end{equation}

\subsection{Correlated operators in angular momentum representation}

For the evaluation of two-body matrix elements it is more convenient
to work with radially correlated operators and tensor correlated
states. Therefore we make use of the unitarity of the radial
correlation operator $\opcr$ and evaluate correlated two-body matrix
elements as
\begin{equation}
  \matrixe{\phi}{\hopcr \hopcom \op{a} \opcom \opcr}{\psi}  =
  \matrixe{\phi}{(\hopcr\opcom\opcr)^{\dagger}(\hopcr\op{a}\opcr)(\hopcr\opcom\opcr)}{\psi} \eqdot
  \label{eq:tensorcentralcorrelated}
\end{equation}
Using this trick radially correlated operators $\hopcr \op{a} \opcr$
can be used in tensor correlated states which are uncorrelated with
respect to the short-range repulsion.

All operators considered in the following are scalar, hence they are
diagonal in $J,M$ and their matrixelements do not depend on the
magnetic quantum numbers $M=-J, \ldots ,J$. Therefore we will omit $M$
and write only $\ket{(LS)J}$. The isospin quantum numbers $T,M_T$ are
omitted as well.

As a consequence of Eq.~(\ref{eq:corrangmomstatenull}) the interaction
in the states $\ket{(JS)J}$ will not be influenced by the tensor
correlations
\begin{equation}
  \matrixe{(JS)J}{\hopcr\hopcom\op{h}\opcom\opcr}{(JS)J} =
  \matrixe{(JS)J}{\hopcr\op{h}\opcr}{(JS)J} \eqdot
\end{equation}

The diagonal matrix elements of the two-body radial kinetic energy
that is fully correlated are calculated using Eqs.~(\ref{eq:ctrad})
and (\ref{eq:tensorcorrtr}) as
\begin{multline}
  \matrixe{(J\!\mp\!1,1)J}{\coptwo{t}_{r}}{(J\!\mp\!1,1)J} = \\
  \begin{split}
    & = \matrixe{(J\!\mp\!1,1)J}{\hopcr \hopcom \op{t}_{r} \opcom
      \opcr - \op{t}_{r}}{(J\!\mp\!1,1)J} \\
    & = \frac{1}{2} \biggl[ \pr^2 \frac{1}{2\cmur(r)} +
    \frac{1}{2\cmur(r)} \pr^2 \biggr] + \frac{J(J+1)}{m} \Bigl(3\,
    \vartheta'(\Rp(r)) \Bigr)^2 + \hat{w}(r)\eqdot
  \end{split} 
  \label{eq:ctradangmom}
\end{multline}
The contribution from the tensor correlation is a centrifugal-like
term originating from the radial dependence of the tensor correlation
function. ``Radial mass'' $\cmur(r)$ and potential term $\hat{w}(r)$
are unchanged from Eq.~(\ref{eq:ctradmass}) and (\ref{eq:ctradu}).

The mixture of different angular momenta by the tensor correlator
leads to the following fully correlated kinetic energy of the
orbital motion in the two-body system
\begin{multline}
  \matrixe{(J\!\mp\!1,1)J}{\coptwo{t}_{\Omega}}{(J\!\mp\!1,1)J} =
  \matrixe{(J\!\mp\!1,1)J}{\hopcr\hopcom\op{t}_{\Omega}\opcom\opcr -
    \op{t}_{\Omega}}{(J\!\mp\!1,1)J}= \\
  \begin{aligned}
    =  \frac{1}{m(\Rp(r))^2}
    \Bigl\{ &\cos^2 \bigl( \theta^{(J)}(\Rp(r)) \bigr)
    \matrixe{(J\!\mp\!1,1)J}{\lsq}{(J\!\mp\!1,1)J}  \\
    & + \sin^2 \bigl( \theta^{(J)}(\Rp(r)) \bigr)
     \matrixe{(J\!\pm\!1,1)J}{\lsq}{(J\!\pm\!1,1)J} \Bigr\}  \\
     - \frac{1}{mr^2}&\matrixe{(J\!\mp\!1,1)J}{\lsq}{(J\!\mp\!1,1)J}\eqdot
\label{eq:ctomangmom}   \end{aligned}
\end{multline}

The central potentials
$\op{v}^{c} = v^{c}(\op{r})$
are not affected by the tensor correlations
\begin{equation}
  \matrixe{(J\!\mp\!1,1)J}{\hopcr \hopcom \op{v}^{c} \opcom
    \opcr}{(J\!\mp\!1,1)J} =
  v^{c}(\Rp(r)) \eqdot
  \label{eq:cromvc}
\end{equation}
Like the above terms the spin-orbit interaction $\op{v}^{b}=v^{b}(r)\:\ls$
has only diagonal contributions from the distinct $L$ channels of
the correlated states have to be considered
\begin{multline}
  \matrixe{(J\!\mp\!1,1)J}{\hopcr \hopcom \op{v}^{b} \opcom
    \opcr}{(J\!\mp\!1,1)J} = \\
  \begin{aligned}
    = v^b(\Rp(r)) \: \Bigl\{ & \cos^2 \bigl(
  \theta^{(J)}(\Rp(r)) \bigr) \:
  \matrixe{(J\!\mp\!1,1)J}{\ls}{(J\!\mp\!1,1)J} \\
  & + \sin^2 \bigl( \theta^{(J)}(\Rp(r)) \bigr) \:
  \matrixe{(J\!\pm\!1,1)J}{\ls}{(J\!\pm\!1,1)J}  \Bigr\} \eqdot
  \label{eq:cromvls}
  \end{aligned}
\end{multline}
The matrix elements of the correlated tensor interactions $\op{v}^{t}
= v^{t}(r) \: \srr$ also include contributions from the
off-diagonal matrix elements between the $L$ channels of the
correlated state
\begin{multline}
  \matrixe{(J\!\mp\!1,1)J}{\hopcr \hopcom \op{v}^{t} \opcom
    \opcr}{(J\!\mp\!1,1)J} = \\
  \begin{aligned}
    v^t&(\Rp(r)) \: \Bigl\{  \cos^2 \bigl(
  \theta^{(J)}(\Rp(r)) \bigr) \:
  \matrixe{(J\!\mp\!1,1)J}{\srr}{(J\!\mp\!1,1)J} \\
  & \pm 2 \cos \bigl( \theta^{(J)}(\Rp(r)) \bigr)
  \sin \bigl( \theta^{(J)}(\Rp(r)) \bigr) \:
  \matrixe{(J\!\mp\!1,1)J}{\srr}{(J\!\pm\!1,1)J} \\
  & + \sin^2 \bigl( \theta^{(J)}(\Rp(r)) \bigr) \:
  \matrixe{(J\!\pm\!1,1)J}{\srr}{(J\!\pm\!1,1)J} \Bigr\} \eqdot
  \label{eq:cromvt}
  \end{aligned}
\end{multline}
The calculation of matrix elements that are off-diagonal in angular
momentum is straightforward using
Eqs.~(\ref{eq:corrangmomstate},\ref{eq:corrangmomstatenull}).


\section{Bonn-A and Argonne-V18 correlators}
\label{sec:bonnargonnecorrelators}

\subsection{Choice of correlation functions}

In this section we discuss the choice of the correlation functions
$\Rp(r)$ and $\vartheta(r)$. We use two general concepts: minimizing
the ground state energy and mapping of uncorrelated onto exact
eigenstates of the Hamiltonian. Both will yield very similar answers.

The ansatz for the correlated many-body state
\begin{equation}
  \ket{\corr{\Psi}} = \opC \ket{\Psi} = \opCom \opCr \ket{\Psi}
\end{equation}
consists of the uncorrelated state $\ket{\Psi}$ and the correlation
operator $\opC$. Both contain degrees of freedom that can be varied
to find the minimum in the energy
\begin{equation}
  \matrixe{\Psi}{\hopC \op{H} \opC}{\Psi}=\mbox{min} \eqdot
  \label{eq:Hmin}
\end{equation}
In $\ket{\Psi}$ they are for example the single-particle states or in
case of a superposition of Slater determinants also their
coefficients.  After having defined the operator structure of $\opCom$
and $\opCr$ the remaining freedom is to choose the correlation
functions $\Rp(r)$ and $\vartheta(r)$. In principle one could do a
functional variation in (\ref{eq:Hmin}) for each nucleus to find the
optimal correlation functions. In practice we parametrize $\Rp(r)$ and
$\vartheta(r)$ and vary only a few parameters.

The other method is to map a typical uncorrelated two-body state at
short distances onto the exact solution. Let us define the exact
eigenvalue problem by
\begin{equation}
  H\ket{\corr{\Psi}_n} = E_n\ket{\corr{\Psi}_n}
\end{equation}
with the exact eigenstates $\ket{\corr{\Psi}_n}$.  The task of
$\opC$ is to map the energetically lowest uncorrelated states as
well as possible onto the exact eigenstates
\begin{equation} \label{eq:Cpsin}
  \opC\ket{\Psi_n} \approx \ket{\corr{\Psi}_n} \eqdot
\end{equation}
As the structure of $\opC$ is given, up to the radial dependences
$\Rp(r)$ and $\vartheta(r)$, in general one representative state
$\ket{\Psi_n}$ in each spin isospin channel is sufficient to determine
those. The physical idea is that the momentum distributions of the
uncorrelated trial states reside in the low momentum regime, while the
short range correlations create high momentum components that are more
or less independent of the long range behaviour described by
$\ket{\Psi_n}$. For the radial correlations this is demonstrated in
Fig.~\ref{fig:argonnerprm_corrwave}. It is obvious that $\opC$ as
defined in the previous sections cannot fulfill Eq.~(\ref{eq:Cpsin})
for all $n$ exactly but it should make the off-diagonal matrix
elements $\bra{\Psi_k}\hopC H \opC\ket{\Psi_n}$ of the
correlated Hamiltonian much smaller than $\bra{\Psi_k} H \ket{\Psi_n}$
of the bare one.

In the uncorrelated many-body states of shell-model or mean-field
configurations two-body states for the relative motion with the lowest
relative angular momenta and relative momenta have the biggest weight.
Therefore the correlators will be determined in the lowest angular
momentum states of the respective spin and isospin channels at zero or
ground state energy.  The influence of the correlations decreases with
increasing relative orbital angular momentum $L$ as the centrifugal
barrier suppresses significantly already the uncorrelated two-body
density at short distances.

In the $S,T=0,0$ and the $S,T=0,1$ channels where we do not have to
deal with tensor correlations the correlators are therefore fixed with
$\ket{(10)1;0}$ and $\ket{(00)0;1}$ trial states. In the $S,T=1,0$
channel our uncorrelated trial state will be $\ket{(01)1;0}$.  The
tensor correlator will admix the $\ket{(21)1;0}$ state. In the
$S,T=1,1$ channel the situation is more complicated. The lowest
orbital angular momentum $L=1$ can be coupled with the spin
$S=1$ to the total angular momenta $J=0,1,2$. Only for $J=2$ we have
to deal with tensor correlations which will admix $\ket{(31)2;1}$
states.

Following these principles we will determine now the radial and tensor
correlators for the Bonn-A and the Argonne~V18 interaction. By
comparing the correlators side-by-side a better understanding of
common and specific properties of the different nuclear interactions
and the correlations can be achieved.

The Argonne~V8' interaction is by construction identical to the
Argonne~V18 interaction in the lowest $L$ channels. As we fix our
correlators in these channels the Argonne~V18 correlators presented
are identical to the Argonne~V8' correlators.

Correlators that minimize the energy of the many-body system will only
be given for the $\Hefour$ nucleus.  For heavier nuclei like
$\Osixteen$ and $\Cafourty$ they turn out to be nearly
indistinguishable from the two-body optimized correlators.  This shows
that the short range correlations are essentially independent of the
size of the many-body system.  The nuclear saturation property limits
the largest possible density so that adding more nucleons has
negligible effect on short range correlations between two nucleons.
This is in accord with the findings of Ref.~\cite{forest96} where the
short-range tensor structure of the deuteron is also found in heavier
nuclei.

The correlation functions determined from mapping the lowest energy
eigenstate in the two-body system will not be used in many-body
calculations but are shown in the graphs for comparison.  Those
resulting from minimizing energies are parametrized in the style of
the ones obtained by mapping the uncorrelated states on the exact
eigenstates.  The form and the three or four parameters which
determine the parametrized correlation functions are summarized in
Appendix~\ref{app:correlators}.

\subsection{$S=0, T=1$ channel}

\begin{figure}[bt]
  \includegraphics[width=0.48\textwidth]{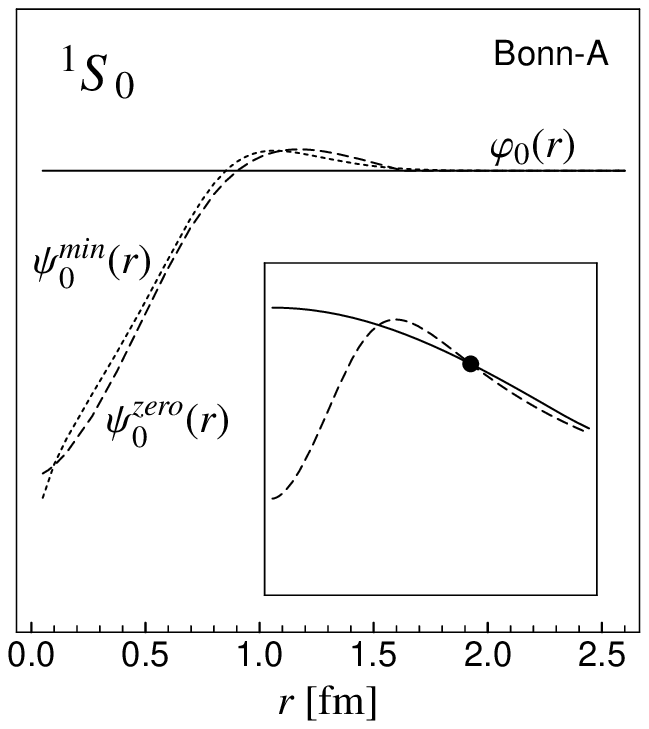}\hfil
  \includegraphics[width=0.48\textwidth]{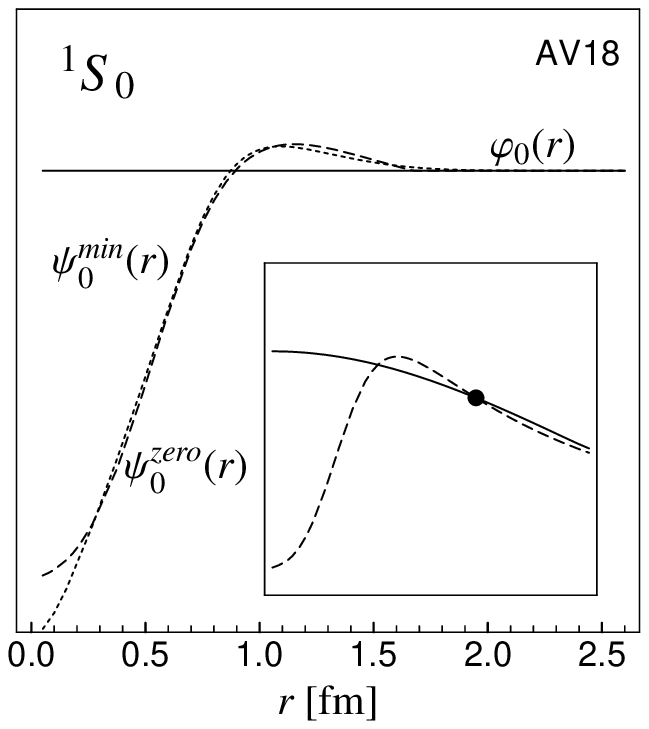}
  \vspace{1ex}
  \includegraphics[width=0.48\textwidth]{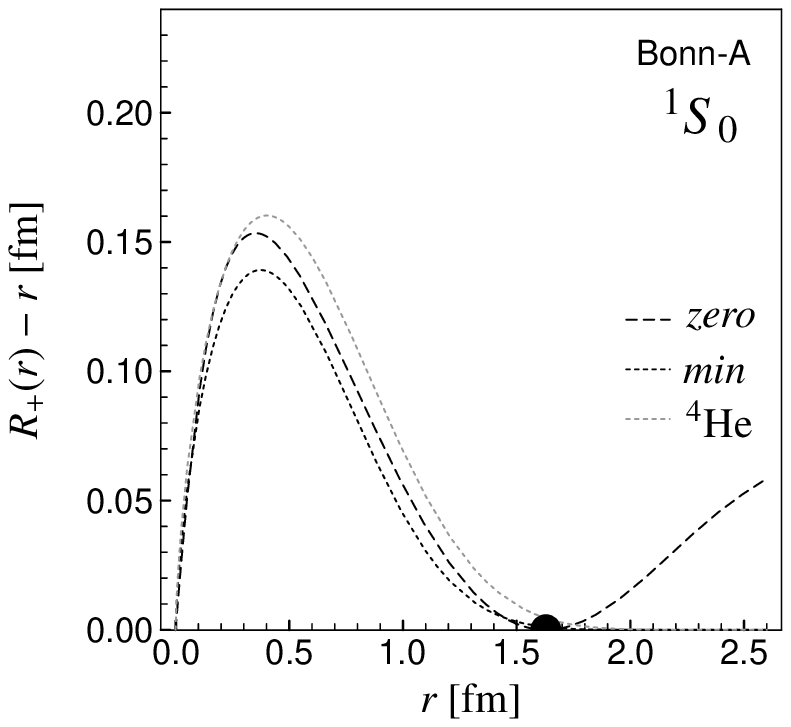}\hfil
  \includegraphics[width=0.48\textwidth]{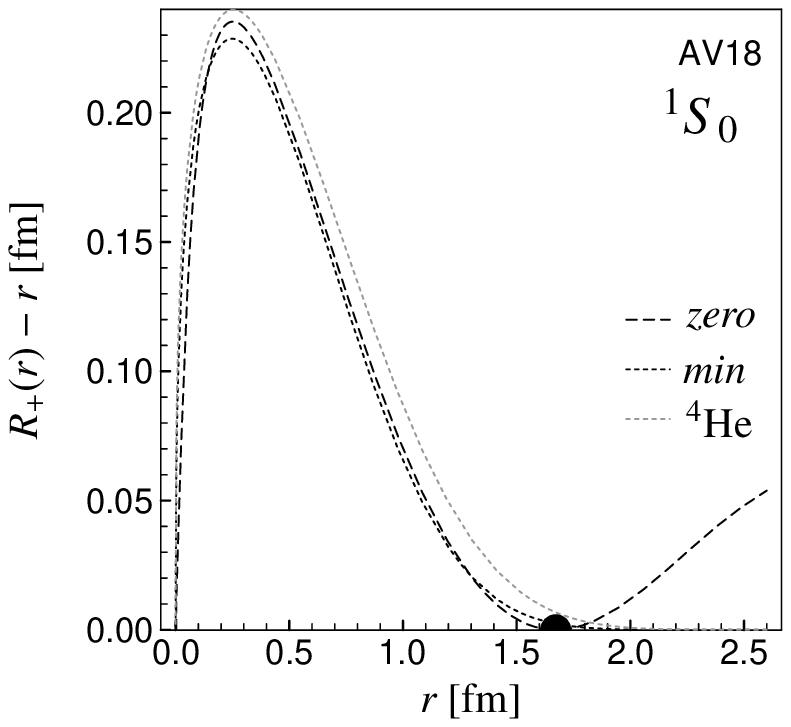}
  \caption{Radial correlations in the $S,T=0,1$ channel
    for the Bonn-A (left hand side) and the Argonne~V18 interaction
    (right hand side). In the upper part $\psi_0^\mathit{min}(r)$ is
    obtained by minimizing the energy with the constant trial function
    $\varphi_0(r)=\mathit{const}$. Mapping a Gaussian trial function
    onto the zero-energy scattering solution as displayed in the
    insets of the upper part defines the correlation function labelled
    ``$\mathit{zero}$''. Applying this correlator to the constant
    trial function $\varphi_0(r)$ results in
    $\psi_0^{\mathit{zero}}(r)$. The correlation functions are
    displayed in the lower part together with the $\Hefour$ optimized
    correlation function.}
  \label{fig:centralcorrelator01}
\end{figure}

In the $S,T=0,1$ channel, where we do not have to deal with tensor
correlations, the radial correlators are fixed with the zero-energy
scattering state $\ket{k=0;(00)0;1}$ state with $L=0$. We obtain
the correlators labelled ``$\mathit{zero}$'' from the mapping of a
Gaussian trial function onto the scattering solution and the
correlators labeled ``$\mathit{min}$'' by minimizing the energy in the
two-body system with a constant trial function \cite{ucom98}. The
resulting correlated radial wave functions are shown in the upper part
of Fig.~\ref{fig:centralcorrelator01} and the corresponding radial
correlation functions in the lower part.  In addition we determine a
correlator optimized for the $\Hefour$ nucleus in two-body
approximation. Here we use a harmonic oscillator shell-model trial
state which reproduces the experimental radius of the $\Hefour$
nucleus. All three correlators turn out to be very similar indicating
that in this channel there is little ambiguity in separating the short
range from the long range behaviour or the high momentum from the low
momentum content in the relative motion.

Comparing the Bonn with the Argonne correlator we observe however
considerable differences. The depletion of the Argonne scattering
solution at small distances $r$ is much stronger than in case of the
Bonn potential. The correlation functions of the Argonne interaction
are correspondingly stronger but of the same range as the Bonn
correlation functions.

\subsection{$S=1, T=0$ channel}
\begin{figure}[b]
  \includegraphics[width=0.48\textwidth]{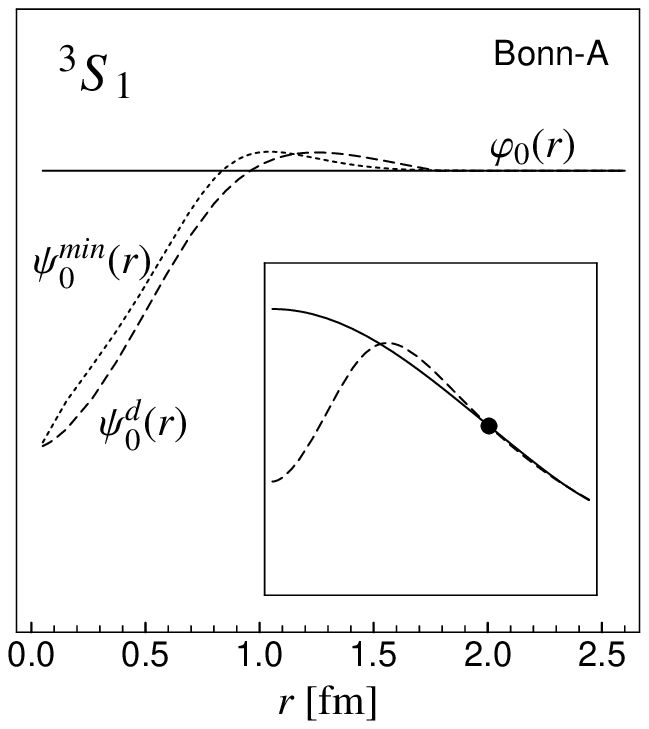}\hfil
  \includegraphics[width=0.48\textwidth]{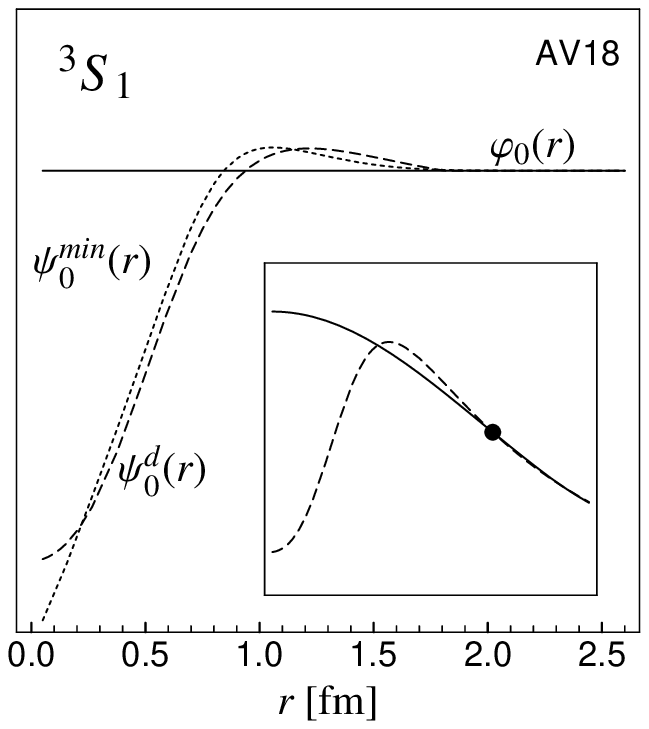}
  \vspace{1ex}
  \includegraphics[width=0.48\textwidth]{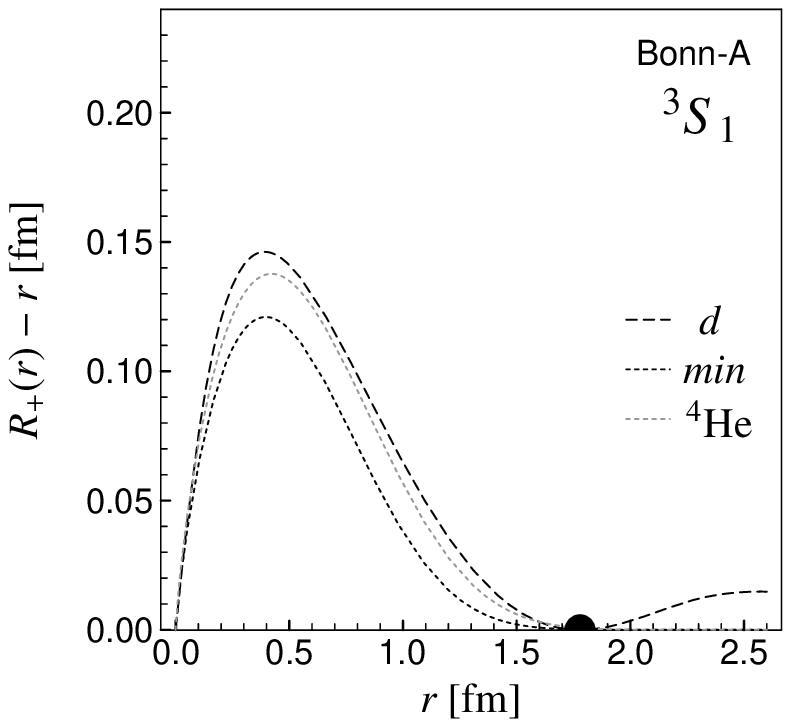}\hfil
  \includegraphics[width=0.48\textwidth]{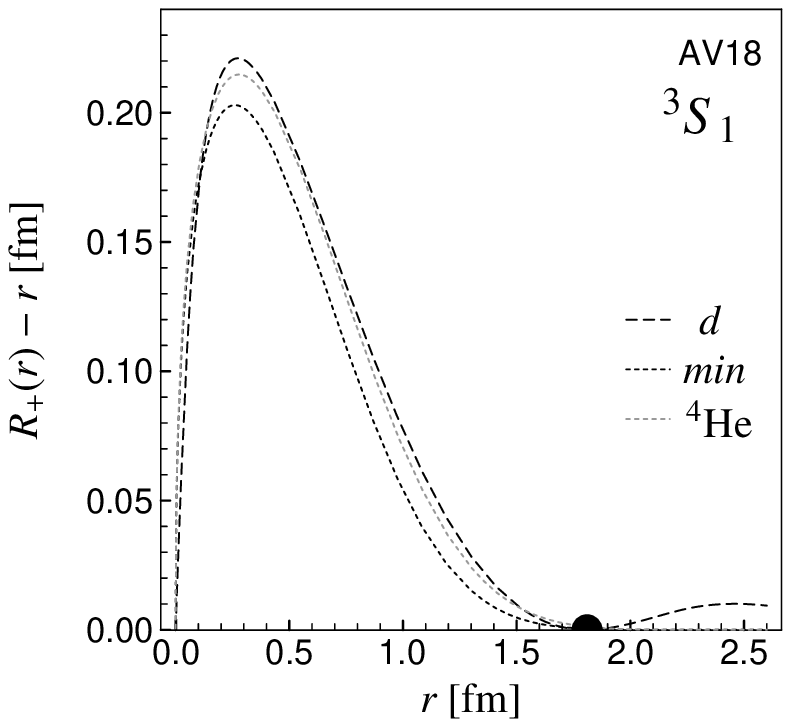}
  \caption{Radial correlation functions $\Rp(r)$ in the $S,T=1,0$
    channel for Bonn-A interaction (left) and the Argonne~AV18
    interaction (right). Applying the radial correlation functions
    labelled ``$\mathit{min}$'' and ``$d$'' to the constant trial
    function $\varphi_0(r)=\mathit{const}$ results in
    $\psi^{\mathit{min}}_0(r)$ and $\psi^d_0(r)$, respectively. In the
    insets the mapping of the Gaussian trial function onto the
    ``tensor-decorrelated'' deuteron solution $\op{c}_r \phi^d_0(r)
    = \psi^d_0(r) =
    \sqrt{\hat{\psi}^d_0(r)^2+\hat{\psi}^d_2(r)^2}$ is indicated. The
    radial correlation functions optimized for the $\Hefour$ are shown
    in addition to the zero-energy scattering and the two-body
    optimized correlation functions in the lower part.}
  \label{fig:centralcorrelator10}
\end{figure}

\begin{figure}[bt]
  \includegraphics[width=0.48\textwidth]{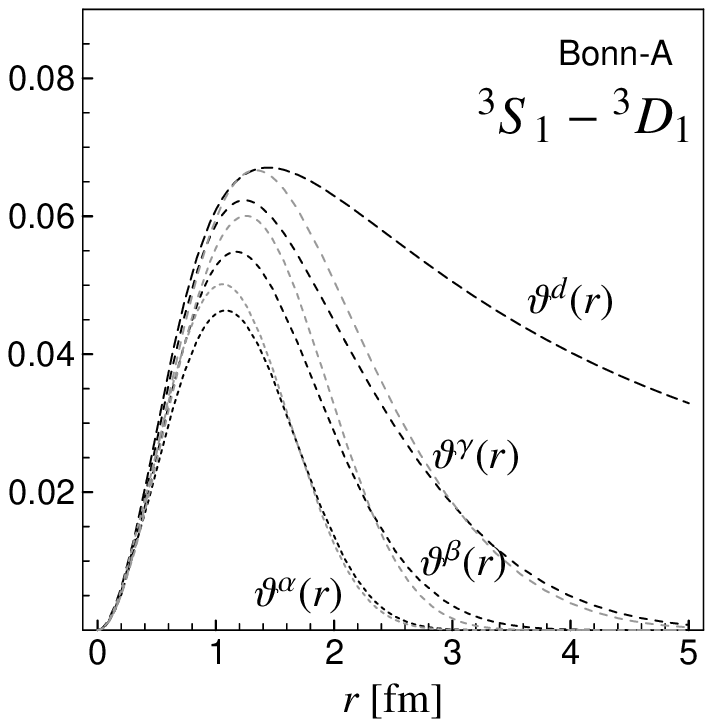}\hfil
  \includegraphics[width=0.48\textwidth]{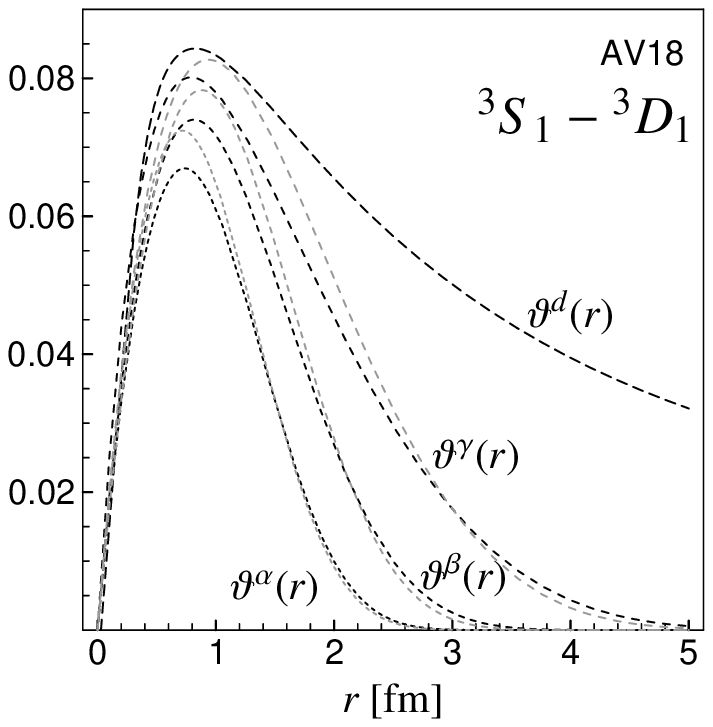}
  \caption{Tensor correlation functions for Bonn-A interaction (left hand
    side) and Argonne~V18 interaction (right hand side). The tensor
    correlation functions $\alpha$, $\beta$ and $\gamma$ are the
    result of a minimization of the energy in the two-body system
    (black short-dashed lines) or in $\Hefour$ trial state (gray
    short-dashed lines) with additional constraints on the range of
    the correlators.}
  \label{fig:tensorcorrelator10}
\end{figure}

In the $S,T=1,0$ channel the lowest energy state is the deuteron
and thus we will use as uncorrelated trial state
\begin{equation}
  \ket{d;(01)1;0}\equiv
  \ket{d;(L=0,S=1)J=1,M;T=0,M_T=0} \eqdot
\end{equation}
As explained in detail in Sec.~\ref{sec:correlations}
we have to deal in this channel with radial and tensor
correlations where the tensor correlator will admix an $L=2$
state.

The energy minimized correlators in the two-body system can be
determined in principle by simultaneously minimizing the energy of
a constant trial function $\varphi_0(r) = \mathit{const}$
\begin{equation}
  \min_{\Rp(r),\vartheta(r)}
  \matrixe{\varphi_0;(01)1;0}
  {\hopcr\hopcom\op{h}\opcom\opcr}{\varphi_0;(01)1;0}
  \label{eq:vartwobody}
\end{equation}
with respect to radial and tensor correlations and additional
constraints concerning the correlation range in case of the tensor
correlator. In practice we proceed in two steps. In the first step we
determine the radial correlation function $\Rp(r)$ using in
Eq.~(\ref{eq:vartwobody}) the deuteron tensor correlation function
$\vartheta^d(r)$, given in Eq.  (\ref{eq:thetadeuteron}). From the
perspective of the radial correlator the long range behavior of the
tensor correlator is not relevant. It therefore makes no real
difference whether we use here the deuteron tensor correlator or a
tensor correlator that has a restricted range. In a second step we
vary the energy in the two-body system with respect to the tensor
correlation function.

We proceed in the same way for the determination of the $\Hefour$
optimized correlators, where the energy in a harmonic oscillator
shell-model trial state is minimized with respect to radial and tensor
correlators
\begin{equation}
  \min_{\Rp(r),\vartheta(r)}
  \matrixe{\Hefour}{\bigl[ \hopCr\hopCom\op{H}\opCom\opCr
    \bigr]^{C2}}{\Hefour} \eqdot
  \label{eq:varhefour}
\end{equation}

We end up with the radial correlators shown in
Fig.~\ref{fig:centralcorrelator10}. The mapping of a Gaussian trial
function on the zero-energy scattering state is indicated in the
insets of Fig.~\ref{fig:centralcorrelator10}.

The radial correlation functions $\Rp(r)$ obtained with the three
methods turn out to be all very similar for the Bonn-A and the
Argonne~V18 interaction, respectively.  But as in the other channels
we can observe that the Argonne interaction induces stronger
correlations, the correlation hole is deeper.

The naive idea to create the whole $L=2$ component only with the
tensor correlator leads to an extremely long ranged correlation
function that extends even outside the range of the interaction.  This
is of course in disaccord with the two-body approximation in the
cluster expansion.  Therefore we construct the tensor correlator by
restricting the range of $\vartheta(r)$ to short distances, where the
correlated states varies rapidly, and presume that the low momentum
part of the $L=2$ state is represented in the uncorrelated trial
state.

Tensor correlation functions $\vartheta^{\alpha,\beta,\gamma}(r)$ with
limited range are obtained by minimizing the energy in the the-body
system with the radially correlated $L=0$ trial state
$\psi^d_0(r)$ under the constraint
\begin{equation} \label{eq:constraint}
  \intd{r} r^2 \: \vartheta^x(r) \overset{!}{=}
  \begin{cases}     0.1\,\fm^3 & ; \quad x=\alpha \\
    0.2\,\fm^3 & ; \quad x=\beta \\
    0.5\,\fm^3 & ; \quad x=\gamma \eqdot
  \end{cases}
\end{equation}
The resulting correlation functions are displayed in
Fig.~\ref{fig:tensorcorrelator10} together with those obtained by
minimizing the energy of the $\Hefour$ trial state in the two-body
approximation. Again both sets of correlation functions are very
similar. Below $r\approx 1\,\fm$ they are almost unaffected by the
limitation in range and even agree with the long ranged deuteron
correlator. Comparing the Bonn-A and Argonne~V18 interaction the
influence of the stronger tensor part in the Argonne~V18 is clearly
visible. The tensor correlator for the zero-energy scattering solution
is not shown because it is practically the same as the one obtained
from the deuteron for distances below $2\,\fm$.

\subsection{$S=0, T=0$ channel}

\begin{figure}[tb]
  \includegraphics[width=0.48\textwidth]{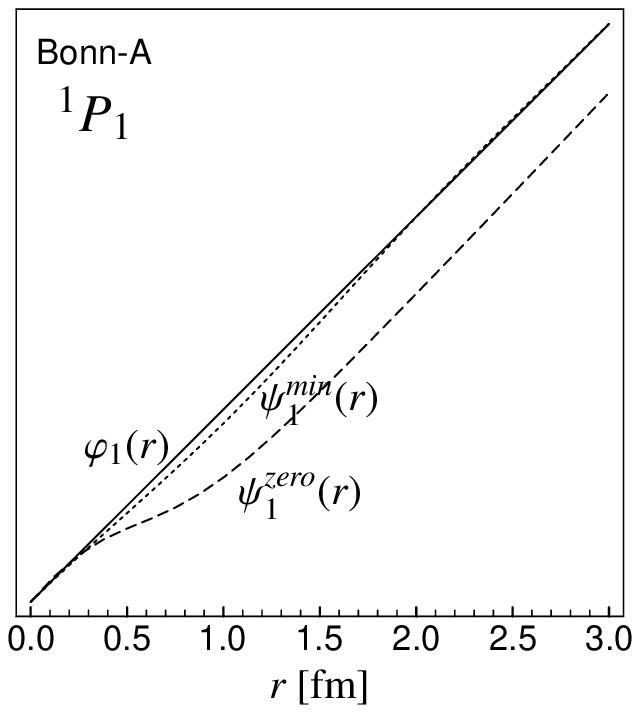}\hfil
  \includegraphics[width=0.48\textwidth]{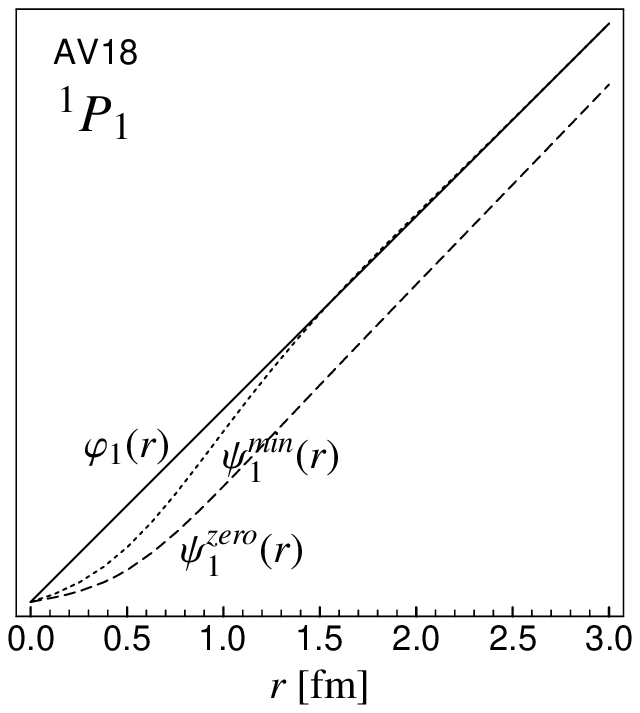}
  \vspace{1ex}
  \includegraphics[width=0.48\textwidth]{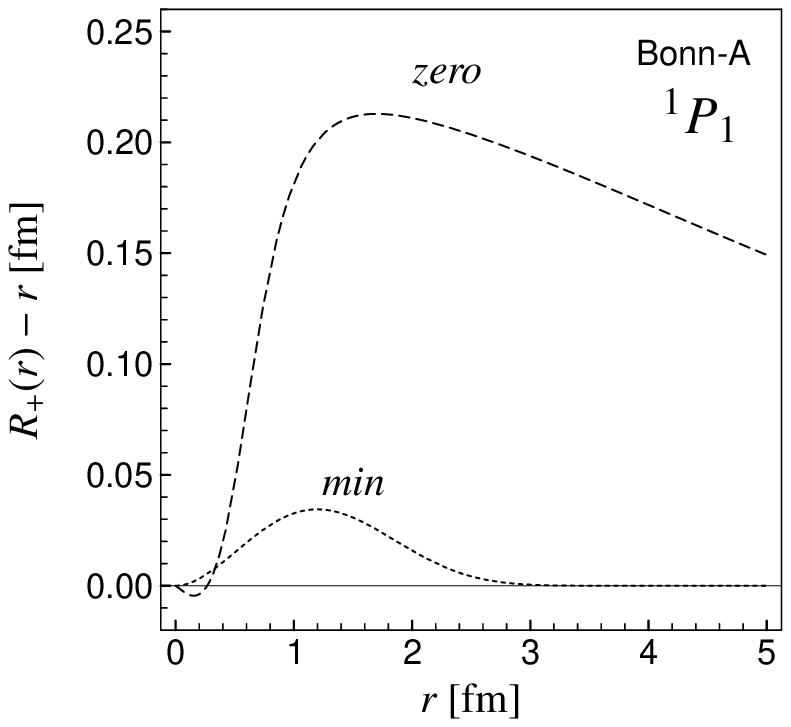}\hfil
  \includegraphics[width=0.48\textwidth]{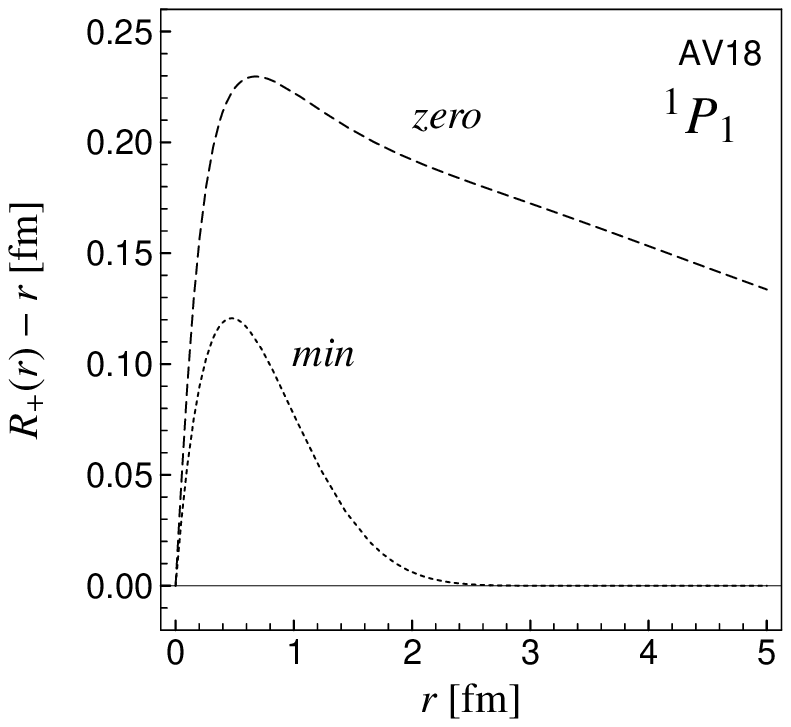}
  \caption{Zero-energy scattering solutions and radial correlation
    functions in the $S,T=0,0$ channel for the Bonn-A interaction
    (left hand side) and the Argonne~V18 interaction (right hand
    side).  The $L=1$ trial state $\varphi_1(r)=r$ is mapped onto the
    zero-energy scattering solution $\psi^{\mathit{zero}}_1(r) =
    \op{c}_r \varphi_1(r)$ with the correlation function
    $\Rp^{\mathit{zero}}(r)$. Minimizing the energy in the two-body
    system with additional constraint on the correlation range leads
    to the correlation functions labelled ``$\mathit{min}$"}
  \label{fig:centralcorrelator00}
\end{figure}

According to our prescription we have to fix the correlator for the
lowest angular momentum, which is $L=1$ in this channel and
hence $J=1$.  Both, the Bonn-A and the Argonne~V18 interaction
show a strong repulsion in the $S,T=0,0$ channel. This can be
seen in the zero-energy scattering solutions
$\psi^{\mathit{zero}}_1(r)$ plotted in the upper part of
Fig.~\ref{fig:centralcorrelator00}.  We observe a remarkable
difference in the Bonn-A and the Argonne~V18 scattering solutions.
Whereas the local Argonne interaction strongly suppresses the
scattering solution at small distances the momentum dependent
repulsion of the Bonn interaction shifts the wave function only
for distances greater than about $0.5\,\fm$. This difference
manifests itself also in the radial correlation functions
$\Rp^{\mathit{zero}}(r)$ derived from mapping the trial function
$\varphi_{L=1}(r) = r$ onto the scattering solutions. The Argonne
correlation function increases steeply for small $r$ whereas the Bonn
correlation function starts to rise at greater distances as seen in
the lower part of Fig.~\ref{fig:centralcorrelator00}. The resulting
correlation functions $\Rp^{\mathit{zero}}(r)$ are extremely long
ranged and cannot be used meaningfully in a many-body calculation.

By minimizing the energy with the trial function $\varphi_1(r)=r$
in the two-body system under the additional constraint
\begin{equation}
\intd{r} r^2 \: (\Rp^\mathit{min}(r)-r) \overset{!}{=}0.1\,\fm^3
\end{equation}
on the correlation range we obtain the correlators indicated by
$\mathit{min}$ that have about the same range as the radial
correlators in even channels. In the upper part of
Fig.~\ref{fig:centralcorrelator00} their effect on the radially
uncorrelated $\varphi_1(r)$ is compared to the zero-energy scattering
solution $\psi^\mathit{zero}_1(r)$.

This odd channel does not occur in the $\Hefour$ shell-model state. In
many-body calculations of larger nuclei we will only present results
obtained with the correlator $\mathit{min}$. Because of the small
weight of the $S,T=0,0$ channel the final many-body results
depend only very weakly on the particular choice of the
radial correlation function in this channel.

\subsection{$S=1, T=1$ channel}

\begin{figure}[bt]
  \includegraphics[width=0.48\textwidth]{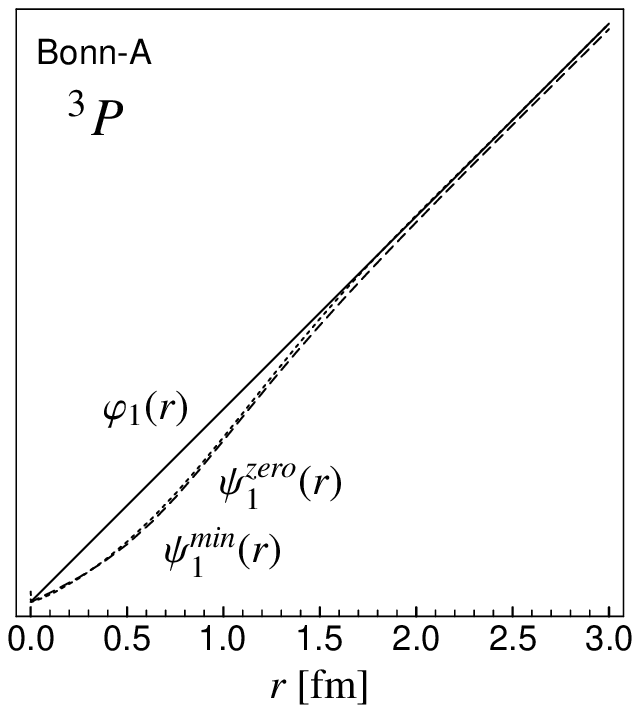}\hfil
  \includegraphics[width=0.48\textwidth]{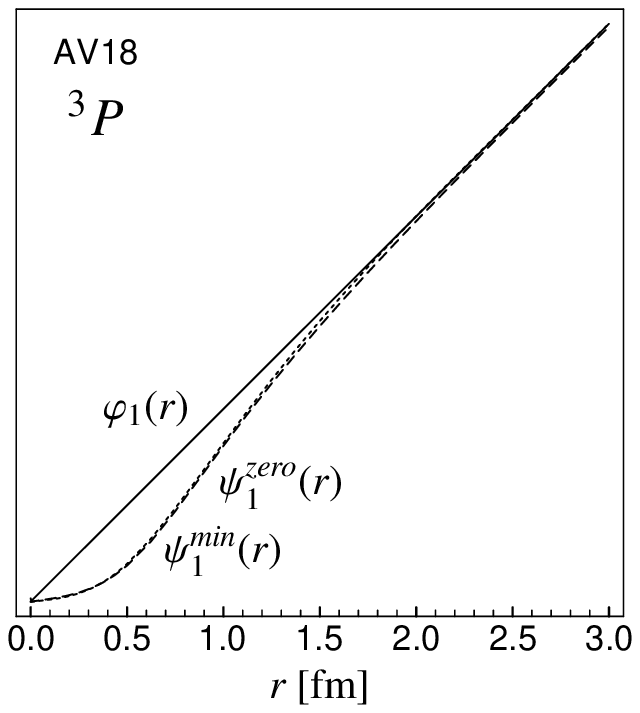}
  \vspace{1ex}
  \includegraphics[width=0.48\textwidth]{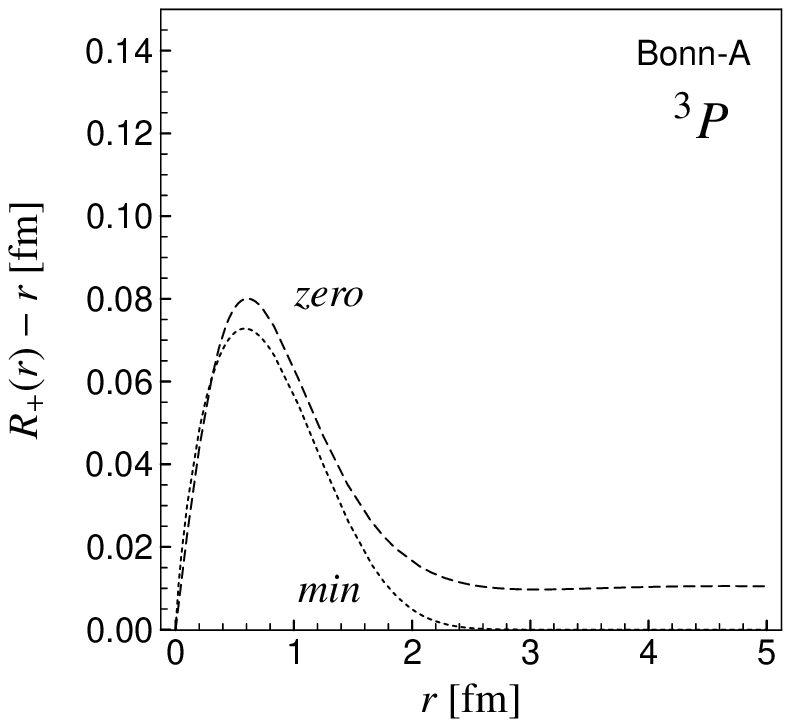}\hfil
  \includegraphics[width=0.48\textwidth]{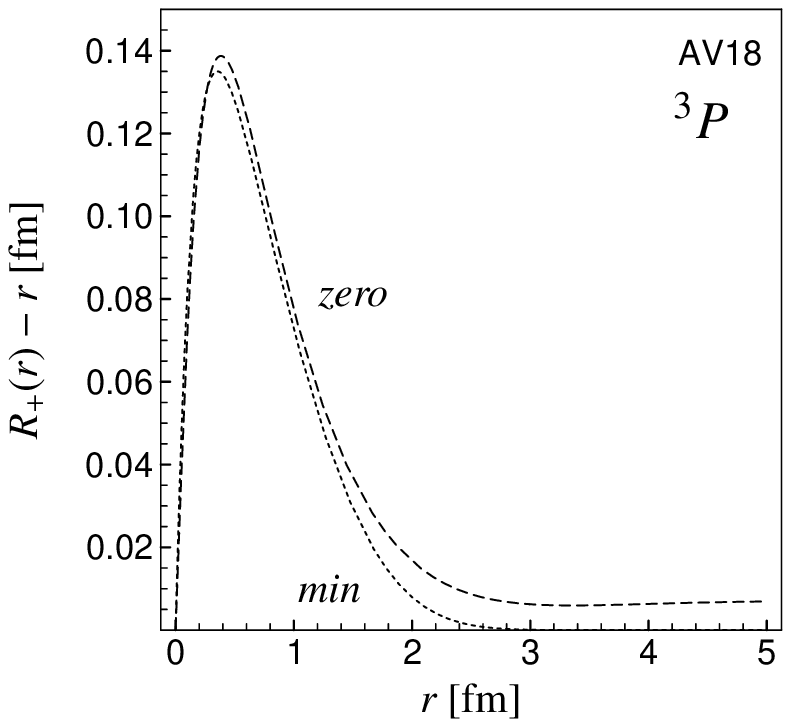}
  \caption{Radial correlation functions $\Rp(r)$ in the $S,T =
    1,1$ channel for Bonn-A interaction (left hand side) and the
    Argonne~V18 interaction (right hand side). The zero-energy
    scattering solution $\psi_1^{\mathit{zero}}(r)$ (dashed line) is
    calculated without the non-central parts of the interaction.
    $\varphi_1(r)=r$ is taken as the uncorrelated zero-energy
    scattering solution. The correlator which results from minimizing
    the averaged energy in the two-body system is labelled
    ``$\mathit{min}$''.}
  \label{fig:centralcorrelator11}
\end{figure}
\begin{figure}[bt]
  \includegraphics[width=0.48\textwidth]{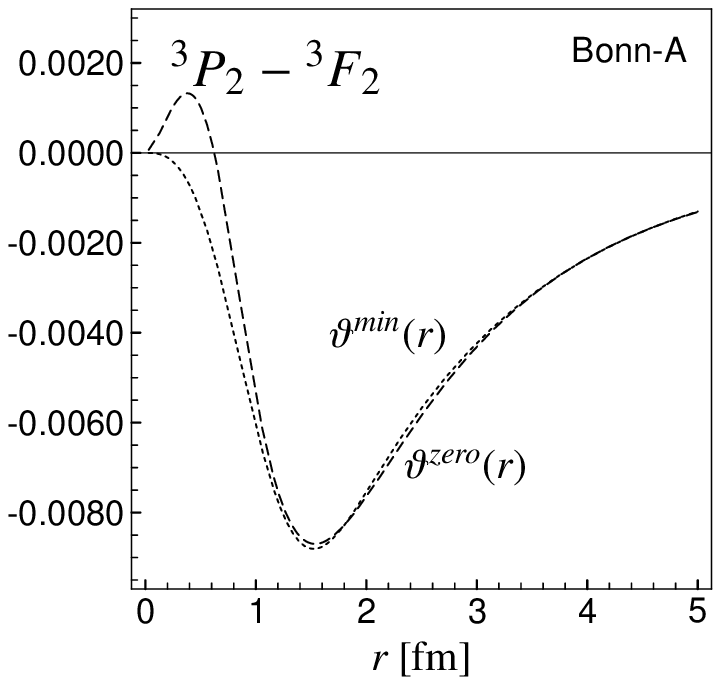}\hfil
  \includegraphics[width=0.48\textwidth]{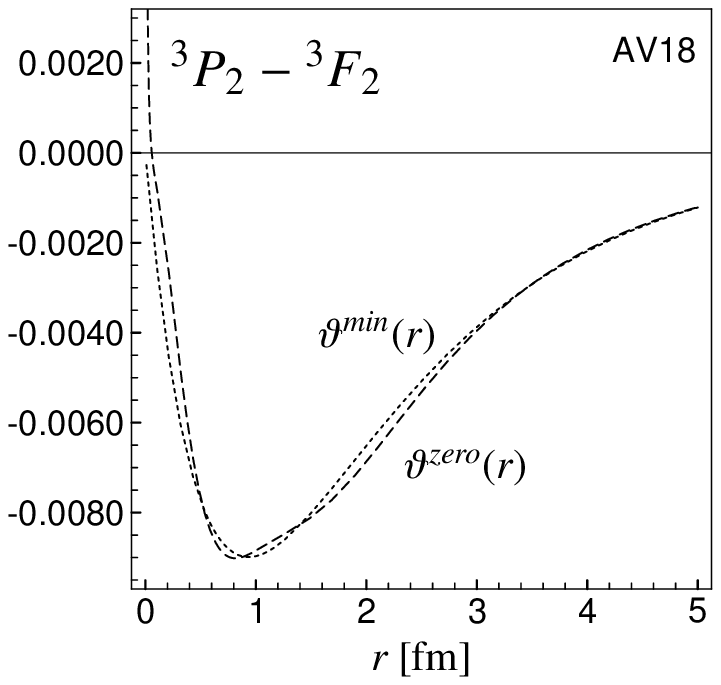}
  \caption{Tensor correlation functions in the $S,T=1,1$
    channel for the Bonn-A (left) and Argonne~V18 (right) interaction.
    Shown are the correlation functions $\vartheta^{\mathit{zero}}(r)$
    (dashed line) determined from the scattering solutions in the
    ${}^3P_2-{}^3F_2$ channel and $\vartheta^{\mathit{min}}(r)$
    (dotted line) determined from the energy minimization in the
    two-body system. Because of the small tensor correlations no
    constraints on the correlation range have been imposed.}
  \label{fig:tensorcorrelator11}
\end{figure}
The lowest orbital angular momentum $L=1$ in the $S,T=1,1$
channel can be coupled to an total angular momentum of $J=
0,1,2$. Tensor correlations affect only the $J=2$ channel. We
can therefore determine the tensor correlator from the zero-energy
scattering solution in the ${}^3P_2-{}^3F_2$ channel. As we want a
radial correlator that is independent of $J$ we fix with the zero-energy
scattering solution obtained with the central part of the interaction
only and not for each $J$ individually.

Minimizing the energy in the two-body system defines our second set of
correlators. Here we minimize the energy for $\varphi_1(r)=r$
\begin{multline}
  E = \frac{1}{9}
  \matrixe{\varphi_1;(11)0;1}{\hopcr \op{h}\opcr}{\varphi_1;(11)0;1}
  +   \frac{3}{9}
  \matrixe{\varphi_1;(11)1;1}{\hopcr \op{h}\opcr}{\varphi_1;(11)1;1} \\
  +\frac{5}{9}
  \matrixe{\varphi_1;(11)2;1}{\hopcr\hopcom\op{h}\opcom\opcr}
  {\varphi_1;(11)2;1}
\end{multline}
that is averaged over the different total angular momenta $J$.  Like
in the $S,T=1,0$ channel we first minimize the energy with
respect to the radial correlator using the tensor correlation
function $\vartheta^{\mathit{zero}}(r)$ derived from the
${}^3P_2-{}^3F_2$ scattering solution. In a second step the energy is
minimized with respect to the tensor correlator.

In contrast to the $S,T=0,0$ channel where the radial correlation
function determined from the zero-energy scattering solution is
extremely long ranged the radial correlator in the $S,T=1,1$ channel
is essentially short-ranged, see Fig.~\ref{fig:centralcorrelator11}.
We therefore do not have to impose a constraint on the correlation
range for the energy minimized correlator.  The tensor correlations
are very weak compared to the $S,T=1,0$ channel and we refrain from
imposing additional constraints on the range of the tensor correlator.
The tensor correlators are displayed in
Fig.~\ref{fig:tensorcorrelator11}.  Like in all the other channels we
can observe stronger correlations in case of the Argonne interaction.

\subsection{Momentum space representation of the interaction}
\label{sec:momentumspace}

To study the effect of unitary correlations on the interaction
in momentum space the
correlated and uncorrelated interactions are evaluated in
eigenstates of momentum and angular momentum as
\begin{align}
  \matrixe{kLM}{\op{V}}{k'L'M'} & = \\
  \I^{L'-L} m \intdt{x}&\!\!\intdt{x'} Y^{\star}_{LM}(\unitvec{x}) j_L(k x)
  \matrixe{\vec{x}}{\op{V}}{\vec{x'}} j_{L'}(k' x')
  Y_{L'M'}(\unitvec{x}') \eqcomma \nonumber \\
  \matrixe{kLM}{\coptwo{H}}{k'L'M'} & = \\i^{L'-L} m \intdt{x}&\!\!\intdt{x'}
  Y^\star_{LM}(\unitvec{x}) j_L(k x)
  \matrixe{\vec{x}}{\coptwo{H}}{\vec{x'}} j_{L'}(k' x')
  Y_{L'M'}(\unitvec{x}') \nonumber \eqdot
\end{align}
The correlated interaction $\coptwo{H}$ consists of the two-body part
of the correlated kinetic energy $\coptwo{T}$ and the correlated
potential $\coptwo{V}$.

\begin{figure}[tb]
  \begin{minipage}[b]{0.48\textwidth}
    \caption{Correlated and uncorrelated Bonn-A and Argonne~V18
      interaction in momentum representation in the $S,T=0,1$; $L=0$
      channel.  Symbols denote the corresponding diagonal elements of
      $\Vlowk$, taken from \cite{kuo01}.}
    \label{fig:vlowk}
    \vspace{3ex}
  \end{minipage}\hfil
  \includegraphics[width=0.48\textwidth]{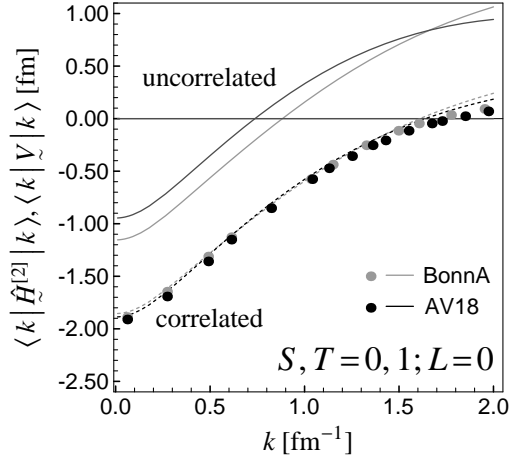}
\end{figure}

The resulting diagonal matrix elements of the interactions in the
$S,T=0,1$ and $L=0$ state are shown in Fig.~\ref{fig:vlowk} for the
uncorrelated and correlated {Bonn-A} and Argonne~V18 interaction.  The
difference in the uncorrelated interactions is mainly due to the
different short range behaviour of $v^c_{01}(r)$ as already seen in
Fig.~\ref{fig:centralpot}. Nevertheless, the correlated interactions
are almost indistinguishable. This shows that the unitary mapping
performed with the respective radial correlators transforms the
different realistic potentials to the same effective interaction in
the low energy regime. In this plot we also show the $\Vlowk$
potential \cite{kuo01} for the Bonn-A and Argonne~V18 interactions. In
case of the $\Vlowk$ renormalization group techniques have been used
to derive the low-momentum potential.  The high relative momentum
modes have been integrated out, while preserving the half-on-shell
T-matrix and bound state properties of the realistic potential.  The
agreement with the correlated interactions reflects that both methods
are based on the same physics, namely treating the short range or high
momentum components by means of an effective interaction while keeping
the low momentum properties of the interactions that are determined by
the low energy phase shifts and bound state properties.

\begin{figure}[b]
  \includegraphics[width=0.48\textwidth]{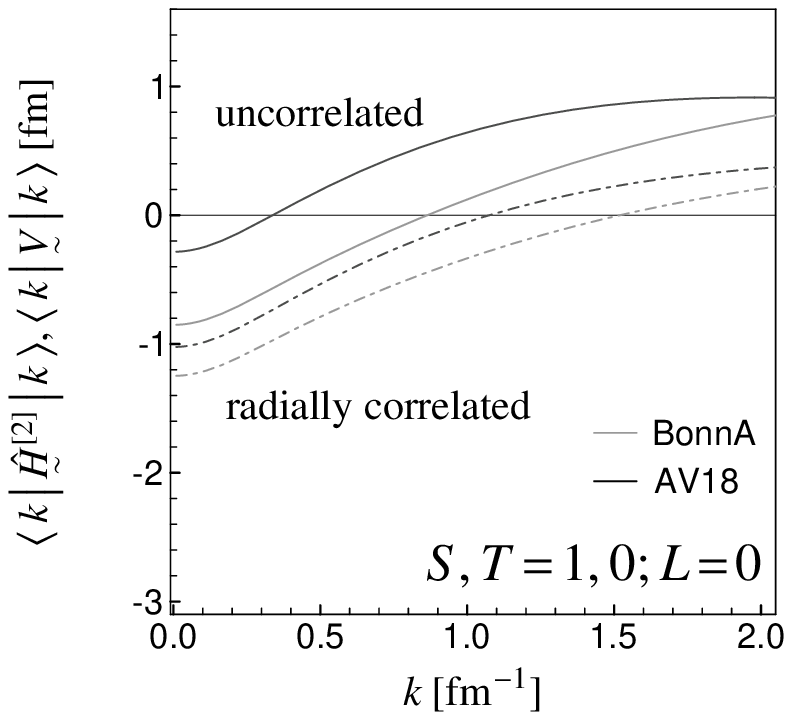}\hfil
  \includegraphics[width=0.48\textwidth]{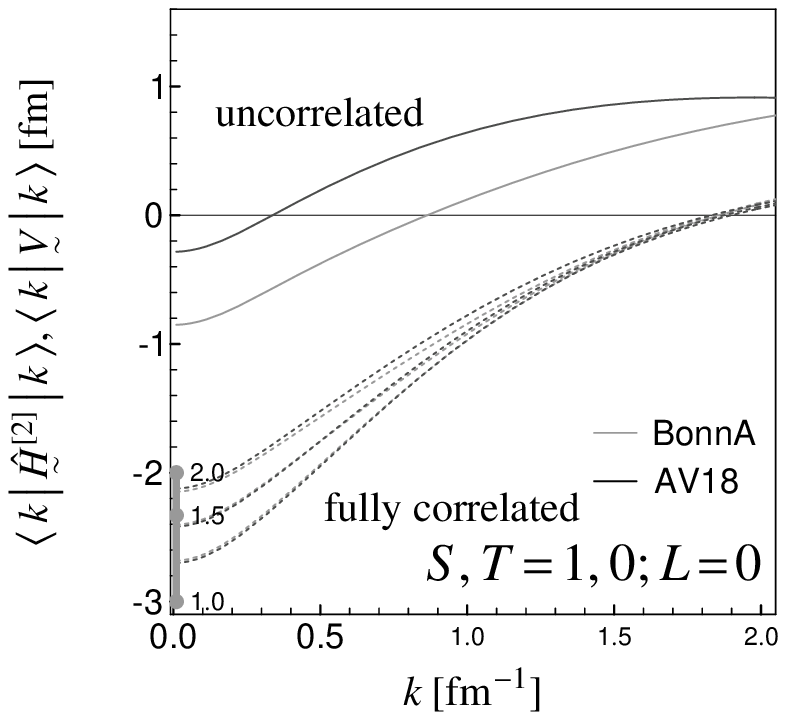}
  \caption{Diagonal matrix elements in momentum space of the
    correlated (dashed) and uncorrelated (solid) Bonn-A and
    Argonne~V18 interaction in the $S,T=1,0$; $L=0$ channel. Left:
    only radial correlations are used.  Right: radial and tensor
    correlations are included.  The three short-dashed curves show the
    results with the three tensor correlators ($\alpha$, $\beta$,
    $\gamma$) of different ranges. The $\Vlowk$ potential depends in
    this channel strongly on the cutoff $\Lambda$. The matrix elements
    for $k=0$ with the cutoffs $\Lambda=1.0,1.5,2.0\:\fm^{-1}$ are
    indicated.}
  \label{fig:vlowk10}
\end{figure}

In Fig.~\ref{fig:vlowk10} the diagonal matrix elements in the
$S,T=1,0$ and $L=0$ channel are shown. Due to the different shapes of
the central potentials $v^c_{10}(r)$ in this channel (see Fig.
\ref{fig:centralpot}) the matrix elements of the uncorrelated Bonn-A
and Argonne~V18 interaction are quite different.  The tensor force
does not contribute in the uncorrelated case because of $L=0$.  The
application of the radial correlator makes both interactions more
attractive but they still differ significantly.  The additional
application of tensor correlator which admixes $L=2$ leads for
all three tensor correlators ($\alpha$, $\beta$, $\gamma$) not only to
strong attraction but also to almost identical correlated
interactions.

This shows that two potentials that describe equally well the low
energy phase shifts may not only differ in their short range behaviour
but also with respect to their relative strength between central and
tensor interaction. Nevertheless the unitary correlator maps on the
same low momentum effective interaction that reflects only the low
momentum properties of nuclear scattering.

For the tensor correlations it is not possible to get a clear
separation of scales between short and medium to long range
correlations in the many-body state.  Therefore in the unitary
correlator method the correlation range is not unique.  In the
language of the $\Vlowk$ potential ($k$ dependence not published for
$S,T=1,0$) this problem is revealed by the fact that the potential
cannot be made independent of the cutoff anymore.

\section{Many-body calculations}
\label{sec:manybody}

In this section many-body calculations of the ground state properties
of the doubly magic nuclei $\Hefour$, $\Osixteen$ and $\Cafourty$ are
performed using the Bonn-A and Argonne~V18 interactions and their
corresponding correlators determined in
Sec.~\ref{sec:bonnargonnecorrelators}. Of particular interest is the
role of the tensor correlator. The tensor correlator derived according
to Eq.~(\ref{eq:thetadeuteron}) from the deuteron or the zero-energy
scattering solution is of very long range.  It includes due to its
naive construction even the non-vanishing $d$-wave outside the range
of the interaction.  As we want to use the two-body approximation we
have to restrict the range of the correlator. To study the influence
of the correlation range on the many-body system we will compare the
results obtained with the three differently ranged tensor correlators
$\alpha$, $\beta$ and $\gamma$.  We investigate the question how long
ranged the correlator must be to successfully describe the tensor
correlations and how short ranged it has to be if we want to restrict
ourselves to the two-body approximation.

The many-body calculations are done with the harmonic oscillator
shell-model trial states
\begin{align}
  \ket{\Hefour} & = \ket{(1s)^4} \\
  \ket{\Osixteen} & = \ket{(1s)^4(1p)^{12}} \\
  \ket{\Cafourty} & = \ket{(1s)^4(1p)^{12}(2s)^4(1d)^{20}} \eqdot
\end{align}
The correlated interactions are used in two-body approximation and we
can use the correlated operators given in
Sec.~\ref{sec:tensorinangmomeigenstates}.

\subsection{The $\Hefour$ Nucleus}

We will discuss the calculations in the $\Hefour$ nucleus in some
detail to illustrate the formalism. The most simple
uncorrelated trial state $\ket{\Hefour}$ is the product of a
harmonic oscillator ground state wave function with different spins
and isospins. The only parameter of the trial state is the oscillator
width $a$, that is related to the radius of the $\Hefour$ nucleus.

This uncorrelated $\Hefour$ trial state has only $s$-wave components
in its relative wave functions and therefore the tensor and spin-orbit
forces do not contribute to the expectation value
$\matrixe{\Hefour}{\op{H}_{\intr}}{\Hefour}$ of the Hamilton
operator. The $d$-wave admixtures in the relative wave functions of
the $\Hefour$ nucleus have to be solely generated by the
tensor correlator.

With the Talmi transformation (\ref{eq:hefourexpect}) we can
calculate the $\Hefour$ expectation value of the Hamilton operator in
two-body approximation
\begin{multline}
  \matrixe{\Hefour}{\bigl[ \hopCr\hopCom\op{H}_{\intr}\opCom\opCr \bigr]^{C2}}{\Hefour} =
  \matrixe{\Hefour}{\op{T}-\op{T}_{\icm}}{\Hefour} \\
  \begin{aligned}
  & + 3  \matrixe{1;(00)0;1}{\hopcr \op{t} \opcr - \op{t}}{1;(00)0;1}
  + 3 \matrixe{1;(01)1;0}{\hopcr\hopcom \op{t} \opcom\opcr -\op{t}}{1;(01)1;0} \\
  & + 3 \matrixe{1;(00)0;1}{\hopcr \op{v}\opcr}{1;(00)0;1}
  + 3 \matrixe{1;(01)1;0}{\hopcr\hopcom \op{v}\opcom\opcr}{1;(01)1;0} .
\end{aligned}
\end{multline}
The expectation value $\matrixe{\Hefour}{\op{T}-\op{T}_{\icm}}{\Hefour}$ can be
calculated analytically, see Eq.~(\ref{eq:tintrshell}).

In the $S=0$ channel, where we only have to deal with radial
correlations, we get for the two-body matrix element of the correlated
kinetic energy
\begin{multline}
  \matrixe{1;(00)0;1}{\hopcr\op{t}\opcr - \op{t}}{1;(00)0;1} = \\
  - \intd{r} \; r\phi_{10}(r)  \frac{1}{2\hat{\mu}_{r01}(r)} (r\phi_{10})''(r)
  + \intd{r} \; (r\phi_{10}(r))^2\ \hat{w}_{01}(r) \eqcomma
\end{multline}
see Eq.~(\ref{eq:ctrad}). The reduced mass $\hat{\mu}_{rST}(r)$ and
local potential $\hat{w}_{ST}(r)$, defined in
Eqs.~(\ref{eq:ctradmass}) and (\ref{eq:ctradu}), respectively, depend
on the radial correlation function $\Rp^{ST}(r)$ of the respective
$S,T$ channel. The relative wave function of the two-body states
$\ket{n;(LS)J,T}$ are harmonic oscillator states with twice the
variance of the single-particle states, thus
\begin{equation}
  \phi_{10}(r) =  \left( \frac{2}{\pi a^3} \right)^{\frac{1}{4}} \exp \left\{
      - \frac{r^2}{4a} \right\}
\end{equation}
is the radial wave function of the relative motion for $n=1$ and $L=0$.

In the $S=1$ channel also the tensor correlations have to be
considered and we get with help of Eqs. (\ref{eq:ctradangmom}) and
(\ref{eq:ctomangmom})
\begin{multline}
  \matrixe{1;(01)1;0}{\hopcr\hopcom\op{t}\opcom\opcr - \op{t}}{1;(01)1;0} =
  - \intd{r} \; r\phi_{10}(r) \frac{1}{2\hat{\mu}_{r10}(r)} (r\phi_{10})''(r) \\
  \begin{aligned}
  & + \intd{r} \; (r\phi_{10}(r))^2 \frac{1}{m}
   \bigl[\theta^{(1)}{}' (\Rp^{10}(r)) \bigr]^2
   + \intd{r} \; (r\phi_{10}(r))^2 \ \hat{w}_{10}(r)
    \\
    & + \intd{r} \; (r\phi_{10}(r))^2 \sin^2 \left(\theta^{(1)}(\Rp^{10}(r))\right)
     \frac{1}{m(\Rp^{10}(r))^2} \matrixe{(21)1}{\oplsq}{(21)1} \eqdot
  \end{aligned}
\end{multline}

In the $S,T=0,1$ channel only the radial part contributes to
the matrix element of the potential
\begin{equation}
  \matrixe{1;(00)0;1}{\hopcr\op{v}\opcr}{1;(00)0;1} =
  \intd{r} \; (r\phi_{10}(r))^2 \: v^c_{01}(\Rp^{01}(r)) \eqcomma
\end{equation}
but in the $S,T=1,0$ channel we have to consider also the spin-orbit
and tensor force. The expectation value of the central force does not
change when including the tensor correlation. The spin-orbit force has
only diagonal matrix elements in the $L=2$ states whereas the tensor
force has strong off-diagonal contributions between the $L=0$ and
$L=2$ states, see Eqs.~(\ref{eq:cromvc}-\ref{eq:cromvt}):
\begin{multline}
  \matrixe{1;(01)1;0}{\hopcr\hopcom\op{v}\opcom\opcr}{1;(01)1;0} =
  \intd{r} \; (r\phi_{10}(r))^2 v^c_{10}(\Rp^{10}(r)) \\
  \begin{aligned}
    & + \intd{r} \; (r\phi_{10}(r))^2 \sin^2\left(\theta^{(1)}(\Rp^{10}(r)\right)
    \, v^b_{10}(\Rp^{10}(r)) \ \matrixe{(21)1;0}{\opls}{(21)1;0} \\
    & + 2 \intd{r} \; (r\phi_{10}(r))^2 \cos\left(\theta^{(1)}(\Rp^{10}(r)\right)
    \sin\left(\theta^{(1)}(\Rp^{10}(r))\right)\, v^t_{10}(\Rp^{10}(r)) \\
    & \qquad\qquad \times \matrixe{(01)1;0}{\opsrr}{(21)1;0} \\
    & + \intd{r} \; (r\phi_{10}(r))^2 \sin^2\left(\theta^{(1)}(\Rp^{10}(r)\right)
    \, v^t_{10}(\Rp^{10}(r))\ \matrixe{(21)1;0}{\opsrr}{(21)1;0} \eqdot
    \end{aligned}
\end{multline}
The other components of the interaction have to be treated accordingly.

The correlations also influence the radius of the nucleus. The
rms-radius defined as
\begin{equation}
  r_{\mathrm{rms}}^2 =
     \frac{1}{A}  \expect{\sum_{i=1}^A(\op{\vec{r}}_i-\op{\vec{R}}_{\icm})^2}
   = \frac{1}{A}\expect{\bigl(\sum_{i=1}^A \op{\vec{r}}_{i}^2-
                                                   \op{\vec{R}}_{\icm}^2\bigr)}
\end{equation}
is obtained as a sum of the uncorrelated mean square radius and a
correction from the radial correlations.
\begin{equation}
  \begin{split}
    r_{\mathrm{rms}}^2 & = \frac{1}{4}  \matrixe{\Hefour}{\hopCr\hopCom\
      \bigl(\sum_{i=1}^4\op{\vec{r}}_{i}^2-\op{\vec{R}}_{\icm}^2\bigr)
      \opCom\opCr}{\Hefour} \\
    & = \frac{1}{4}  \matrixe{\Hefour}{\bigl(\sum_{i=1}^4\op{\vec{r}}_{i}^2-
      \op{\vec{R}}_{\icm}^2\bigr)}{\Hefour}  + \Delta r_{\mathrm{rms}}^2
  \end{split}
  \label{eq:rms}
\end{equation}
For calculating $\Delta r_{\mathrm{rms}}^2$ one uses that the center
of mass operator $\op{\vec{R}}_{\icm}=\frac{1}{A}\sum_{i=1}^A
\op{\vec{r}}_{i}$ commutes with $\opCr$ and $\opCom$ and
$r^2_{ij}=(\op{\vec{r}}_i - \op{\vec{r}}_j)^2$ commutes with $\opCom$.
In two-body approximation it is given as
\begin{equation}
  \begin{split}
    \Delta r_{\mathrm{rms}}^2 & = 
    \frac{3}{4} \matrixe{1;(00)0;1}{\frac{1}{2}\left(\hopcr\op{r}^2\opcr-r^2\right)}{1;(00)0;1} \\
    & \phantom{=} + \frac{3}{4}
    \matrixe{1;(01)1;0}{\frac{1}{2}\left(\hopcr\op{r}^2\opcr-r^2\right)}{1;(01)1;0} \\
    & =  \frac{3}{4}
    \intd{r} \;(r\phi_{10}(r))^2 \biggl\{ \frac{1}{2}\left(\Rp^{01}(r)^2-r^2\right)
    + \frac{1}{2}\left(\Rp^{10}(r)^2-r^2\right) \biggr\} \eqdot
  \end{split}
\end{equation}
The contribution of $\Delta r_{\mathrm{rms}}^2$ is however less than one percent.

\subsubsection{Results - Argonne~V8'}

The Argonne~V8' is simpler in its operator structure than the full
Argonne~V18 interaction and has been used recently as a benchmark
potential for many-body calculations of the $\Hefour$ nucleus
\cite{kamada01}. All the quasi-exact many-body methods presented in
Ref. \cite{kamada01} agree with each other. We can therefore use their
results as  a reliable reference for testing our approximation.

\begin{figure}[htb]
  \begin{minipage}[b]{0.48\textwidth}
    \caption{Contributions of the Argonne~V8' interaction in $\Hefour$.
      For the intrinsic kinetic energy $\expect{\op{T}-\op{T}_{\icm}}$,
      the spin-orbit $\expect{\op{V}^b}$ and the tensor potential
      $\expect{\op{V}^t}$ three curves for the tensor correlators
      $\alpha$, $\beta$ and $\gamma$ are shown. The central
      interaction $\expect{\op{V}^c}$ (full line and grey dot) is not
      influenced by the tensor correlations.  $\Hefour$ optimized
      correlators are used.  Reference values \cite{kamada01} are
      given by the dots.  }
    \label{fig:av8pcontributions}
    \vspace{3ex}
  \end{minipage}\hfil
  \includegraphics[width=0.48\textwidth]{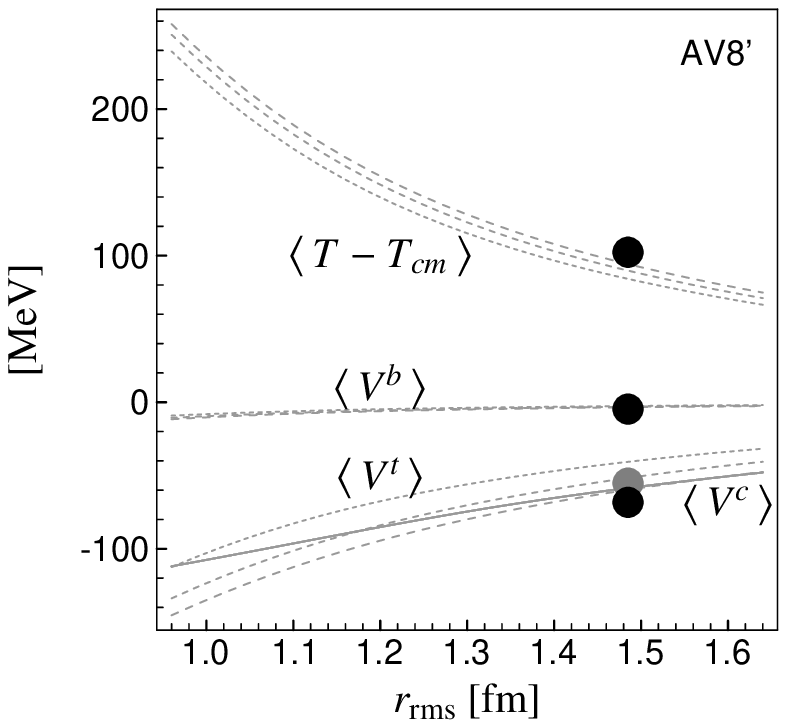}
\end{figure}

\begin{figure}[htb]
  \begin{minipage}[b]{0.48\textwidth}
    \caption{$\Hefour$ binding energies with the Argonne~V8'
      interaction (without Coulomb interaction) as a function of the
      matter radius $r_{\mathrm{rms}}$.  Black lines: two-body
      minimized radial correlators and tensor correlators with
      restricted range (labeled with $\alpha$, $\beta$, $\gamma$).
      Gray lines: $\Hefour$ optimized correlators instead.  The
      result of the reference calculations \cite{kamada01} is
      included.}
    \label{fig:he4av8p}
    \vspace{3ex}
  \end{minipage}\hfil
  \includegraphics[width=0.48\textwidth]{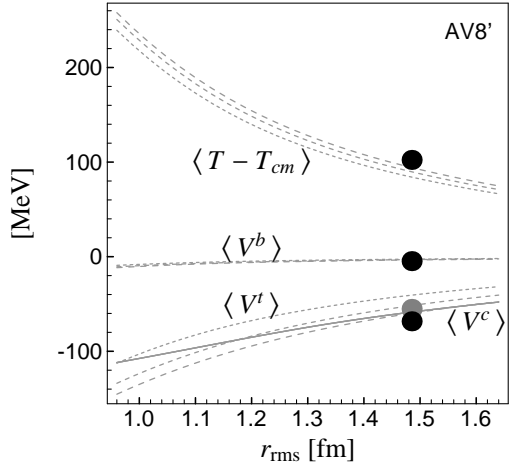}
\end{figure}

In Fig.~\ref{fig:av8pcontributions} the contributions of the kinetic
energy, the central, spin-orbit and tensor interaction to the total
energy are shown as a function of the rms-radius when the oscillator
parameter $a$ is varied.  The correlation functions are obtained by
minimizing the total energy for a given parameter $a$ under the
constraints (\ref{eq:constraint}) that restrict the range of the
tensor correlation with $\alpha$ (dotted line) denoting the shortest
one. The expectation value $\expect{\op{V}^c}$ (full line and grey
point) is not influenced by the tensor correlations as $\op{V}^c$
commutes with $\opCom$.  The spin-orbit contribution
$\expect{\op{V}^b}$ dependence is rather weak.  As expected the
variations in the tensor correlator have the largest impact on the
expectation value of the tensor interaction $\expect{\op{V}^t}$.
Without tensor correlations, i.e. $\opCom=1$, the expectation value
$\expect{\op{V}^t}$ would vanish completely because we use only the
most simple uncorrelated shell-model state of four nucleons in the
$s$-shell. The more it is surprising that the correlation function
$\vartheta^\gamma(r)$ results in almost the exact tensor energy.  The
correlated kinetic energy increases for shorter ranged correlations
because the relative wave functions vary more rapidly and contain
higher momenta.  The gain in binding in $\expect{\op{V}^t}$ is however
larger than the loss in $\expect{\op{T}-\op{T}_{\icm}}$.

In Fig.~\ref{fig:he4av8p} the sum of all contributions to the binding
energy of $\Hefour$ is shown.  The additional black lines denote the
results when radial and tensor correlation functions are used that are
obtained by minimizing the energy in the two-body system (see
Sec.~\ref{sec:bonnargonnecorrelators}).  The corresponding energies
are almost the same as the ones where the correlation functions are
optimized for $\Hefour$ (gray lines), which indicates that the short
range correlations are not very sensitive to the size of the nuclear
system but are determined by the nuclear interaction at short
distances. We will therefore use in the following only the correlators
obtained at $r_{\mathrm{rms}}=1.48\,\fm$.

Fig.~\ref{fig:he4av8p} also shows that our uncorrelated trial state
$\ket{\Hefour}$ is too simple for a precise reproduction of the exact
values of binding energy and radius. The delicate balance between
large positive and large negative contributions leads to a minimum in
the total energy that is at a too small radius.  The correlation
function $\vartheta^\beta(r)$ gives about the correct energy but a
radius which is 0.17 fm too small.  In order to keep the three- and
higher-body contributions small we should use the short-range
correlator labelled $\alpha$ and admix $L=2$ states of the
relative motion in the trial state explicitly.  In that case we would
start at the minimum of the dotted black curve at about
$r_{\mathrm{rms}}\approx 1.35\ \fm$ and the long range part of the tensor force
would admix those states and give the missing binding.

This perception is also supported by Ref. \cite{kamada01} where a
probability of 13.9\% is quoted to find the nucleus in an $L=2$
eigenstate of the total angular momentum.  The uncorrelated $\Hefour$
trial state has only an $L=0$ component but the tensor correlator
induces an $L=2$ admixture.  The square of the total orbital
angular momentum of the $\Hefour$ nucleus can be expressed as
\begin{equation}
  \expect{ \opLsq } = \expect{\bigl(\sum_{i=1}^A
    \op{\vec{r}}_i\times\op{\vec{p}}_i -\op{\vec{L}}_{\icm}\bigr)^2} =  
  \sum_{L=0} L(L+1) P_L \eqcomma
\end{equation}
where the percentage of an $L$-component is denoted by $P_L$.  In the
two-body approximation this is given by
\begin{multline}
  \matrixe{\Hefour}{\left[
      \hopCr\hopCom\opLsq\opCom\opCr\right]^{C2}}{\Hefour} = \\
  \begin{split}
  & = 3 \matrixe{1;(01)1;0}{\hopcr\hopcom\oplsq\opcom\opcr - \oplsq}{1;(01)1;0} \\
  & = 2(2+1)\ 3 \intd{r} \; (r\phi_{10}(r))^2 \sin^2\left(\theta^{(1)}
    (\Rp^{10}(r))\right) \eqdot
  \end{split}
\end{multline}
There are no contributions from the one-body part of the cluster
expansion because all one-body states have $L=0$. From the two-body
part the only non-vanishing contribution is in the tensor correlated
$S,T=1,0$ channel with its $L=2$ admixture, see
Eq.~(\ref{eq:corrangmomstate}).  In Table \ref{tab:percentages} the
probalities resulting from the differently ranged correlators are
displayed. The correlator $\gamma$ which reproduces the tensor
interaction best, see Fig.~\ref{fig:av8pcontributions}, gives also an
$L=2$ contribution that agrees well with the exact one. However,
it has also the longest range and hence causes the largest three-body
contributions to the correlated interactions which we have neglected.
Therefore, we prefer a shorter ranged correlator, for example between
$\alpha$ and $\beta$, and an improved trial state that allows for
explicit admixture of $d$-waves.  In our approximation we do not get
any probability for $L=1$, this appears only due to higher
particle orders of the correlated Hamiltonian or additional components
in the trial state. In any case it is very small as the exact
calculations show.

\begin{table}\label{tab:percentages}
\begin{center}
\begin{tabular}{l|ccc}  \toprule
  correlator                 & $P_{L=0}$ & $P_{L=1}$ & $P_{L=2}$  \\ \midrule
  $\mathit{min}^{\alpha}$    & 0.959       & 0.0       & 0.041      \\
  $\mathit{min}^{\beta}$     & 0.926       & 0.0       & 0.074      \\
  $\mathit{min}^{\gamma}$    & 0.881       & 0.0       & 0.119      \\
  \midrule
  $\mathit{min}^{\alpha}-\Hefour$& 0.956   & 0.0       & 0.044     \\
  $\mathit{min}^{\beta} -\Hefour$& 0.916   & 0.0       & 0.084     \\
  $\mathit{min}^{\gamma}-\Hefour$& 0.858   & 0.0       & 0.142     \\
  \midrule
  Ref. \cite{kamada01}       & 0.857     & 0.0037      & 0.139      \\
  \bottomrule
\end{tabular}
\end{center}
\caption{Probabilities of total orbital angular momentum in $\Hefour$ for the AV8' interaction at $r_{\mathrm{rms}}=1.48\;\fm$.}
\end{table}

It should be emphasized again that the correlations are absolutely
essential for a successful description of the nuclei. As shown in
Fig.~\ref{fig:energies} and discussed in the introduction the
existence of bound nuclei can only be explained with the combination
of central and tensor correlations.

\subsubsection{Results -- Bonn-A and Argonne~V18}

\begin{figure}[tb]
  \begin{center}
  \includegraphics[width=0.48\textwidth]{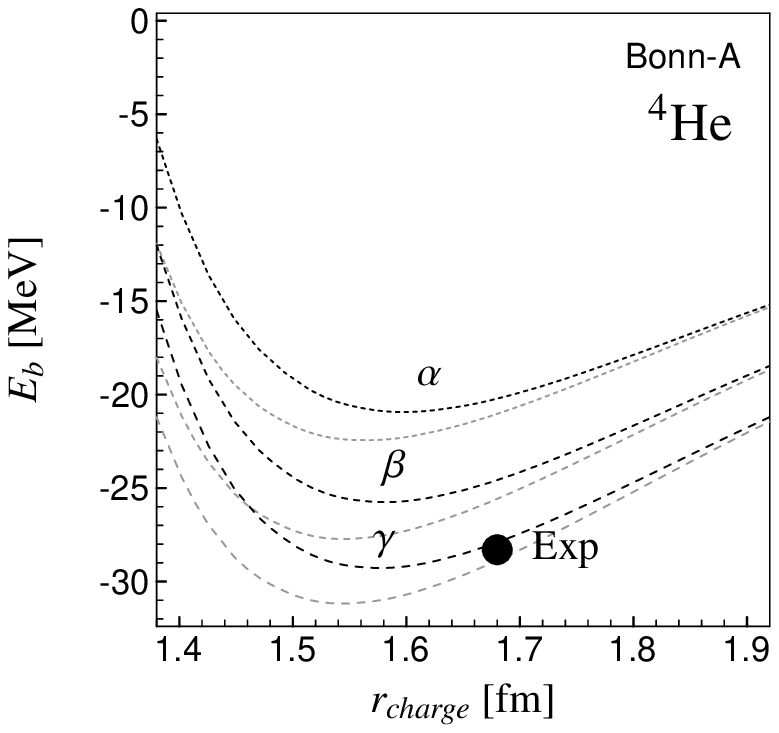}\hfil
  \includegraphics[width=0.48\textwidth]{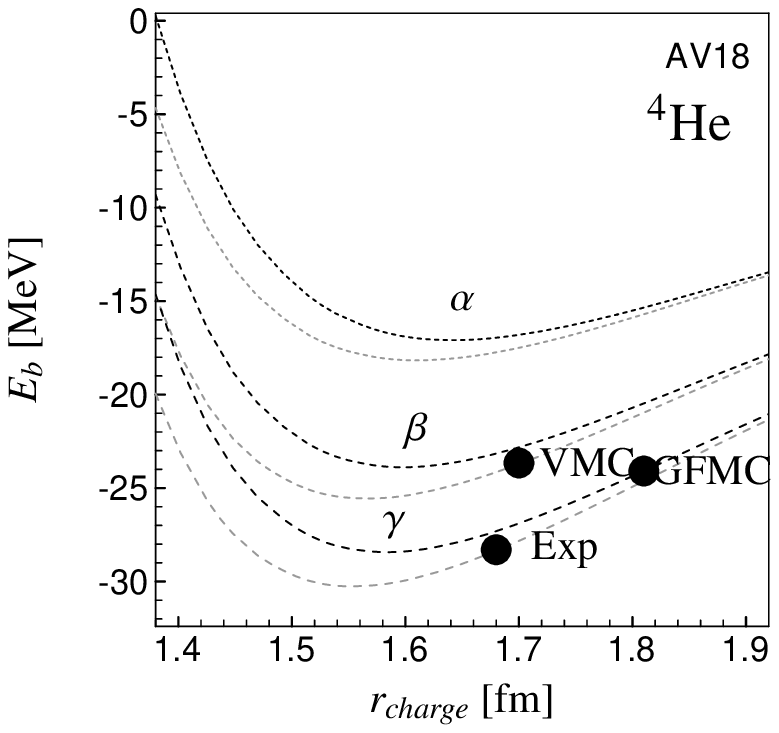}
  \caption{$\Hefour$ binding energy as function of the rms-charge-radius
    (including proton size) calculated with the Bonn-A (left hand
    side) and the Argonne~V18 (right hand side) interaction including
    the Coulomb interaction.  Black lines: two-body minimized radial
    correlators and tensor correlators with restricted range (labeled
    with $\alpha$, $\beta$, $\gamma$).  Gray lines: $\Hefour$
    optimized correlators instead.  For the Argonne~V18 we the results
    of VMC and GFMC calculations \cite{wiringa92} are included.  }
  \label{fig:nucleihe4}
  \end{center}
\end{figure}

In Fig.~\ref{fig:nucleihe4} we show our results for the $\Hefour$
nucleus using the Bonn-A and the Argonne~V18 interactions. The
Argonne~V18 nuclear interaction is identical to the Argonne~V8' in the
channels relevant in the two-body approximation except for the Coulomb
interaction. The energies are plotted against the charge radius that
is obtained by folding the many-body charge distribution with the
charge distribution of the proton.

For reference we show results of VMC and GFMC calculations
\cite{wiringa92} with the Argonne~V18 interaction. As in the case of
the Argonne~V8' interactions we observe that tensor correlator $\beta$
gives the correct energy but at a somewhat smaller radius.  One should
however also note the large difference in the radii of the VMC and the
GFMC calculations.  At the reference radius of the GFMC calculation
the long range tensor correlator $\gamma$ reproduces the GFMC binding
energy.  The differences between the GFMC calculation and the
experimental binding energy and radius are interpreted as an
indication for the necessity of genuine three-body forces.

For the Bonn-A interaction there are no reference calculations but the
results are very similar to the Argonne~V18 calculations, except that
the Bonn-A interaction is more attractive with the shorter tensor
correlators $\alpha$ and $\beta$.

\subsection{The $\Osixteen$ and the $\Cafourty$ nucleus}

The doubly magic nuclei are calculated with the harmonic oscillator
shell-model trial states given in App.~\ref{app:talmi} using
the two-body approximation. The evaluation of the two-body matrix
elements is done with the correlated interaction in angular momentum
eigenstates given in Sec.~\ref{sec:tensorinangmomeigenstates} and as
it was illustrated for the $\Hefour$ nucleus.

\subsubsection{Results -- Argonne~V8'}

\begin{figure}[tb]
  \includegraphics[width=0.48\textwidth]{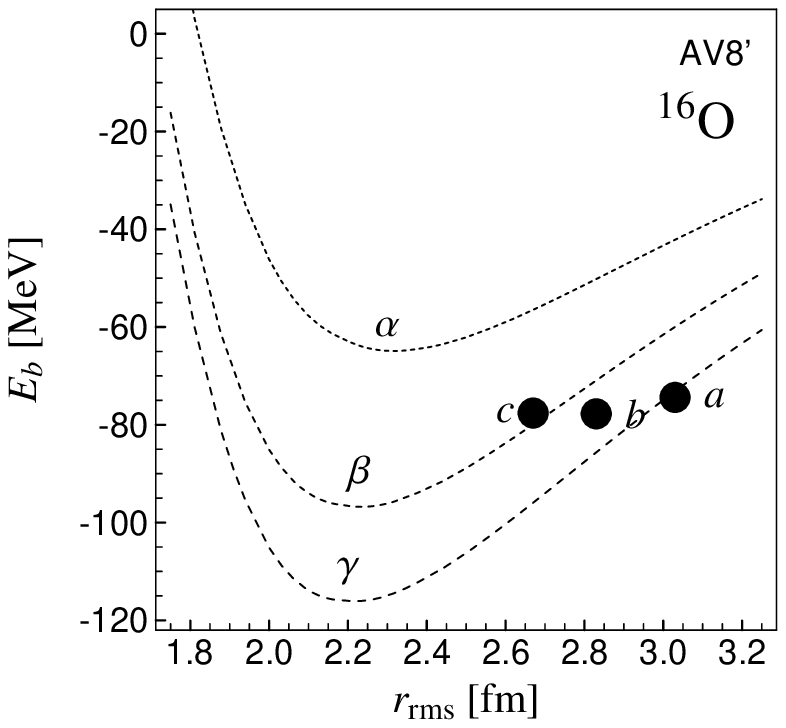}\hfil
  \includegraphics[width=0.48\textwidth]{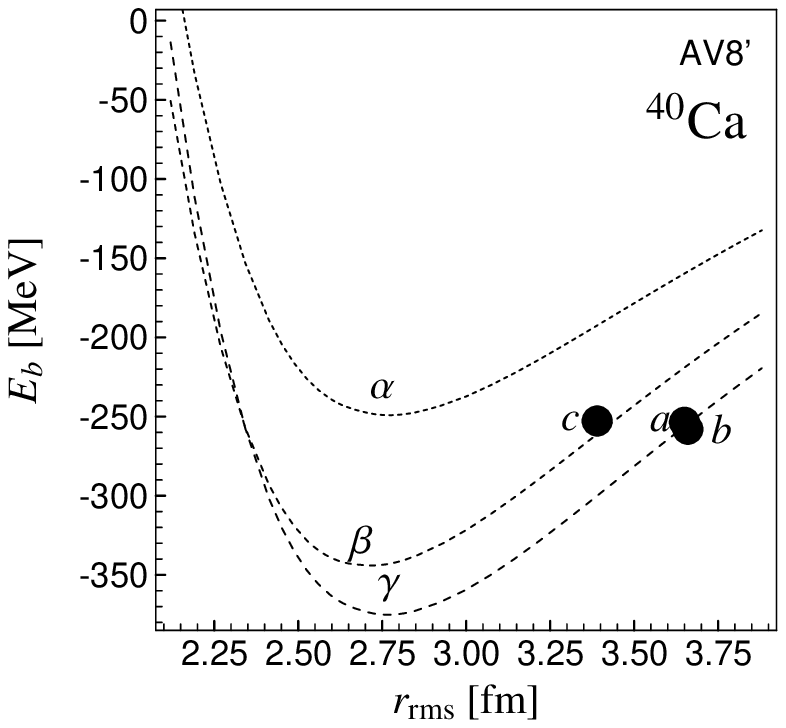}
  \caption{Binding energy of $\Osixteen$ and $\Cafourty$ as function of
    matter radius $r_{\mathrm{rms}}$ using the Argonne~V8' interaction (Coulomb
    included).  Results for two-body optimized radial correlators and
    tensor correlators with restricted range (labelled $\alpha$,
    $\beta$ and $\gamma$). Points denote FHNC/SOC \cite{fabrocini98}
    calculations. The radii for the points $a$ and $b$ represent
    minima of the energy calculated with additional three-body forces.
    Calculation $a$ used harmonic oscillator states, $b$ and $c$
    Woods-Saxon states.  Calculation $c$ used a trial state that
    reproduces the experimental radius.}
  \label{fig:av8pheavynuclei}
\end{figure}

In Fig.~\ref{fig:av8pheavynuclei} our results with the Argonne~V8'
interaction are displayed.  The FHNC/SOC calculations
\cite{fabrocini98} shown for reference were done with a Hamiltonian
consisting of the Argonne~V8' potential together with the Urbana~IX
three-body force.  Therefore we can only compare the expectation
values of the Argonne~V8' potential at the radii obtained in the
FHNC/SOC calculations. At those radii the $\beta$ and $\gamma$ tensor
correlators yield very similar energies.

\subsubsection{Results -- Bonn-A and Argonne~V18}

\begin{figure}[tb]
  \includegraphics[width=0.48\textwidth]{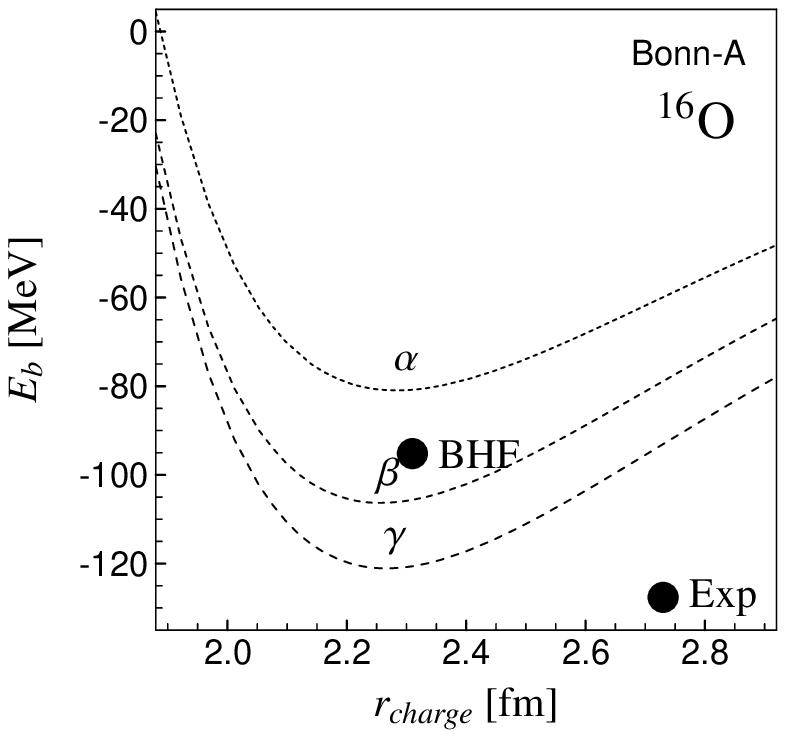}\hfil
  \includegraphics[width=0.48\textwidth]{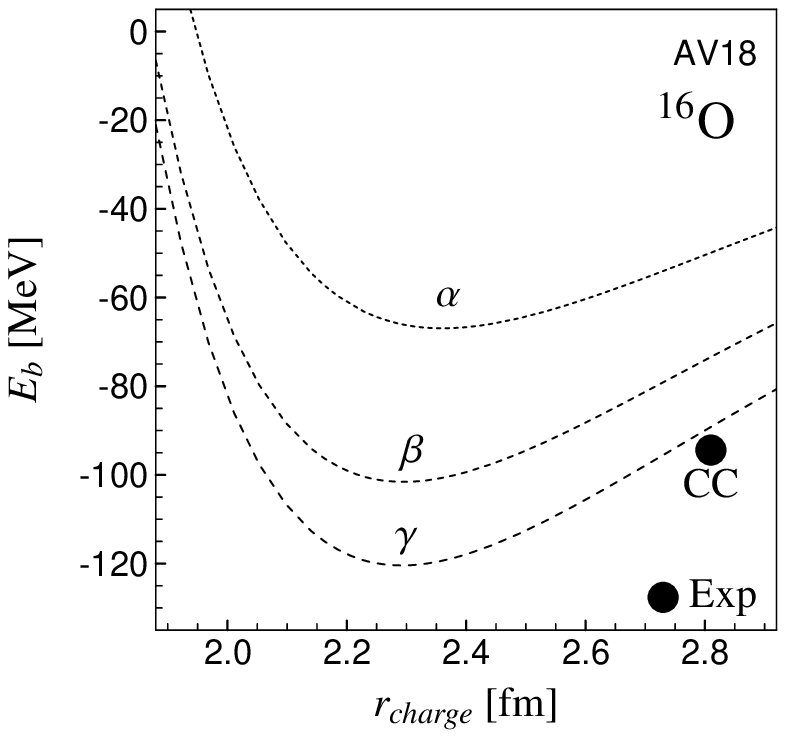}
  \caption{Binding energy of $\Osixteen$ as function of charge radius
    (including proton size) using the Bonn-A (left) and the
    Argonne~V18 (right) interaction.  Results for two-body optimized
    radial correlators and tensor correlators with restricted range
    (labelled $\alpha$, $\beta$ and $\gamma$).  Result of a Brueckner
    Hartree Fock (BHF) calculation \cite{muether00} and a Coupled
    Cluster (CC) calculation \cite{heisenberg99} are included.}
  \label{fig:nucleio16}
\end{figure}

\begin{figure}[tb]
  \includegraphics[width=0.48\textwidth]{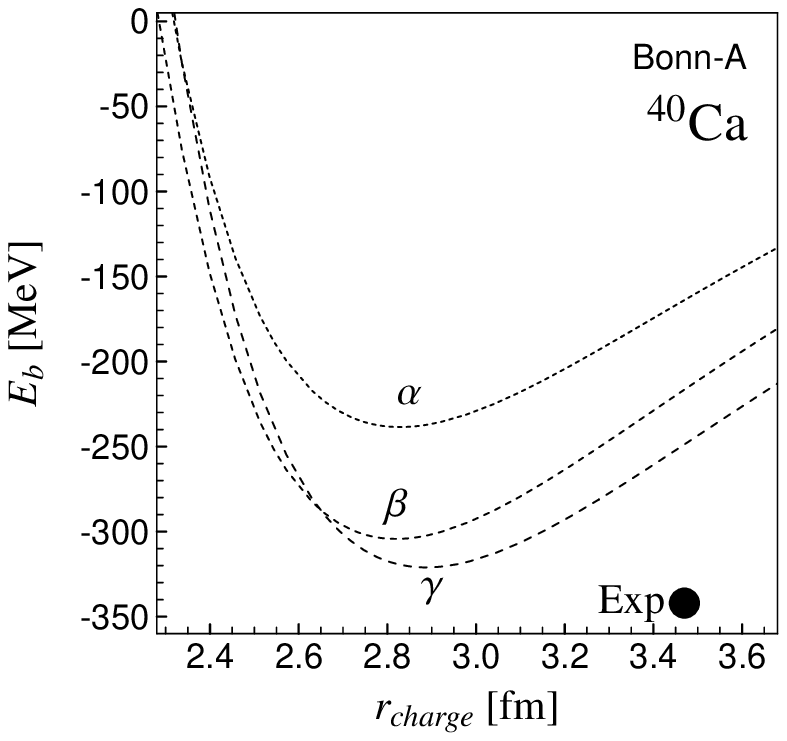}\hfil
  \includegraphics[width=0.48\textwidth]{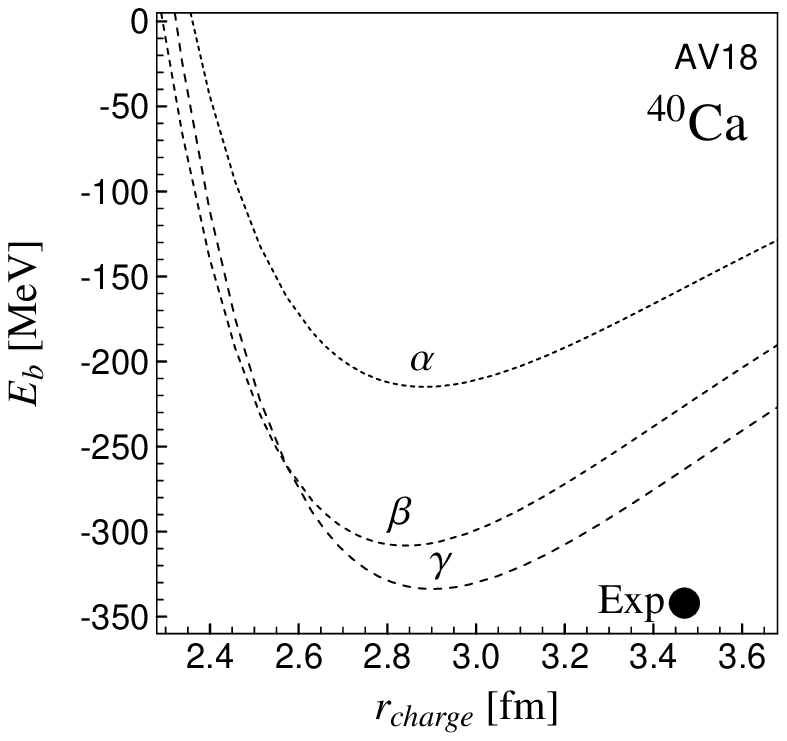}
  \caption{Binding energy of $\Cafourty$ as function of the charge radius
    (including proton size) using the Bonn-A (left) and the
    Argonne~V18 (right) interaction.  Results for two-body optimized
    radial correlators and tensor correlators with restricted range
    (labelled $\alpha$, $\beta$ and $\gamma$).  }
  \label{fig:nucleica40}
\end{figure}

In Fig.~\ref{fig:nucleio16} and Fig.~\ref{fig:nucleica40} the results
of many-body calculations using the Bonn-A and the Argonne~V18
interaction are presented. In case of the Bonn-A interaction the
results of a Brueckner Hartree-Fock (BHF) calculation \cite{muether00}
and for the Argonne~V18 a Coupled Cluster Calculation
\cite{heisenberg99} are included for reference.  It is interesting to
note that the Brueckner Hartree-Fock calculation gives the same result
as our calculation with a correlator in between $\alpha$ and $\beta$,
both in energy and radius. Like in our approximation the BHF method
uses an effective two-body Hamiltonian and results in a too small
radius. As discussed in the case of the $\Hefour$ this could be an
indication that a single Slater determinant is a too simple trial
state and admixtures of particle-hole excitations should describe the
long range part of the tensor correlation.

The similarity in the results obtained with the Bonn-A and Argonne~V18
interactions as displayed in Figs.~\ref{fig:nucleihe4},
\ref{fig:nucleio16} and \ref{fig:nucleica40} is astonishing. As shown
in Fig.~\ref{fig:energies} the energies obtained with the bare
energies are quite different.  It seems that the unitary correlators,
which are interaction specific, map the two different interactions on
the same correlated interaction. This is illustrated further in
Sec.~\ref{sec:momentumspace} where the low-momentum behavior of the
correlated interaction is discussed.

We can further notice that neither the correlated Argonne~V18 nor the
correlated Bonn-A interaction can reproduce the experimental binding
energies at the experimental radii.  It is known from GFMC
calculations of light nuclei ($A<8$)\cite{pieper01mp} that the
Argonne~V18 interaction does not provide enough binding and additional
genuine three-body forces are needed to reproduce the experimental
radii and binding energies.

\subsection{Momentum distributions}
\label{sec:momentumdistribution}

Short range correlations induce high momentum components in the
nuclear many-body state that can be seen in the one-body momentum
distribution $\hat{n}(\vec{k})$,
\begin{equation} 
  \hat{n}(\vec{k})= \sum_{\chi,\xi}
  \expect{\conop{a}_{\chi\xi} (\vec{k}) \desop{a}_{\chi\xi}(\vec{k})}
  \approx  \sum_{\chi,\xi}
  \matrixe{\Phi}{[\hopCr\hopCom \conop{a}_{\chi\xi}(\vec{k})
    \desop{a}_{\chi\xi}(\vec{k}) \opCom\opCr]^{C2}}{\Phi} \eqcomma
  \label{eq:momentumdistr}
\end{equation}
where $\conop{a}_{\chi\xi} (\vec{k})$ creates a nucleon with
spin $\chi$, isospin $\xi$ and momentum $\vec{k}$.  This is another
example that, once the unitary correlators $\opCom$ and $\opCr$ are
determined, not only the Hamiltonian but also any other observable can
easily be correlated and although one deals with simple many-body
trial states $\ket{\Phi}$ that do not contain short range correlations
these correlations are not lost like in mean-field models.

\begin{figure}[hbt]
  \includegraphics[width=0.48\textwidth]{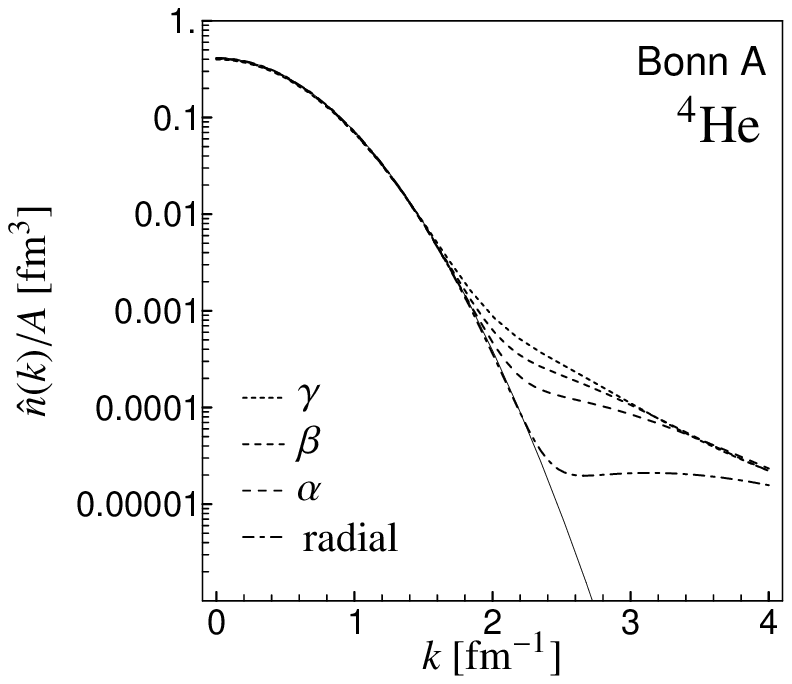} \hfil
  \includegraphics[width=0.48\textwidth]{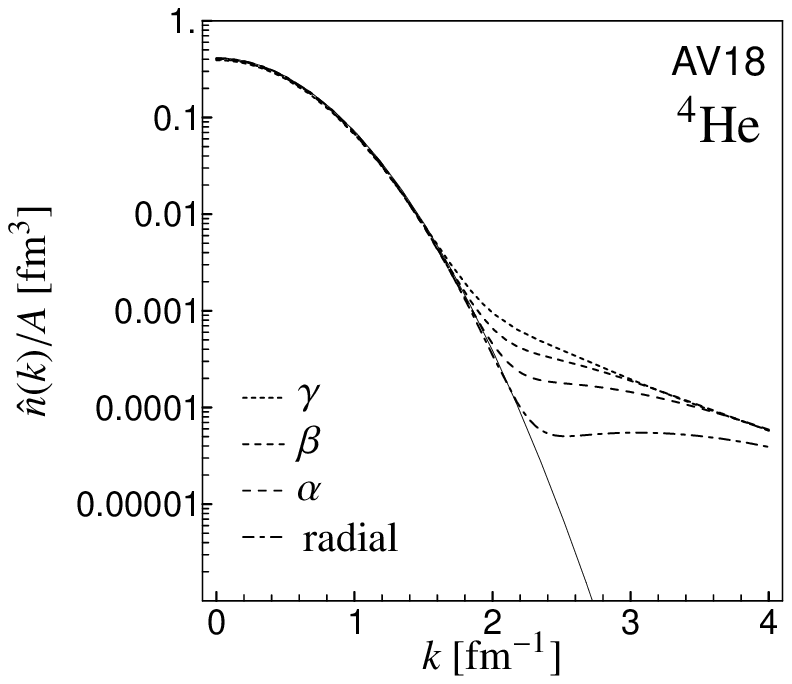}\\
  \includegraphics[width=0.48\textwidth]{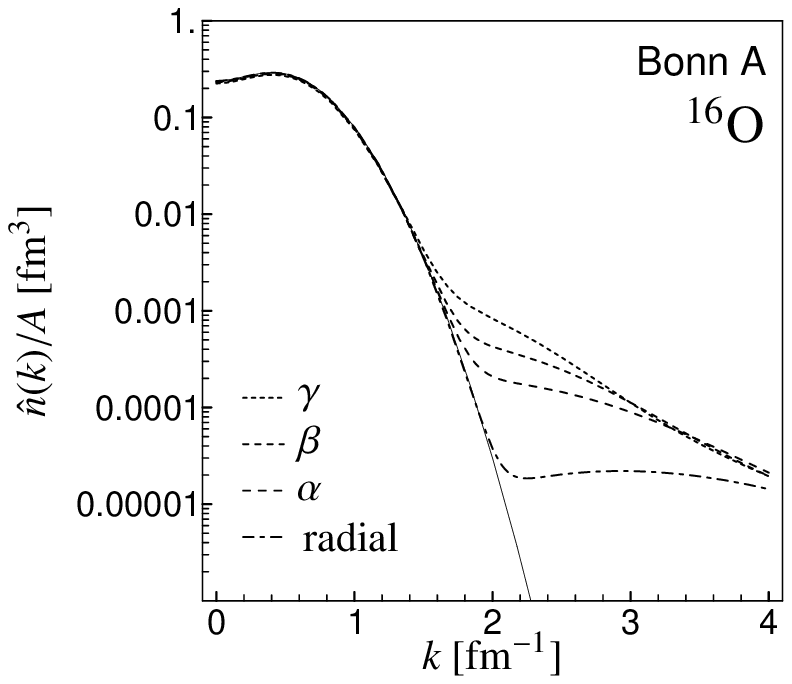}\hfil
  \includegraphics[width=0.48\textwidth]{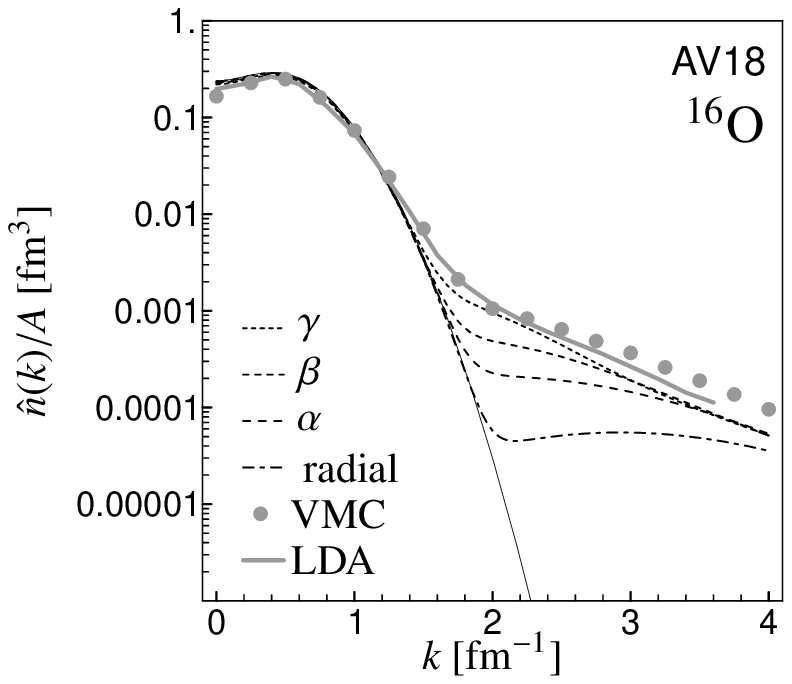}
  \caption{Momentum distribution of $\,\Hefour$ and $\Osixteen$ using
    the Bonn-A (left) and the Argonne~V18 (right) interaction for
    uncorrelated trial state (thin solid), two-body optimized radial
    correlators (dashed dotted) and tensor correlators with restricted
    range ($\alpha$, $\beta$ and $\gamma$).  Results of a variational
    Monte-Carlo (VMC) calculation (grey dots) but using Argonne~V14
    plus Urbana VII three-nucleon interaction \cite{pieper92} and of a
    spectral function analysis (grey line) \cite{benhar94} are
    included.  }
  \label{fig:nk}
\end{figure}

Figure \ref{fig:nk} displays the momentum distributions of $\Hefour$
and $\Osixteen$ for the Bonn-A and Argonne~V18 potential. The thin
solid lines denote the uncorrelated shell-model distribution that
reproduces the experimental radius. When only the radial correlations
are introduced we obtain a contribution at high momenta that is small
but reaches out to very high momenta. This is due to the correlation
hole at short distances seen in the two-body density in Figs.
\ref{fig:correlationhole} and \ref{fig:CrCom}. The more repulsive AV18
leads to a high momentum tail about twice the one obtained with the
Bonn~A potential.  Introducing also the tensor correlators $\opCom$
with different ranges $\alpha$, $\beta$, $\gamma$ we obtain a
substantial enhancement around $k=2\,\fm^{-1}$.  As expected the
longer ranged correlator $\gamma$ gives larger momentum distributions
than the shorter ranged $\alpha$ and $\beta$, but they all merge for
$k > 3.5\, \fm^{-1}$.  In the $\Osixteen$ AV18 frame we include the
results of a variational Monte-Carlo calculation by Pieper et al.
\cite{pieper92} (grey dots) which are however for the Argonne~V14
interaction supplemented by the Urbana VII three-nucleon interaction.
Although the interaction is not the same the comparison shows the
importance of the tensor correlator for filling up the momentum
distribution around $k=2\,\fm^{-1}$. The distribution corresponding to
the spectral function of Benhar et al. \cite{benhar94} which they used
successfully to describe inclusive electron-nucleus scattering at
large momentum transfer results in the momentum distribution shown as
grey full line.

One should however keep in mind that particle-hole excitations in the
trial state also contribute in the region of the Fermi surface.  As we
discussed in Sec.~\ref{sec:manybody} it is better to use a
short-ranged correlator $\alpha$ or $\beta$ and describe the long
range tensor correlations by the uncorrelated many-body state in order
to avoid three- and higher-body terms in the correlated operators.

\appendix
\begin{appendix}
\section{Algebra of operators}
\label{app:ucom}

In the following sections of the appendix we summarize the properties
and the algebra of the
operators that appear in processing central and tensor correlations.
The symbol $\eqrrep$ writes the operator in coordinate representation.

\subsection{Radial and orbital momentum operator}
\label{app:radialmomentum}

The momentum operator $\opvecp$ is decomposed in the radial
momentum $\opvecpr$ and the so called orbital momentum $\opvecpom$.
\begin{equation}
  \opvecp = \opvecpr + \opvecpom \eqcomma
\end{equation}
The radial momentum is the component in radial direction given by
\begin{equation}
  \opvecpr = 
  \frac{\vecr}{r} \pr = \frac{1}{2} \biggl[
   \biggl( \vecp \cdot \frac{\vecr}{r} \biggr)  \frac{\vecr}{r} +
  \frac{\vecr}{r} \biggl( \frac{\vecr}{r} \cdot \vecp \biggr)
  \biggr]
\end{equation}
and its absolute value is
\begin{equation}
  \oppr = \frac{1}{2} \biggl\{ \op{\vecp} \cdot \frac{\op{\vecr}}{\op{r}} +
  \frac{\op{\vecr}}{\op{r}}\cdot \vecp  \biggr\}  \eqdot
\end{equation}
Useful coordinate representations of the radial momentum operator are
\begin{equation}
  \pr \eqrrep \frac{1}{i} \biggl( \frac{1}{r} + \partd{}{r} \biggr) =
  \frac{1}{i} \frac{1}{r} \partd{}{r} \bigl( r \circ \bigr) \eqcomma
\end{equation}
\begin{equation}
  \pr^2 \eqrrep - \frac{1}{r} \partd{}{r} \biggl( r \frac{1}{r}
  \partd{}{r} r \circ \biggr) = - \frac{1}{r} \partdd{}{r}
  \bigl( r \circ \bigr) \eqdot
\end{equation}

The radial correlator $\opcr$, which shifts a pair of nucleons only
radially in the relative coordinate, is constructed with the generator
\begin{equation}
  \begin{split}
  \opgr &= \frac{1}{2} \Bigr\{ \pr s(r) + s(r) \pr \Bigr\}  =
  \frac{1}{2} \biggl\{ \vecp \cdot \frac{\vecr}{r} s(r)
  + s(r) \frac{\vecr}{r} \cdot \vecp \biggr\} \\
  & \eqrrep \frac{i}{2} \biggl\{
  \overleftarrow{\vec{\nabla}} \cdot \frac{\vecr}{r} s(r) -
    s(r) \frac{\vecr}{r} \cdot \overrightarrow{\vec{\nabla}}
   \biggr\}
   =\frac{1}{i} \biggl\{\frac{s'(r)}{2}+s(r) \partd{}{r} \biggr\}
   \eqdot
  \end{split}
\end{equation}

Using the above relations the momentum dependent interaction terms
can be expressed in different ways
\begin{align}
  \frac{1}{2} \Bigl( \pr^2 m(r) + m(r) \pr^2
  \Bigr) & = \pr m(r) \pr - \frac{1}{2} m''(r) - \frac{m'(r)}{r} \\
  & =
  \left(\vecp\cdot\frac{\vecr}{r}\right) m(r)
  \left(\frac{\vecr}{r}\cdot\vecp\right) -
  \frac{1}{2} m''(r) \\
  & =
  \vecp\cdot m(r) \vecp - m(r)
  \frac{\lsq}{r^2} - \biggl( \frac{m'(r)}{r} + \frac{1}{2}
  m''(r) \biggr) \eqdot
\end{align}

The {\it orbital momentum} $\opvecpom$ is the remaining component
that is perpendicular to the distance vector $\op{\vecr}$
\begin{equation}
\opvecpom =
  \frac{1}{2 r} \biggl( \vecl\times\frac{\vecr}{r} -
  \frac{\vecr}{r} \times \vecl \biggr) \tag{\ref{eq:opvecpom}} \eqdot
\end{equation}
It is not to be confused with the orbital angular momentum $\opvecl$.
From the definition it is obvious that $\vecpom$ commutes with
functions which depend only on the relative distance $\op{r}$
because $\op{\vecl}$ commutes with $\op{r}$
\begin{equation}
  \comm{\vecpom}{f(r)} = 0 \eqcomma
\end{equation}
and hence $\vecpom$ acts only on the angular degrees of freedom.

The radial momentum operator $\opvecpr$ and the orbital momentum operator
$\opvecpom$ do not commute. Using the elementary commutator relations of
position and momentum we can verify
\begin{equation}
  \comm{\pr}{\vecpom} = \frac{i}{r} \vecpom
\end{equation}
and for the scalar product commutator
\begin{equation}
  \comm{\vecpr \cdot}{\vecpom} = \vecpr\cdot\vecpom - \vecpom\cdot\vecpr
  = i \left( \pr \frac{1}{r} + \frac{1}{r} \pr
  \right) \eqrrep -\frac{1}{r^2} + \frac{2}{r} \partd{}{r} \eqdot
\end{equation}

We can further calculate the scalar product commutator with the
relative distance operator $\opvecr$
\begin{equation}
  \comm{\vecr \cdot}{\vecpom} = 2i
\end{equation}
and using
\begin{equation}
    \vecp^2 = \vecpr^2 + \vecpr\cdot\vecpom + \vecpom\cdot\vecpr +
    \vecpom^2
    \eqcomma
\end{equation}
we derive the properties
\begin{equation}
  \vecpr\cdot\vecpom + \vecpom\cdot\vecpr = - \frac{1}{r^2} \eqcomma
\end{equation}
\begin{equation}
  \vecpom^2 = \frac{1}{r^2} \bigl(\lsq + 1\bigr) \eqdot
\end{equation}

\subsection{Algebra and matrix elements of tensor operators}
\label{app:metensorops}

We need the algebra of the scalar two-body operators for
calculating correlated operators. The determination of the algebra is
performed using the irreducible spherical tensor representation of
the operators. The conventions of Ref.~\cite{Varshalovich88} are used.

\subsubsection{Spherical tensor operators}
\label{sec:tensors}

An irreducible spherical tensor operators of rank $k$ is noted as
\begin{equation}
  \op{A}^{(k)}_{q} \eqdot
\end{equation}
The transformation from cartesian to spherical tensors of first rank
is given by
\begin{equation}
  \op{A}^{(1)}_{1} = - \frac{\op{A}_{x} + i \op{A}_{y}}{\sqrt{2}} \qquad
  \op{A}^{(1)}_{0} = \op{A}_{z}
  \qquad
  \op{A}^{(1)}_{-1} = \frac{\op{A}_{x} - i \op{A}_{y}}{\sqrt{2}} \eqdot
\end{equation}

In general two tensor operators of rank $j_1$ and $j_2$ can be coupled to a
tensor operator of rank $j$ using the Clebsch-Gordan coefficients
\begin{equation}
  \couple{\op{A}^{(j_1)}}{\op{B}^{(j_2)}}{j}_{q} = \sum_{m_1,m_2}
  \cg{j_1}{m_1}{j_2}{m_2}{j}{q} \, \op{A}^{(j_1)}_{m_1} \;
  \op{B}^{(j_2)}_{m_2} \eqdot
\end{equation}

A shorthand notation is used for the coupled (and symmetrized)
product of vector operators $\vec{a}$ and $\vec{b}$ in coordinate
space
\begin{equation}
  \couplev{\op{a}}{\op{b}}{j}_{q} = \frac{1}{2} \left(
  \couple{\op{a}^{(1)}}{\op{b}^{(1)}}{j}_{q} +
    \couple{\op{b}^{(1)}}{\op{a}^{(1)}}{j}_{q} \right) \eqdot
\end{equation}

The scalar product of spherical tensor operators (acting in different
Hilbert spaces) of first rank is given by
\begin{equation}
  \op{A}^{(1)}\cdot\op{T}^{(1)} = - \sqrt{3}
  \coupletensor{\op{A}^{(1)}}{\op{T}^{(1)}}{0} \eqcomma
\end{equation}
and the scalar product of spherical tensor operators of second rank by
\begin{equation}
  \op{A}^{(2)}\cdot\op{T}^{(2)} = \sqrt{5}
  \coupletensor{\op{A}^{(2)}}{\op{T}^{(2)}}{0} \eqdot
\end{equation}

Using the spherical tensors we can write for example the
spin-orbit operator $\opls$ as
\begin{equation}
  \opls = \lone \cdot \opSone = - \sqrt{3} \coupletensor{\lone}{\opSone}{0}
\end{equation}
and the tensor operator
$\op{s}_{\!12}(\vec{a},\vec{b})$ can be expressed using the spherical
tensors as
\begin{equation}
  \begin{split}
    \op{s}_{\!12}(\vec{a},\vec{b}) & =
    3 (\op{\vec{\sigma}}_{1}\cdot\op{\vec{a}})(\op{\vec{\sigma}}_{2}\cdot{\op{\vec{b}}})-
    (\op{\vec{\sigma}}_{1}\cdot\op{\vec{\sigma}}_{2})(\op{\vec{a}}\cdot\op{\vec{b}}) \\
    & =
    3 \couple{\op{a}^{(1)}}{\op{b}^{(1)}}{2}\cdot \op{S}^{(2)} = 
    3 \sqrt{5} \coupletensor{ (\op{a}\op{b})^{(2)} }{\opStwo}{0}
    \eqcomma
  \end{split}
\end{equation}
with the operators $\opSone$ and $\opStwo$ in the two-body spin space.

\subsubsection{Matrix elements in coordinate space}

In coordinate space matrix elements of $\vec{r}$, $\vecpom$ and
$\vecl$ and tensor products of these operators occur. The reduced
matrix elements of these elementary operators are given by
\begin{equation}
  \matrixered{L'}{\op{r}^{(1)}}{L} = \bigl( \sqrt{L+1}\,\delta_{L',L+1} -
  \sqrt{L}\,\delta_{L',L-1} \big) r \eqcomma
  \label{eq:meredr}
\end{equation}
\begin{equation}
  \matrixered{L'}{\pom^{(1)}}{L} = \bigl( (L+1)^{\frac{3}{2}}\,
  \delta_{L',L+1} + L^{\frac{3}{2}}\,\delta_{L',L-1} \bigr)
  \frac{\I}{r}
  \label{eq:meredpom}
\end{equation}
and
\begin{equation}
  \matrixered{L'}{\op{L}^{(1)}}{L} = \sqrt{L(L+1)(2L+1)} \,
  \delta_{L',L} \eqdot
  \label{eq:meredl}
\end{equation}

The tensor product of two tensor operators can be expressed by the
reduced matrixe elements of these operators
\begin{multline}
  \matrixered{L'}{\couple{a^{(k_1)}}{b^{(k_2)}}{k}}{L} = \\
  (-1)^{L'+L-k} \sqrt{2k+1} \sum_{L''}
  \sixj{k_1}{L}{k_2}{L'}{k}{L''} \matrixered{L'}{a^{(k_1)}}{L''}
  \matrixered{L''}{b^{(k_2)}}{L} \eqdot
\end{multline}
With this equation the matrix elements of $\srpom$ can be
calculated. We get with (\ref{eq:meredr}) and (\ref{eq:meredpom})
\begin{equation}
  \matrixered{L}{\rpomtwo}{L} = 0
\end{equation}
and
\begin{align}
  \matrixered{L+2}{\rpomtwo}{L} & = \frac{1}{2}
  \sqrt{(L+1)(L+2)(2L+3)} \eqcomma \notag\\
  \matrixered{L-2}{\rpomtwo}{L} & = \frac{1}{2}
  \sqrt{(L-1)L(2L-1)} \eqdot
  \label{eq:merpom}
\end{align}
As $\srpom$ is a tensor operator of rank~2 all matrix elements
with $|L-L'| \neq 0,2$ vanish.

\subsubsection{Matrix Elements in spin space}

In the two-body spin space we have the projectors $\opPinot$ onto total
spin $0$ and $\opPione$ onto total spin $1$,
\begin{equation}
  \op{\Pi}_0 = \frac{1}{4} (1 - \op{\vec{\sigma}} \otimes
  \op{\vec{\sigma}}) \eqcomma \qquad
  \op{\Pi}_1 = \frac{1}{4} (3 + \op{\vec{\sigma}} \otimes
  \op{\vec{\sigma}}) \eqcomma
\end{equation}
the total spin operator
$\opSone$ which is a tensor operator of rank~1
\begin{equation}
  \opSone = \frac{1}{2} \bigl( \op{\sigma}^{(1)} \otimes \op{1} +
  \op{1} \otimes \op{\sigma}^{(1)} \bigr)
  \label{eq:mesone}
\end{equation}
and the tensor operator $\opStwo$ of rank~2
\begin{equation}
  \opStwo = \couple{\opSone}{\opSone}{2} \eqdot
  \label{eq:mestwo}
\end{equation}

The reduced matrix elements of $\opSone$ and $\opStwo$ are given by
\begin{equation}
  \matrixered{1}{\opSone}{1} = \sqrt{6}
\end{equation}
and
\begin{equation}
  \matrixered{1}{\opStwo}{1} = 2 \sqrt{5} \eqdot
\end{equation}

\subsubsection{Matrix elements in angular momentum eigenstates}
\label{app:mesrpomangmom}

The matrix elements of a scalar product of tensor operators of rank~k
in coordinate space $\op{R}^{(k)}$ and spin space $\op{S}^{(k)}$ is
given by
\begin{multline}
  \matrixe{(L'1)JM}{\op{R}^{(k)}\cdot \op{S}^{(k)}}{(L1)JM} = \\
  (-1)^{J+L+1}
  \sixj{L'}{1}{L}{1}{k}{J} \matrixered{L'}{\op{R}^{(k)}}{L}
  \matrixered{1}{\op{S}^{(k)}}{1} \eqdot
\end{multline}

With Eqs.~(\ref{eq:merpom}) and (\ref{eq:mestwo}) we get the matrix
elements of $\opsrpom$ in the angular momentum basis 
\begin{align}
  \matrixe{(J+1,1)JM}{\opsrpom}{(J-1,1)JM} & = 3 \I \sqrt{J(J+1)}
  \eqcomma \\
  \matrixe{(J,1)JM}{\opsrpom}{(J,1)JM} & = 0 \eqcomma \\
  \matrixe{(J-1,1)JM}{\opsrpom}{(J+1,1)JM} & = - 3 \I \sqrt{J(J+1)}
\end{align}

\subsubsection{Products of scalar two-body operators}
\label{sec:stbproducts}

The products, commutators and anti-commutators of scalar two-body
operators can be calculated by recoupling the tensor operators
with the help of 9j-symbols
\begin{multline}
  \coupletensor{\op{A}^{(J_1)}}{\op{S}^{(J_1)}}{0}
  \coupletensor{\op{B}^{(J_2)}}{\op{T}^{(J_2)}}{0} = \\
  \sum_{K=0}^{2} (2K+1) \ninej{J_1}{J_2}{K}{J_1}{J_2}{K}{0}{0}{0}
  \coupletensor{\couple{\op{A}^{(J_1)}}{\op{B}^{(J_2)}}{K}}
  {\couple{\op{S}^{(J_1)}}{\op{T}^{(J_2)}}{K}}{0} \eqdot
  \label{eq:stboprods}
\end{multline}
Here $\op{A}$ and $\op{B}$ are operators in the two-body coordinate
space and $\op{S}$ and $\op{T}$ are operators in the two-body spin
space.  Because the two-body spin space is only four dimensional there
is no tensor operator with a rank higher than two.

\subsection{Cartesian tensor operator relations}

Using (\ref{eq:stboprods}) the following relations for the tensor
operators in cartesian representation are obtained
\begin{equation}
  \oplssq = \frac{2}{3} \oplsq \opPione - \frac{1}{2} \opls +
  \frac{1}{6} \opsll \eqcomma
  \label{eq:lssq}
\end{equation}
\begin{equation}
  \opsrpom^2 = 6 (\oplsq + 3) \opPione + \frac{45}{2} \opls +
  \frac{3}{2} \opsll \eqdot
  \label{eq:srpomsq}
\end{equation}

The commutators needed for the calculation of the correlated
interaction are given by
\begin{align}
  \comm{\opsrpom}{\opsrr} & = - 24 i \opPione - 18 i \opls + 3i \opsrr
  \\
  \comm{\opsrpom}{\opls} & = -i \opsbarpompom \\
  \comm{\opsrpom}{\oplsq} & = 2 i \opsbarpompom \\
  \comm{\opsrpom}{\opsll} & = 7 i \opsbarpompom \\
  \comm{\opsrpom}{\opsbarpompom} & = i (96 \oplsq + 108) \opPione + i
  (36 \oplsq + 153) \opls + 15 i \opsll \\
  \comm{\opsrpom}{\oplsq \opls} & = - i (\oplsq+3) \opsbarpompom \\
  \comm{\opsrpom}{\oplsq \opsbarpompom} & = i (144
  \op{\vec{l}}^4+600\lsq+324)\opPione \notag \\
  & \quad + i (36 \op{\vec{l}}^4 + 477
  \oplsq + 477) \opls + i (27 \oplsq + 51) \opsll \eqcomma
\end{align}
where the abbreviation
\begin{equation}
  \opsbarpompom = 2 r^2 \opspompom + \opsll - \frac{1}{2} \opsrr \eqcomma
\end{equation}
is used.

\section{Harmonic oscillator shell-model states and Talmi transformation}
\label{app:talmi}

The single-particle harmonic oscillator states used in the shell model
\cite{sitenko:nucleus}
\begin{equation}
  \begin{split}
  \phi^{a}_{nlm}(\vec{r})
  & = \frac{R_{nl}(a)(r)}{r} Y^l_m(\hat{\vec{r}}) \\
  & = \frac{1}{\Gamma(l+\frac{3}{2})}
  \frac{\sqrt{2\Gamma(l+n+\frac{1}{2})}}{\sqrt{\Gamma(n)}}
    \frac{1}{\sqrt{a}^{l+\frac{3}{2}}} e^{-\frac{r^2}{2a}} r^l
    F\Bigl(1-n,l+\frac{3}{2};\frac{r^2}{a}\Bigr) Y^l_m(\hat{\vec{r}})
  \end{split}
  \label{eq:oscillatorwavefunction}
\end{equation}
have the unique feature of allowing the separation of center-of-mass and
internal motion in a two-body product state. This property can be used in the
calculation of interaction matrix elements with the help of the Talmi
transformation.

\subsection{Talmi coefficients}

The Talmi coefficients \cite{sitenko:nucleus,irvine:nuclearstructure}
provide the transformation of a product of two single-particle
oscillator wave functions to the product of a function of the
center-of-mass motion $\vec{R}=\frac{1}{2}(\vec{r}_{1}+\vec{r}_2)$ and
a function of the relative motion depending on the relative position
vector $\vec{r} = \vec{r}_{1} - \vec{r}_{2}$
\begin{equation}
  \phi^{a}_{n_1 l_1 m_1}(\vec{r}_1) \phi^{a}_{n_2 l_2 m_2}(\vec{r}_2) =
  \sum_{NLMnlm} \talmi{n_1 l_1 m_1}{n_2 l_2 m_2}{NLM}{nlm}
  \phi^{a/2}_{NLM}(\vec{R}) \, \phi^{2a}_{nlm}(\vec{r}) \eqdot
  \label{eq:talmicoefficient}
\end{equation}
We calculate the Talmi coefficient by explicitly performing the
integrals over the oscillator wave functions.

With the help of the Talmi transformation matrix elements of operators
depending only on the relative motion can be calculated as
\begin{multline}
  \matrixe{n_1 l_1 m_1; n_2 l_2 m_2}{\op{v}}{n'_1 l'_1 m'_1; n'_2 l'_2
    m' _2} = \\
  \sum_{\substack{NLM\\nlmn'l'm'}} \talmi{n_1 l_1 m_1}{n_2 l_2 m_2}{NLM}{nlm}
  \talmi{n'_1 l'_1 m'_1}{n'_2 l'_2 m'_2}{NLM}{n'l'm'}
  \matrixe{nlm}{\op{v}}{n'l'm'} \eqcomma
  \label{eq:homatrixelementl}
\end{multline}
with the relative wave functions
\begin{equation}
  \braket{\vec{r}}{nlm} = \phi^{2a}_{nlm}(\vec{r}) \eqdot
\end{equation}

Including the spin $\chi$ and isospin $\xi$ degrees of freedom we get
for scalar and isoscalar operators the following result
\begin{multline}
  \matrixe{n_1 l_1 m_1 \chi_1 \xi_1; n_2 l_2 m_2 \chi_2
    \xi_2}{\op{v}}{n'_1 l'_1 m'_1 \chi'_1 \xi'_1; n'_2 l'_2
    m'_2 \chi'_2 \xi'_2} = \\
  \begin{aligned}
  & \sum_{\substack{n n'\\ j m l l' s\\ t m_t}} \sum_{\substack{m_l m'_l\\m_s m'_s}} \sum_{NLM}
  \talmi{n_1 l_1 m_1}{n_2 l_2 m_2}{NLM}{nlm_l}
  \talmi{n'_1 l'_1 m'_1}{n'_2 l'_2 m'_2}{NLM}{n'l'm'_l}
  \cg{\half}{\xi_1}{\half}{\xi_2}{t}{m_t}
  \cg{\half}{\xi'_1}{\half}{\xi'_2}{t}{m_t} \\
  & \times \cg{\half}{\chi_1}{\half}{\chi_2}{s}{m_s}
  \cg{\half}{\chi'_1}{\half}{\chi'_2}{s}{m'_s}
  \cg{l}{m_l}{s}{m_s}{j}{m}
  \cg{l'}{m'_l}{s}{m'_s}{j}{m}
    \matrixe{n(ls)j,t}{\op{v}}{n'(l's)j,t} \eqdot
  \end{aligned}
  \label{eq:homatrixelementj}
\end{multline}

Using this formula the matrix elements of arbitrary shell-model
configurations can be calculated.

\subsection{Doubly-magic nuclei}
\label{app:doublymagiccoefficients}

For closed-shell nuclei whose ground states are described in the shell
model by completely occupied shells we get the following results.

Given are the expectation values for a two-body operator $\op{A}$ that
is defined in the four spin-isospin channels
\begin{equation}
  \op{a} = \sum_{ST} \op{a}_{ST} \; \op{\Pi}_{ST} \eqdot
\end{equation}

In the case of $\Hefour$ all nucleons are in the $1s$ one-particle
state. There are only contributions from the even channels in the
$\Hefour$ expectation value (the isospin labels have been omitted in
the two-body states $\ket{n(LS)J}$ for brevity)
\begin{equation}
  \matrixe{\Hefour}{\op{A}}{\Hefour} =
  3 \matrixe{1(00)0}{\op{a}_{01}}{1(00)0} +
  3 \matrixe{1(01)1}{\op{a}_{10}}{1(01)1} \eqdot
  \label{eq:hefourexpect}
\end{equation}

In $\Osixteen$ the $1s$ and $1p$ shells are fully
occupied. We have contributions from the odd channels and
contributions with higher relative angular momentum in the even channels
\begin{multline}
  \matrixe{\Osixteen}{\op{A}}{\Osixteen} = \\
  \begin{aligned}
  & 6 \matrixe{1(10)1}{\op{a}_{00}}{1(10)1} \\
  & + 21 \matrixe{1(00)0}{\op{a}_{01}}{1(00)0} +
  \frac{3}{2} \matrixe{2(00)0}{\op{a}_{01}}{2(00)2}+
  \frac{15}{2} \matrixe{1(20)2}{\op{a}_{01}}{1(20)2} \\
  & + 21 \matrixe{1(01)1}{\op{a}_{10}}{1(01)1} +
  \frac{3}{2} \matrixe{2(01)1}{\op{a}_{10}}{2(01)1} \\
  & + \frac{3}{2} \matrixe{1(21)1}{\op{a}_{10}}{1(21)1} +
  \frac{5}{2} \matrixe{1(21)2}{\op{a}_{10}}{1(21)2} +
  \frac{7}{2} \matrixe{1(21)3}{\op{a}_{10}}{1(21)3} \\
  & + 6 \matrixe{1(11)0}{\op{a}_{11}}{1(11)0} +
  18 \matrixe{1(11)1}{\op{a}_{11}}{1(11)1} +
  30 \matrixe{1(11)2}{\op{a}_{11}}{1(11)2} \eqdot
  \end{aligned}
  \label{eq:osixteenexpect}
\end{multline}

In $\Cafourty$ the $1s$, $1p$, $2s$ and $1d$ shells are fully
fully occupied and we obtain
\begin{multline}
  \matrixe{\Cafourty}{\op{A}}{\Cafourty} = \\
  \begin{aligned}
    & 30 \matrixe{1(10)1}{\op{a}_{00}}{1(10)1} +
    \frac{9}{2} \matrixe{2(10)1}{\op{a}_{00}}{2(10)1} +
    \frac{21}{2} \matrixe{1(30)3}{\op{a}_{00}}{1(30)3} \\
    & + \frac{555}{8} \matrixe{1(00)0}{\op{a}_{01}}{1(00)0} +
    \frac{105}{8} \matrixe{2(00)0}{\op{a}_{01}}{2(00)0} +
    \frac{9}{8} \matrixe{3(00)}{\op{a}_{01}}{3(00)0} \\
    & + \frac{525}{8} \matrixe{1(20)2}{\op{a}_{01}}{1(20)2} +
    \frac{45}{8} \matrixe{2(20)2}{\op{a}_{01}}{2(20)2} +
    \frac{81}{8} \matrixe{1(40)4}{\op{a}_{01}}{1(40)4} \\
    & + \frac{555}{8} \matrixe{1(01)1}{\op{a}_{10}}{1(01)1} +
    \frac{105}{8} \matrixe{2(01)1}{\op{a}_{10}}{2(01)1} +
    \frac{9}{8} \matrixe{3(01)1}{\op{a}_{10}}{3(01)1} \\
    & + \frac{105}{8} \matrixe{1(21)1}{\op{a}_{10}}{1(21)1} +
    \frac{175}{8} \matrixe{1(21)2}{\op{a}_{10}}{1(21)2} +
    \frac{245}{8} \matrixe{1(21)3}{\op{a}_{10}}{1(21)3} \\
    & + \frac{9}{8} \matrixe{2(21)1}{\op{a}_{10}}{2(21)1} +
    \frac{15}{8} \matrixe{2(21)2}{\op{a}_{10}}{2(21)2} +
    \frac{21}{8} \matrixe{2(21)3}{\op{a}_{10}}{2(21)3} \\
    & + \frac{21}{8} \matrixe{1(41)3}{\op{a}_{10}}{1(41)3} +
    \frac{27}{8} \matrixe{1(41)4}{\op{a}_{10}}{1(41)4} +
    \frac{33}{8} \matrixe{1(41)5}{\op{a}_{10}}{1(41)5} \\
    & + 30 \matrixe{1(11)0}{\op{a}_{11}}{1(11)0} +
    90 \matrixe{1(11)1}{\op{a}_{11}}{1(11)1} +
    150 \matrixe{1(11)2}{\op{a}_{11}}{1(11)2} \\
    & + \frac{9}{2} \matrixe{2(11)0}{\op{a}_{11}}{2(11)0} +
    \frac{27}{2} \matrixe{2(11)1}{\op{a}_{11}}{2(11)1} +
    \frac{45}{2} \matrixe{2(11)2}{\op{a}_{11}}{2(11)2} \\
    & + \frac{49}{2} \matrixe{1(31)2}{\op{a}_{11}}{1(31)2} +
    \frac{63}{2} \matrixe{1(31)3}{\op{a}_{11}}{1(31)3} +
    \frac{77}{2} \matrixe{1(31)4}{\op{a}_{11}}{1(31)4} \eqdot
  \end{aligned}
  \label{eq:cafourtyexpect}
\end{multline}

\subsection{Explicit operator expectation values}

For some operators the matrix elements in the
harmonic oscillator basis can be calculated analytically.

\subsubsection{Intrinsic kinetic energy (uncorrelated)}

The expectation value of the uncorrelated intrinsic kinetic energy
$\expect{\op{T}-\op{T}_{\icm}}$ is given by
\begin{equation}
  \Hefour: \: \frac{1}{m} \frac{9}{4a} \qquad
  \Osixteen: \: \frac{1}{m} \frac{69}{4a} \qquad
  \Cafourty: \: \frac{1}{m} \frac{237}{4a} \eqdot
  \label{eq:tintrshell}
\end{equation}

\subsubsection{Coulomb interaction (uncorrelated)}

If we neglect the effect of the central correlations on the
expectation values of the Coulomb interaction we obtain
\begin{equation}
  \Hefour: \: 2 \frac{e^2}{\sqrt{2\pi a}} \qquad
  \Osixteen: \: \frac{83}{2} \frac{e^2}{\sqrt{2\pi a}} \qquad
  \Cafourty: \: \frac{7905}{32} \frac{e^2}{\sqrt{2\pi a}} \eqdot
\end{equation}

\subsubsection{Radius $r_{\mathrm{rms}}$ (uncorrelated)}
\label{app:magicradii}

The uncorrelated $r_{\mathrm{rms}}$ radius of the doubly-magic nuclei
(or the one-body contribution of the correlated radius) is given by
\begin{equation}
  \Hefour: \: \frac{3}{2\sqrt{2}} \sqrt{a} \qquad
  \Osixteen: \: \frac{\sqrt{69}}{4\sqrt{2}} \sqrt{a} \qquad
  \Cafourty: \: \frac{\sqrt{237}}{4\sqrt{5}} \sqrt{a} \eqdot
  \label{eq:rrmsshell}
\end{equation}

\section{Correlator parameters}
\label{app:correlators}

\subsection{Correlator parameterizations}

The correlation functions $\Rp(r)$ and $\vartheta(r)$ are used in
parameterized form. For short distances $r$ they resemble the ones obtained
by mapping the constant trial wave function onto the lowest eigenstate in
the two-body system. At larger distances they fall off by either an
exponential or a double exponential.

In this work we employ the following parameterizations for the radial
correlation functions:
\begin{equation}
  \Rp(r) = r + \alpha' \left( \frac{r}{\beta} \right)^{\eta}
  \exp \biggl\{ - \exp \biggl\{ \frac{r}{\beta} \biggr\} \biggr\} \eqcomma
  \label{eq:RpBpara}
\end{equation}
\begin{equation}
  \Rp(r) = r + \alpha'  \biggl( 1 - \exp \biggl\{ - \biggl(
  \frac{r}{\gamma} \biggr)^{\eta} \biggr\} \biggr) \exp \biggl\{ -
  \frac{r}{\beta} \biggr\}
  \label{eq:RpEpara}
\end{equation}
\begin{equation}
  \Rp(r) = r + \alpha'  \biggl( 1 - \exp \biggl\{ - \biggl(
  \frac{r}{\gamma} \biggr)^{\eta} \biggr\} \biggr) \exp \biggl\{ - \exp
  \biggl\{ \frac{r}{\beta} \biggr\} \biggr\} \eqdot
  \label{eq:RpFpara}
\end{equation}

For the tensor correlation functions $\vartheta(r)$ the parameterizations
\begin{equation}
  \vartheta(r) = \alpha'  \left( 1-\exp
    \biggl\{- \biggl(\frac{r}{\gamma}\biggr)^{\eta} \biggr\} \right) \exp \biggl\{ -
  \frac{r}{\beta} \biggr\}  \eqcomma
  \label{eq:thetaEpara}
\end{equation}
\begin{equation}
  \vartheta(r) = \alpha'  \left( 1-\exp
    \biggl\{- \biggl(\frac{r}{\gamma}\biggr)^{\eta} \biggr\}\right) \exp \biggl\{ - \exp \biggl\{
  \frac{r}{\beta} \biggr\} \biggr\}
  \label{eq:thetaFpara}
\end{equation}
are used.

\subsection{Bonn-A potential}
\label{app:bonnacorrelators}

\begin{center}
\begin{tabular}{@{}lc|c|cccc@{}} \toprule
\multicolumn{7}{c}{Radial correlation function $\Rp(r)$}\\ \midrule
  correlator & ST & type & $\alpha'$ [fm] & $\beta$ [fm] & $\gamma$ [fm] & $\eta$ \\ \midrule
  $\mathit{min}$ & 01 & (\ref{eq:RpBpara}) & 1.199 & 0.8082 & & 0.7340 \\
  $\mathit{min}$ & 10 & (\ref{eq:RpBpara}) & 1.132 & 0.7788 & & 0.8481 \\
  $\mathit{min}$ & 00 & (\ref{eq:RpFpara}) & 250232 & 1.406 & 1000 & 2 \\
  $\mathit{min}$ & 11 & (\ref{eq:RpBpara}) & 0.6581 & 1.198 & & 0.7981 \\ \midrule
  $\mathit{min} - \Hefour$ & 01 & (\ref{eq:RpBpara}) & 1.344 & 0.8992 & & 0.6990 \\
  $\mathit{min} - \Hefour$ & 10 & (\ref{eq:RpBpara}) & 1.256 & 0.8526 & & 0.8107 \\
  \bottomrule
\end{tabular}
\end{center}

\begin{center}
\begin{tabular}{@{}lc|c|cccc@{}} \toprule
\multicolumn{7}{c}{Tensor correlation function $\vartheta (r)$}\\ \midrule
  correlator & ST & type & $\alpha'$ & $\beta$ [fm]& $\gamma$ [fm] & $\eta$ \\ \midrule
  $\mathit{min}^{\alpha}$ & 10 & (\ref{eq:thetaFpara}) & 10.2655 & 1.287 & 4.994 & 2 \\
  $\mathit{min}^{\beta}$  & 10 & (\ref{eq:thetaFpara}) & 0.60234 & 1.834 & 1.215 & 2 \\
  $\mathit{min}^{\gamma}$ & 10 & (\ref{eq:thetaFpara}) & 0.35938 & 2.745 & 0.9249 & 2 \\
  $\mathit{min}$          & 11 & (\ref{eq:thetaEpara}) & -0.024735 & 1.699 & 1.197 & 3 \\ \midrule
  $\mathit{min}^{\alpha} - \Hefour$ & 10 & (\ref{eq:thetaFpara}) & 4677.8 & 1.241 & 100 & 2 \\
  $\mathit{min}^{\beta} - \Hefour$  & 10 & (\ref{eq:thetaFpara}) & 3979.5 & 1.472 & 100 &  2 \\
  $\mathit{min}^{\gamma} - \Hefour$ & 10 & (\ref{eq:thetaFpara}) & 0.47077 & 2.551 & 1.109 & 2\\
  $\mathit{min} - \Hefour$          & 10 & (\ref{eq:thetaEpara}) & 0.08402& 7.201 & 0.7413 & 2 \\
  \bottomrule
\end{tabular}
\end{center}

\subsection{Argonne V8' and Argonne V18 potential}
\label{app:argonnecorrelators}

\begin{center}
\begin{tabular}{@{}lc|c|cccc@{}} \toprule
\multicolumn{7}{c}{Radial correlation function $\Rp(r)$}\\ \midrule
  correlator & channel & type & $\alpha'$ [fm]& $\beta$ [fm]& $\gamma$ [fm]& $\eta$ \\ \midrule
  $\mathit{min}$ & 01 & (\ref{eq:RpBpara}) & 1.379 & 0.8854 &        & 0.3723 \\
  $\mathit{min}$ & 10 & (\ref{eq:RpBpara}) & 1.296 & 0.8488 &        & 0.4187 \\
  $\mathit{min}$ & 00 & (\ref{eq:RpFpara}) & 0.76554 & 1.272 & 0.4243 & 1 \\
  $\mathit{min}$ & 11 & (\ref{eq:RpFpara}) & 0.57947 & 1.3736 &  0.1868 & 1 \\ \midrule
  $\mathit{min} - \Hefour$ & 01 & (\ref{eq:RpBpara}) & 1.380 & 0.9805 & & 0.3362 \\
  $\mathit{min} - \Hefour$ & 10 & (\ref{eq:RpBpara}) & 1.372 & 0.9072 & & 0.4190 \\
  \bottomrule
\end{tabular}
\end{center}

\begin{center}
\begin{tabular}{@{}lc|c|cccc@{}} \toprule
\multicolumn{7}{c}{Tensor correlation function $\vartheta (r)$}\\ \midrule
  correlator & channel & type & $\alpha'$ & $\beta$ [fm]& $\gamma$ [fm]& $\eta$ \\ \midrule
  $\mathit{min}^{\alpha}$ & 10 & (\ref{eq:thetaFpara}) & 530.38 & 1.298 & 1000.0 & 1 \\
  $\mathit{min}^{\beta}$  & 10 & (\ref{eq:thetaFpara}) & 0.92094 & 1.717 & 1.590 & 1 \\
  $\mathit{min}^{\gamma}$ & 10 & (\ref{eq:thetaFpara}) & 0.383555 & 2.665 & 0.4879 & 1 \\
  $\mathit{min}$          & 11 & (\ref{eq:thetaEpara}) & -0.023686 & 1.685 & 0.8646 & 1 \\ \midrule
  $\mathit{min}^{\alpha} - \Hefour$ & 10 & (\ref{eq:thetaFpara}) & 59.026 & 1.266 & 100.0 & 1 \\
  $\mathit{min}^{\beta} - \Hefour$  & 10 & (\ref{eq:thetaFpara}) & 54.817 & 1.554 & 105.42 &  1 \\
  $\mathit{min}^{\gamma} - \Hefour$ & 10 & (\ref{eq:thetaFpara}) & 0.54833 & 2.446 & 0.883 &  1\\
  $\mathit{min} - \Hefour$          & 10 & (\ref{eq:thetaEpara}) & 0.10965 & 4.017 & 0.3034 & 1 \\
  \bottomrule
\end{tabular}
\end{center}
\end{appendix}

\bibliography{tensorcorr}

\begin{thebibliography}{10}

\bibitem{entem01}
D. Entem and R. Machleidt,
\newblock Phys. Lett. B524 (2001) 93, nucl-th/0108057.

\bibitem{machleidt89}
R. Machleidt,
\newblock Adv. Nucl. Phys. 19 (1989) 189.

\bibitem{machleidt01}
R. Machleidt,
\newblock Phys. Rev. C63 (2001) 024001.

\bibitem{wiringa84}
R. Wiringa, R. Smith and T. Ainsworth,
\newblock Phys. Rev. C29 (1984) 1207.

\bibitem{wiringa95}
R. Wiringa, V. Stoks and R. Schiavilla,
\newblock Phys. Rev. C51 (1995) 38.

\bibitem{pieper01mp}
S. Pieper and R. Wiringa,
\newblock Ann. Rev. Nucl. Part. Sci. 51 (2001) 53, nucl-th/0103005.

\bibitem{barrett00}
P. Navr{\'a}til, J. Vary and B. Barrett,
\newblock Phys. Rev C62 (2000) 054311.

\bibitem{fabrocini98}
A. Fabrocini et~al.,
\newblock Phys. Rev. C57 (1998) 1668.

\bibitem{fabrocini00}
A. Fabrocini, F.A. de~Saavedra and G. Co',
\newblock Phys. Rev. C61 (2000) 044302.

\bibitem{fmd00}
H. Feldmeier and J. Schnack,
\newblock Rev. Mod. Phys. 72 (2000) 655.

\bibitem{neff:diss}
T. Neff,
\newblock {S}hort-{R}anged {C}entral and {T}ensor {C}orrelations in {N}uclear
  {M}any-{B}ody {S}ystems,
\newblock PhD thesis, TU Darmstadt, 2002.

\bibitem{ucom98}
H. Feldmeier et~al.,
\newblock Nucl. Phys. A632 (1998) 61.

\bibitem{kuo01}
S. Bogner et~al.,
\newblock (2001), nucl-th/0108041.

\bibitem{pieper92}
S. Pieper, R. Wiringa and V. Pandharipande,
\newblock Phys. Rev. C46 (1992) 1741.

\bibitem{benhar94}
O. Benhar et~al.,
\newblock Nuc. Phys. A579 (1994) 493.

\bibitem{pieper01}
S. Pieper et~al.,
\newblock Phys. Rev. C64 (2001) 014001.

\bibitem{weinberg90}
S. Weinberg,
\newblock Phys. Lett. B251 (1990) 288.

\bibitem{weinberg91}
S. Weinberg,
\newblock Nuc. Phys. B363 (1991) 3.

\bibitem{pudliner97}
B. Pudliner et~al.,
\newblock Phys. Rev. C56 (1997) 1720.

\bibitem{wiringa00}
R.B. Wiringa et~al.,
\newblock Phys. Rev. C62 (2000) 014001.

\bibitem{roth:doktor}
R. Roth,
\newblock {E}ffektive {W}echselwirkungen f{\"u}r {Q}uantenfl{\"u}ssigkeiten und
  {Q}uantengase: {K}ernmaterie, fl{\"u}ssiges {H}elium und ultrakalte atomare
  {F}ermigase,
\newblock PhD thesis, TU Darmstadt, 2000.

\bibitem{brown71}
G. Brown,
\newblock {U}nified {T}heory of {N}uclear {M}odels and {F}orces, third ed.
  (North-Holland, 1971).

\bibitem{forest96}
J. Forest et~al.,
\newblock Phys. Rev. C54 (1996) 646.

\bibitem{kamada01}
H. Kamada et~al.,
\newblock Phys. Rev. C64 (2001) 044001.

\bibitem{wiringa92}
R. Wiringa,
\newblock Nucl. Phys. A543 (1992) 199.

\bibitem{muether00}
H. M{\"u}ther and A. Polls,
\newblock Prog. Part. Nucl. Phys 45 (2000) 243.

\bibitem{heisenberg99}
J.H. Heisenberg and B. Mihaila,
\newblock Phys. Rev. C59 (1999) 1440.

\bibitem{Varshalovich88}
D.A. Varshalovich, A.N. Moskalev and V.K. Khersonskii,
\newblock {Q}uantum {T}heory of {A}ngular {M}omentum (World Scientific, 1988).

\bibitem{sitenko:nucleus}
A. Sitenko and V. Tartakovskii,
\newblock {L}ectures on the {T}heory of the {N}ucleus (Pergamon Press, 1975).

\bibitem{irvine:nuclearstructure}
J. Irvine,
\newblock {N}uclear {S}tructure {T}heory, First ed. (Pergamon Press, 1972).

\end{thebibliography}
\bibliographystyle{h-elsevier}

\end{document}